\documentclass[%
reprint,
superscriptaddress,
preprintnumbers,
nofootinbib,
amsmath,amssymb,
aps%
]{revtex4-2}

\usepackage{geometry}
\usepackage[colorlinks]{hyperref}
\usepackage{graphicx}% Include figure files
\usepackage{dcolumn}% Align table columns on decimal point
\usepackage{bm}% bold math
\usepackage{color}
\usepackage[dvipsnames]{xcolor}
\usepackage{placeins}
\usepackage{slashed}
\usepackage{multirow}
\usepackage{subfigure}
\usepackage{tabularx}
\usepackage{mathtools}
\usepackage{floatrow}
\usepackage{blindtext}
\usepackage{soul}
\usepackage{mathrsfs} 
\usepackage{comment}
\usepackage{nicefrac}
\usepackage[normalem]{ulem}
\definecolor{red}{rgb}{0.9, 0,0}
\definecolor{cerulean}{rgb}{0., 0.62,0.9}
\definecolor{navy}{rgb}{0.05, 0.05,0.8}
\definecolor{orange}{rgb}{0.1, 0.9,0.08}

\renewcommand{\eqref}[1]{Eq.~\ref{#1}}

\def\mc#1{{\bf  \textcolor{magenta}{[MC: {#1}]}}}

\newcommand{\mev}{\text{ MeV}}
\newcommand{\gev}{\text{ GeV}}
\newcommand{\tev}{\text{ TeV}}
\newcommand{\mdm}{M_\text{DM}}
\newcommand{\hc}{\text{h.c.}}
\newcommand{\chidm}{\chi_\text{DM}}
\newcommand{\vrel}{v_\text{rel}}
\newcommand{\permil}{\textperthousand}
\newcommand{\Leff}{\Lambda_{\text{UV}}}

\newcommand{\omix}{\mathcal{O}_{0}}
\newcommand{\Qbar}{ {Q_M} }
\newcommand{\ochgd}{\mathcal{O}_{+}}
\newcommand{\pbs}{P_\text{BS}}
\newcommand{\alphaeff}{\alpha_\text{eff}}
\newcommand{\taulcp}{\tau_{\mathrm{LCP}}}
\newcommand{\shat}{\hat{S}}
\newcommand{\that}{\hat{T}}
\newcommand{\yhat}{\hat{Y}}
\newcommand{\what}{\hat{W}}

\begin{document}
\title{The last Complex WIMPs standing}

\author{Salvatore Bottaro}
\affiliation{Scuola Normale Superiore, Piazza dei Cavalieri 7, 56126 Pisa, Italy}
\affiliation{INFN, Sezione di Pisa, Largo Bruno Pontecorvo 3, I-56127 Pisa, Italy}
\author{Dario Buttazzo}
\affiliation{INFN, Sezione di Pisa, Largo Bruno Pontecorvo 3, I-56127 Pisa, Italy}
\author{Marco Costa}
\affiliation{Scuola Normale Superiore, Piazza dei Cavalieri 7, 56126 Pisa, Italy} \affiliation{INFN, Sezione di Pisa, Largo Bruno Pontecorvo 3, I-56127 Pisa, Italy}
\author{Roberto Franceschini}
\affiliation{Universit\`a degli Studi and INFN Roma Tre, Via della Vasca Navale 84, I-00146, Rome}
\author{Paolo Panci}
\affiliation{Dipartimento di Fisica E. Fermi, Universit\`a di Pisa, Largo B. Pontecorvo 3, I-56127 Pisa, Italy}
\affiliation{INFN, Sezione di Pisa, Largo Bruno Pontecorvo 3, I-56127 Pisa, Italy}
\author{Diego Redigolo}
\affiliation{CERN, Theoretical Physics Department, Geneva, Switzerland.}
\affiliation{INFN, Sezione di Firenze Via G. Sansone 1, 50019 Sesto Fiorentino, Italy}
\author{Ludovico Vittorio}
\affiliation{Scuola Normale Superiore, Piazza dei Cavalieri 7, 56126 Pisa, Italy}
\affiliation{INFN, Sezione di Pisa, Largo Bruno Pontecorvo 3, I-56127 Pisa, Italy}

\preprint{CERN-TH-2022-080}

\date{\today}

\begin{abstract}
{We continue the study of weakly interacting massive particles (WIMP) started in~\cite{Bottaro:2021snn}, focusing on a single complex electroweak $n$-plet with non-zero hypercharge added to the Standard Model. The minimal  splitting between the Dark Matter and its electroweak neutral partner required to circumvent direct detection constraints allows only multiplets with hypercharge smaller or equal to 1. We compute for the first time \emph{all the calculable WIMP masses} up to the largest multiplet allowed by perturbative unitarity. For the minimal allowed splitting, most of these multiplets can be fully probed at future large-exposure direct detection experiments, with the notable exception of the doublet with hypercharge 1/2. We show how a future muon collider can fully explore the parameter space of the complex doublet combining missing mass, displaced track and long-lived track searches. In the same spirit, we study how a future muon collider can probe the parameter space of complex WIMPs in regions where the direct detection cross section drops below the neutrino floor. Finally, we comment on how precision observables can provide additional constraints on complex WIMPs.}
\end{abstract}

\maketitle

\section{Introduction}
Despite the intense experimental activity in the past 50 years, the possibility that Dark Matter (DM) is a new weakly interacting massive particle (WIMP) remains open today. Within this framework, the most minimal and predictive possibility is that the DM is the lightest neutral component of a single
electroweak (EW) multiplet. In this simplified setup, the experimentally viable DM candidates can be fully classified and their thermal masses predicted once the relic abundance from freeze-out is required to match the observed DM density today $\Omega_{\text{DM}}h^2= 0.11933 \pm 0.00091$~\cite{Planck:2018vyg}.    

Experimentally, direct detection (DD) experiments robustly excluded the possibility that DM has tree-level $Z$-mediated interactions with the nuclei. This experimental milestone left only two possible scenarios for EW WIMPs wide open: i) WIMPs with zero hypercharge ($Y=0$), interacting with nucleons only trough loop-induced interactions, ii) inelastic WIMPs with non-zero hypercharge ($Y\neq0$), where the elastic scattering between DM and the nucleons is forbidden kinematically. These two possibilites have been historically incarnated by the Wino and the Higgsino DM scenarios in the Minimal Supersymmetric Standard Model, but a full classification of the viable EW WIMPs from an effective field theory standpoint is still lacking.  

Attempting to fill up this gap in the literature, we studied EW WIMPs with $Y=0$ which are naturally embedded in real EW representations in Ref.~\cite{Bottaro:2021snn}. Here we focus instead on inelastic WIMPs which are embedded in complex EW representations with $Y\neq0$.\footnote{Complex EW representations include also multiplets with $Y=0$. Some of these were discussed in Ref.~\cite{DelNobile:2015bqo}, their phenomenology is similar to the one discussed in Ref.~\cite{Bottaro:2021snn}. For completeness we give the full predictions for the thermal masses up to the limit of perturbative unitarity in Appendix~\ref{table:summarymilli}.}

Similarly to our previous study, our goal here is to precisely determine and fully classify all the \emph{calculable} freeze-out predictions for complex inelastic WIMPs, including Sommerfeld enhancement (SE) and Bound State Formation (BSF). The final purpose is to use these predictions to establish if and how the future experimental program will be able to test all the viable WIMP scenarios. In particular, we will show how future large exposure direct detection experiments together with a future high energy muon collider will probe complementary portions of the complex WIMP parameter space being potentially able to fully test the theoretically allowed WIMP scenarios. 

\begin{table*}[t]
\begin{center}
\begin{tabular}{ c | c |  c  c  c c c c}
    
            DM spin             & $n_Y$ & $\mdm$ (TeV) & $\Lambda_{\text{Landau}}/\mdm$ & $(\sigma v)^{J=0}_{\text{tot}}/(\sigma v)^{J=0}_{\text{max}}$ & $\delta m_0$ [MeV]& $\Lambda_{\text{UV}}^{\max}/\mdm$  & $\delta m_{\Qbar}$ [MeV]  \\ \hline 
            \multirow{9}{*}{Dirac fermion}& $2_{1/2}$ & $1.08\pm 0.02$ &  $> M_{\text{Pl}}$&- & 0.22 - $2 \times 10^4$& $10^7$& 4.8 -   $10^4$\\
     &$3_1$ & $2.85\pm 0.14$ &  $> M_{\text{Pl}}$&- & 0.22 - 40& 60& 312 - 1.6 $\times 10^4$\\
                                                          & $4_{1/2}$ & $4.8\pm0.3$ & $\simeq M_{\text{Pl}}$&0.001& 0.21 - $3\times 10^4$& $5\times 10^6$& 20 - 1.9 $\times 10^4$ \\
                                                          & $5_1$ & $9.9\pm0.7$ & $3\times 10^6$&0.003& 0.21 - 3& 25 & $10^3 - 2 \times 10^3$ \\
                                                          & $6_{1/2}$ & $31.8\pm 5.2$ & $2\times 10^4$& 0.01 &0.5 - $2\times 10^4$ &  $4\times 10^5$& 100 - 2 $\times~ 10^4$\\
                                                           &$8_{1/2}$ & $82\pm 8$ & $15$& 0.05 &0.84 - $10^4$& $10^5$ & 440 - $10^4$\\
                                                           &$10_{1/2}$ & $158\pm 12$ & $3$& 0.16 &1.2 - 8 $\times~ 10^3$& 6 $\times 10^4$ & 1.1  $\times~ 10^3$ - 9 $\times~ 10^3$\\
                                                           &$12_{1/2}$ & $253\pm 20$ & $2$& 0.45 &1.6 - 6 $\times~ 10^3$& 4 $\times 10^4$ & 2.3 $\times~ 10^3$ - 7 $\times~ 10^3$\\\hline
   \multirow{9}{*}{Complex scalar} & $2_{1/2}$ & $0.58\pm 0.01$ &  $> M_{\text{Pl}}$& - &  4.9 - 1.4 $\times~ 10^4$ & -& 4.2 - 7 $\times~ 10^3$\\
   & $3_1$ & $2.1\pm 0.1$ &  $> M_{\text{Pl}}$& - &  3.7 - 500 & 120 & 75 -  1.3 $\times 10^4$\\
                                               &  $4_{1/2}$ & $4.98\pm0.25$ & $> M_{\text{Pl}}$ & 0.001 & 4.9 - 3 $\times~ 10^4$& -& 17 - 2 $\times 10^4$\\
                                                &  $5_1$ & $11.5\pm 0.8$ & $> M_{\text{Pl}}$ & 0.004 & 3.7 - 10& 20 &650 - 3 $\times 10^3$\\
                                               &  $6_{1/2}$ & $32.7\pm 5.3$ &$\simeq 6\times 10^{13}$& 0.01 & 4.9 - 8$\times 10^4$& - & 50 - 5 $\times~ 10^4$\\
                                               &  $8_{1/2}$ & $84\pm 8$ & $2\times 10^4$& 0.05 &$4.9$ - 6 $\times10^4$& - & 150 - 6 $\times~ 10^4$ \\
                                               &  $10_{1/2}$ & $162 \pm 13$ & $20 $& 0.16 & 4.9 - $4 \times 10^4$ & -& 430 - 4 $\times~ 10^4$\\
                                               &  $12_{1/2}$ & $263 \pm 22$ & $4 $& 0.4 & 4.9 - $3 \times 10^4$ & -& $10^3$ -  $3\times 10^4$ \\
                                                        \end{tabular}
 \caption{Thermal masses of complex WIMPs with $Y\neq 0$, obtained including Sommerfeld enhancement and BSF. The upper bound on $n$ for even multiplets comes from the perturbative unitarity bound, as can be seen from the $(\sigma v)^{J=0}_{\text{tot}}/(\sigma v)^{J=0}_{\text{max}}$, where $(\sigma v)^{J=0}_{\text{max}}$ is the maximal allowed annihilation cross section~\cite{Griest:1989wd}. The loss of perturbativity is also signaled by the Landau pole $\Lambda_\text{Landau}$ progressively approaching the DM mass. The upper bound on odd $n$ with $Y=1$ comes from the perturbativity of the higher dimensional operators generating 
 $\delta m_0$ . For multiplets with $n>5$ the largest UV cutoff $\Lambda_{\text{UV}}^{\text{max}}$ required to generate the minimal  viable splitting is smaller than $10\mdm$. For each candidate we provide the allowed range for the mass splittings. The lower limit on $\delta m_0$ comes from strongest bound between direct detection and BBN as shown in Fig.~\ref{fig:bbn}. The upper bound from the most stringent condition between DD constraint from PandaX-4T \cite{PandaX:2021osp} and the perturbativity of the coupling of $\omix$ in \eqref{eq:ren_lag}. Similarly, the lower limit on $\delta m_{\Qbar}$ comes from the BBN bound on the charged state decay rate, while the upper limit from the strongest limit between DD and the perturbativity of the coupling of $\ochgd$ in \eqref{eq:ren_lag}.} 
 \label{tab:TableI}
\end{center}
\end{table*}

Our results are summarized in Table~\ref{tab:TableI}, whose logic can be explained as follows. Once the DM mass is fixed from the freeze-out predictions, the phenomenology of complex WIMPs depends essentially on two parameters: i) the ``inelastic'' splitting  between the-next-to lightest neutral component and the DM; ii) the ``charged'' splitting between the charged components and the DM. In our setup these splittings are generated by its (non-renormalizable) interactions with the SM Higgs (generated by uspecified UV dynamics).   

The inelastic splitting $\delta m_0$ is bounded from below by DD constraints~\cite{Bramante:2016rdh,Song:2021yar} and BBN constraints on the decay of the next to lightest neutral component. Interestingly, this requrement alone selects a limited number of complex WIMPs: i) scalar and fermionic WIMPs with $Y=1/2$ and even $n$ up to the unitarity bound of the freeze-out annihilation cross section~\cite{Griest:1989wd}; ii) scalar and fermionic WIMPs with $Y=1$ and $n=3,5$. At the same time, the inelastic splitting is bounded from above by DD constraints on Higgs-mediated nuclear recoils. This leaves a finite window for the inelastic splitting of every multiplet which we report in Table~\ref{tab:TableI}. This window will be further probed by large exposure DD experiments such as LZ~\cite{Mount:2017qzi}, Xenon-nT~\cite{XENON:2020kmp}, and ultimately by DARWIN/G3~\cite{Aalbers:2016jon,Aalbers:2022dzr}. 

The natural value of the charged splitting $\delta m_Q$ is fixed by the radiative EW contributions~\cite{Cheng:1998hc,Feng:1999fu,Gherghetta:1999sw} but (non-renormalizable) interactions with the SM Higgs can induce large deviation from this value. In particular, for all the $n$-plets with non-maximal hypercharge, these interactions are required to make the DM stable. 

The allowed range of the two splittings above controls the hierarchy of the states within the EW multiplet. In this parameter space one can map out the expected signals in a future hypothetical muon collider~\cite{AlAli:2021let}. Depending on the lifetime of the charged states we can have different signatures at colliders: i) long lived charged tracks; ii) disappearing tracks (DT); iii) missing energy accompanied by an EW bosons. While the first two searches rely on the macroscopic decay length of the charged states, the last is directly related to the DM pair production recoiling against one (or more) EW boson. In general, a future muon collider could complement the large exposure DD experiments in probing complex WIMPs in regions of the parameter space where the DD drops below the neutrino floor of xenon experiments~\cite{OHare:2021utq}.

The doublet with $Y=1/2$ ($2_{1/2}$) and the triplet with $Y=1$ ($3_1$) stand out as the only two complex candidates for which the lightest component of the multiplet is guaranteed to be electrically neutral, which is a necessary requirement for DM. It is well known that the $2_{1/2}$ DD cross section lies well below the neutrino floor because of an accidental cancellation between the different contribution to the elastic cross section with the nuclei~\cite{Hisano:2011cs}. We find that a muon collider of $\sqrt{s}=3\text{ TeV}$ would only be able to exclude the fermionic EW doublet with a large luminosity from around  $\mathcal{L}=5\text{ ab}^{-1}$ to $  \mathcal{L}=50 \text{ ab}^{-1}$, depending on the assumed systematics, by a combination of charged tracks, disappearing, tracks and missing mass searches. Somewhat less high luminosity may suffice at  a $\sqrt{s}=6\text{ TeV}$ collider, which can exclude the $2_{1/2}$ with 4 $\text{ ab}^{-1}$ and make a discovery with a more demanding 20~$\text{ab}^{-1}$.\footnote{Note that our results on the DT differs from the ones in Ref.~\cite{Capdevilla:2021fmj}. This discrepancy can be ascribed to their overestimation of the efficiency for DTs with short lifetimes. We thank the authors for correspondence on this. We believe that our projections on the mono-$\gamma$ are stronger than the ones derived on Ref.~\cite{Han:2020uak} because of our  optimization of the kinematical cuts.} 

The DD cross section of the $3_1$ can only be probed with large exposure Xenon experiment like DARWIN, while a muon collider of $\sqrt{s}=10 \text{ TeV}$ would be able to make a discovery at the thermal mass for the fermionic $3_1$ using a luminosity around 10~$\text{ab}^{-1}$, that is considered as a benchmark for currently discussed project for this type of collider~\cite{Palmer_2014,2203.07224v1}. This is clearly a major theoretical motivation for a high energy muon collider which could be in the unique position of testing the few WIMP scenarios whose DD cross section lies well below (or in proximity) of the neutrino floor.

%The discrepancy with the latter is due to the fact that systematic are completely ignored    different thermal masses and a more thorough analysis including systematics\DR{to complete!}}

%In the same spirit, we discuss the  4-plet with $Y=1/2$ for which center of mass energies a future muon collider could complement the large exposure direct detection experiments in probing light complex WIMPs in region of the parameter space where the direct detection drops below the neutrino floor. 

This paper is organized as follows: in Sec.~\ref{sec:WIMP} we summarize the main features of EW WIMP paradigm focusing on fermionic DM and relegating many details to Appendix~\ref{sec:operators} and analogous formulas for scalar DM to Appendix~\ref{sec:scalarWIMP}. This section provides a full explanation on the results of Table~\ref{tab:TableI} while the more technical aspects of the freeze-out computation are given in  Appendix~\ref{app:boundstates}. In Sec.~\ref{sec:minimal splitting} we show how large exposure DD experiments can probe most of the complex WIMPs with the minimal inelastic and charged splittings. In Sec.~\ref{sec:nonminimal splitting} we study the full parameter space of the complex WIMPs, in the charged vs neutral splitting plane. We take as examples the doublet and 4-plet with $Y=1/2$ and the triplet and 5-plet with $Y=1$. In Sec.~\ref{sec:indirect} we summarize the indirect probes of complex WIMPs in electroweak observables and the electron dipole moment (EDM). In Appendix~\ref{app:millichargeWIMPs} we discuss the freeze-out prediction for complex WIMPs with $Y=0$ and in Appendix~\ref{app:collider} we illustrate the collider reach for all the WIMPs.

\section{Which Complex WIMP?}\label{sec:WIMP}

We recall here the logic of our WIMP classification as described in \cite{Bottaro:2021snn} (see also \cite{Cirelli:2005uq, Cirelli:2007xd, Cirelli:2009uv, Hambye:2009pw, DelNobile:2015bqo} for previous work on the subject). Requiring the neutral DM component to be embedded in a representation of the EW group imposes that $Q=T_3+Y$, where $T_3=\text{diag}\left(\frac{n+1}{2}-i\right)$ with $i=1,\dots, n$, and $Y$ is the hypercharge. At this level, we can distinguish two classes of WIMPs: i) real EW representations with $Y=0$ and odd $n$; ii) complex EW representations with arbitrary $n$ and $Y=\pm\left( \frac{n+1}{2}-i\right)$ for $i=1,\dots, n$. Within class ii), we can further distinguish two subclasses of complex WIMPs: a) complex representations $Y=0$ and odd $n$; b) complex representation with $Y\neq0$ with even $n$ (odd $n$) for half-integer $Y$ (integer $Y$). The first subclass is a straightforward generalization of the models analyzed in \cite{Bottaro:2021snn} where the stability of the DM is guaranteed by an unbroken dark fermion number which can be gauged as was first done in Ref.~\cite{DelNobile:2015bqo}. For completeness, we give the freeze-out prediction of these scenarios in Appendix \ref{app:millichargeWIMPs}.

In this paper we focus on complex WIMPs with $Y\neq0$ whose phenomenology differs substantially from the one with $Y=0$. We focus here on the fermionic case and leave the discussion about the scalar WIMPs to the Appendix~\ref{sec:scalarWIMP}. The minimal Lagrangian for a fermionic complex WIMP with $Y\neq 0$ is:
%\begin{widetext}
\begin{align}
&\mathscr{L}_{\text{D}} =\overline{\chi} \left (i\slashed{ D}-M_\chi\right)\chi +\frac{y_0}{\Leff^{4Y-1}}\mathcal{O}_0 + \frac{y_+}{\Leff}\mathcal{O}_{+}+\hc\ ,\notag\\
&\omix=\frac{1}{2(4Y)!}\left(\overline{\chi}(T^a)^{2Y}\chi^c\right)\left[(H^{c\dagger})\frac{\sigma^a}{2} H\right]^{2Y}\ ,\label{eq:ren_lag}\\ 
&\ochgd=-\overline{\chi}T^a\chi H^\dagger\frac{\sigma^a}{2} H\ ,\notag
\end{align}
%\end{widetext}
where $T^a$ is a SU(2)$_L$ generator in the DM representation.  The main difference with respect to real WIMPs is that the renormalizable Lagrangian is no longer sufficient to make the DM model viable. In \eqref{eq:ren_lag} we write only the minimal amount of UV operators required to make the DM model viable. As we discuss in Appendix~\ref{sec:operators} these operators are also unique and possibly accompanied by their axial counterparts. These are obtained from the ones in \eqref{eq:ren_lag} by adding a $\gamma_5$ inside the DM bilinear. We now illustrate the physical consequence of $\mathcal{O}_0$ and $\mathcal{O}_+$ in turn in Sec.~\ref{sec:inelasticsplit} and Sec~\ref{sec:chargedsplit} and derive the implication for complex WIMPs in Sec.~\ref{subsec:neutral_splitting}. We comment on DM stability in Sec.~\ref{sec:stability}.   

\subsection{Inelastic splitting}~\label{sec:inelasticsplit}

The non-renormalizable operator $\omix$ is required to remove the sizeable coupling to the $Z$ boson of the neutral component $\chi_N$ of the EW multiplet
\begin{equation}
\label{eq:Z_vectorial}
\begin{aligned}
    %&\mathcal{L}_S^Z=\frac{ieY}{\sin\theta_W\cos\theta_W}\chi_N^\dagger\partial_\mu\chi_N Z^\mu\\
    &\mathscr{L}_Z=\frac{ieY}{\sin\theta_W\cos\theta_W}\overline{\chi}_N\slashed{Z}\chi_N\ .
\end{aligned}
\end{equation}
This coupling would lead to an elastic cross section with nuclei already excluded by many orders of magnitude by present DD experiments~\cite{Goodman:1984dc}. After the EWSB, $\omix$ induces a mixing between $\chi_N$ and $\chi_N^c$. Replacing the Higgs with its VEV, $(H^{c\dagger})\frac{\sigma^a}{2} H$ is non-zero only if we pick $\sigma^a=\sigma^+$, so that the new (pseudo Dirac) mass term in the Lagrangian reads
\begin{equation}
\begin{aligned}
\label{eq:neutral_splitting}
   &\mathscr{L}_m=M_\chi\overline{\chi}_N\chi_N+\frac{\delta m_0}{4}\left[\overline{\chi}_N\chi_N^c +\overline{\chi^c}_N\chi_N\right],\\
    &\delta m_0 = 4y_0 c_{nY0}\Leff\left(\frac{v}{\sqrt{2}\Leff}\right)^{4Y}\ .
    \end{aligned}
\end{equation}
$c_{nYQ}=\frac{1}{2^{Y+1}(4Y)!}\prod_{j=-Y-|Q|}^{Y-1-|Q|}\sqrt{\frac{1}{2}\left(\frac{n+1}{2}+j\right)\left(\frac{n-1}{2}-j\right)}$ contains the normalization of $\omix$ and the matrix elements of the generators. The mass eigenstates are Majorana fermions, $\chi_0$ and $\chidm$, with masses $M_0=M_\chi+\delta m_0/2$ and $\mdm=M_\chi-\delta m_0/2$, whose coupling to the $Z$ boson is
%\begin{figure*}[t]
%\begin{center}
%\includegraphics[scale=.65]{BBN_bound}
%\caption{Summary of the lower bounds on the neutral mass splitting. The {\bf dark green} shaded region is excluded by tree-level \textit{Z}-exchange in Xenon1T~\cite{XENON:2018voc} for both scalar and fermionic DM. The {\bf dashed green} line shows what Xenon1T could prbe by analyzing high recoil energy data. The {\bf blue} and {\bf red} lines are the BBN bounds on the splitting for fermionic and scalar DM respectively.}
%\end{center}
%\label{fig:bbn}
%\end{figure*}

\begin{figure*}[t]
\begin{center}
\includegraphics[scale=.6]{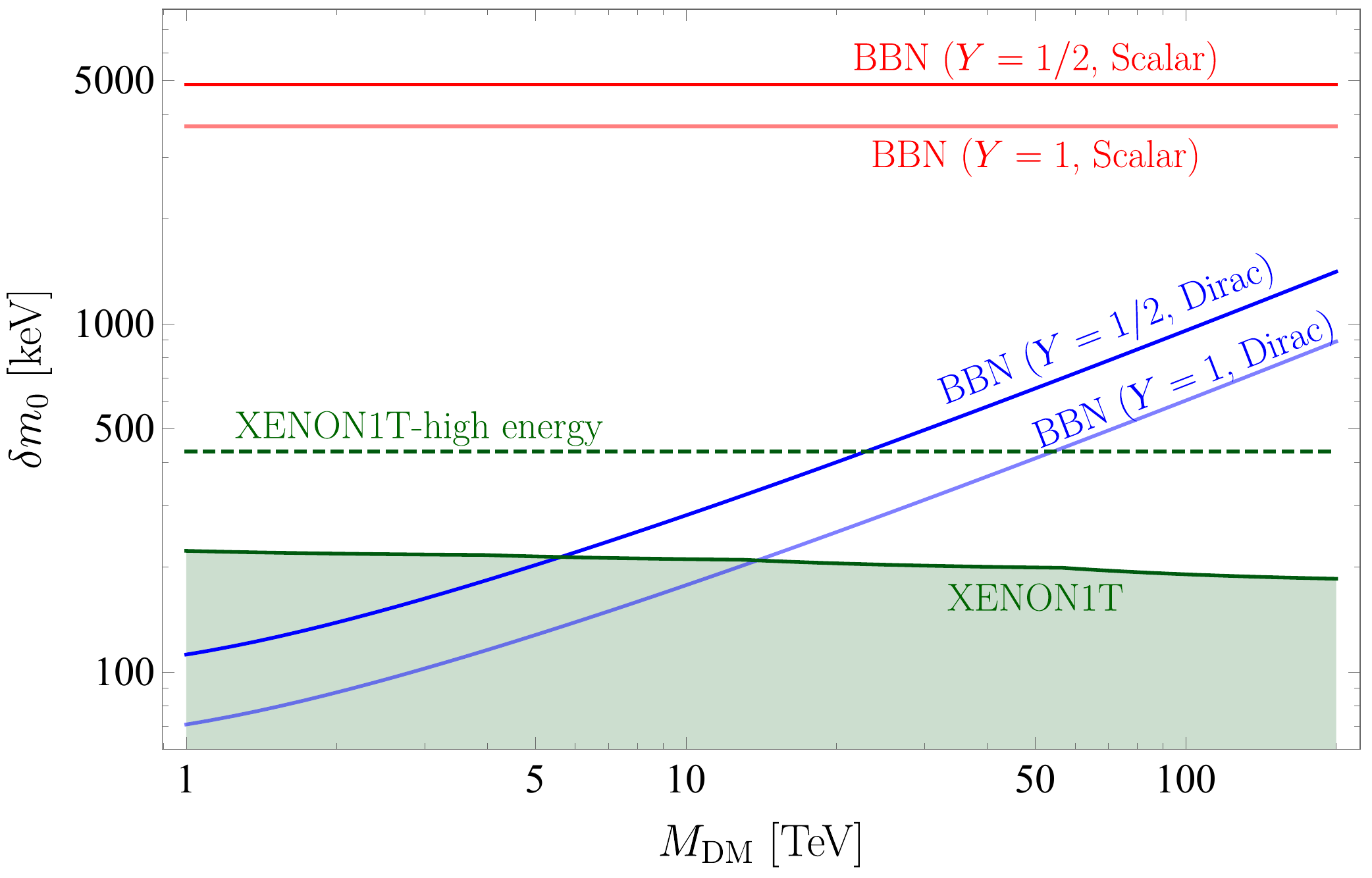}
\caption{Summary of the lower bounds on the neutral mass splitting. The {\bf dark green} shaded region is excluded by tree-level $Z$-exchange in Xenon1T~\cite{XENON:2018voc} for both scalar and fermionic DM. The {\bf dashed green} line shows what Xenon1T could probe by analyzing high recoil energy data. The {\bf blue} and {\bf red} lines are the BBN bounds on the splitting for fermionic and scalar DM respectively.}
\label{fig:bbn}
\end{center}
\end{figure*}

\begin{equation}
\label{eq:Z_nonvectorial}
\begin{aligned}
    %&\mathcal{L}_S^Z=\frac{ieY}{\sin\theta_W\cos\theta_W}\chi_0^\dagger\partial_\mu\chidm Z^\mu\\
    &\mathscr{L}_Z=\frac{ieY}{\sin\theta_W\cos\theta_W}\overline{\chi}_0\slashed{Z}\chidm\ .
\end{aligned}
\end{equation}
The $Z$-mediated scattering of DM onto nucleons is no longer elastic and the process is kinematically forbidden if the kinetic energy of the DM-nucleus system in the center-of-mass frame is smaller than the mass splitting 

\begin{equation}
    \frac{1}{2}\mu \vrel^2<\delta m_0\ ,\quad \mu=\frac{\mdm m_N}{\mdm+m_N}\ ,
\end{equation}
where $m_N$ is the mass of the nucleus, $\mu$ is the reduced mass and $\vrel$ is DM-nucleus relative velocity. In particular, given the upper bound on the relative velocity $\vrel<v_\text{E}+v_\text{esc}$, where $v_\text{E}=240$ km/sec is the Earth's velocity and $v_\text{esc}=600$ km/sec is the assumed escape velocity of DM in the Milky Way, the largest testable mass splitting is $\delta m_0^{\max}=1/2\mu (v_\text{E}+v_\text{esc})^2$ which for xenon nuclei gives $\delta m_0^{\max}\simeq 450$ keV. The splitting for a given recoil energy is
\begin{equation}
    \delta m_0(E_R)=\sqrt{2m_N E_R}(v_\text{E}+v_\text{esc})-E_R\frac{m_N}{\mu}\ ,
\end{equation}
which explain why the maximal constrained splitting experimentally is $\delta m_0^{\max,\exp}\simeq 240 \text{ keV}$ as shown in Fig.~\ref{fig:bbn}, given that Xenon1T~\cite{XENON:2018voc} analyzed data only for $E_R<40\text{ keV}$. Extending the range of Xenon1T to higher recoil energies would be enough to probe splitting up to $\delta m_0^{\max}$ as already noticed in Ref.~\cite{Bramante:2016rdh,Song:2021yar}.

In principle, larger mass splittings can be reached using heavier recoil targets than xenon such as iodine in PICO-60 \cite{PICO:2015pux}, tungsten in CRESST-II \cite{CRESST:2015txj},  CaWO$_4$~\cite{Munster:2014mga}, PbWO$_4$~\cite{Beeman:2012wz}, $^{180}$Ta \cite{Lehnert:2019tuw}, Hf \cite{Broerman:2020hfj} and Os \cite{Belli:2020vnc}. However, these experiments currently do not have enough exposure to probe EW cross-sections.

A complementary bound on $\delta m_0$ comes from requiring that the decay $\chi_0\rightarrow \chidm+$SM happens well before BBN. The leading decay channels are $\chi_0\rightarrow \chidm\gamma$, $\chi_0\rightarrow \chidm \bar{\nu}\nu$ and $\chi_0\rightarrow\chidm\bar{e}e$ with decay widths
\begin{align}
&\Gamma_\gamma=\left(1+\frac{1}{2}\log\left(\frac{m_W^2}{\mdm^2}\right)\right)^2\frac{Y^2\alpha_2^2\alpha_{\text{em}}}{\pi^2}\frac{\delta m_0^3}{\mdm^2}\ ,\\
&\Gamma_{\bar{\nu}\nu}\simeq 6 \Gamma_{\bar{e}e}=\frac{G_F^2\delta m_0^5Y^2}{5\pi^3}\ .
\end{align}

The first process is induced by a dipole operator generated at 1-loop for fermionic DM as computed in \cite{Haber:1988px}. The three body decays are instead induced at tree-level by the EW interactions both for fermionic and scalar DM. For fermionic DM, the dipole induce decay dominates the width in the mass range of interest. In order for these processes not to spoil BBN, we have to impose the following condition on the decay rate of $\chi_0$:

\begin{equation}
\label{eq:bbn_req}
\Gamma_{\chi_0}\equiv \Gamma_{\bar{\nu}\nu}+\Gamma_{\bar{e}e}+\Gamma_{\gamma}>\tau_\text{BBN}^{-1}\ ,
\end{equation}
where $\tau_\text{BBN}^{-1}=6.58 \times 10^{-25}\text{ GeV}$. The lower bounds on the neutral mass splitting for fermions are shown in Fig.~\ref{fig:bbn} together with those for scalars computed in Appendix \ref{sec:scalarWIMP}. The main difference between scalars and fermions is that the former are typically more long lived due to the suppression of $\chi_0\rightarrow \chidm\gamma$. As a consequence the BBN bouns are stronger for scalar WIMPs. 

\begin{figure*}[htbp]
\begin{center}
    \includegraphics[width=0.48\textwidth]{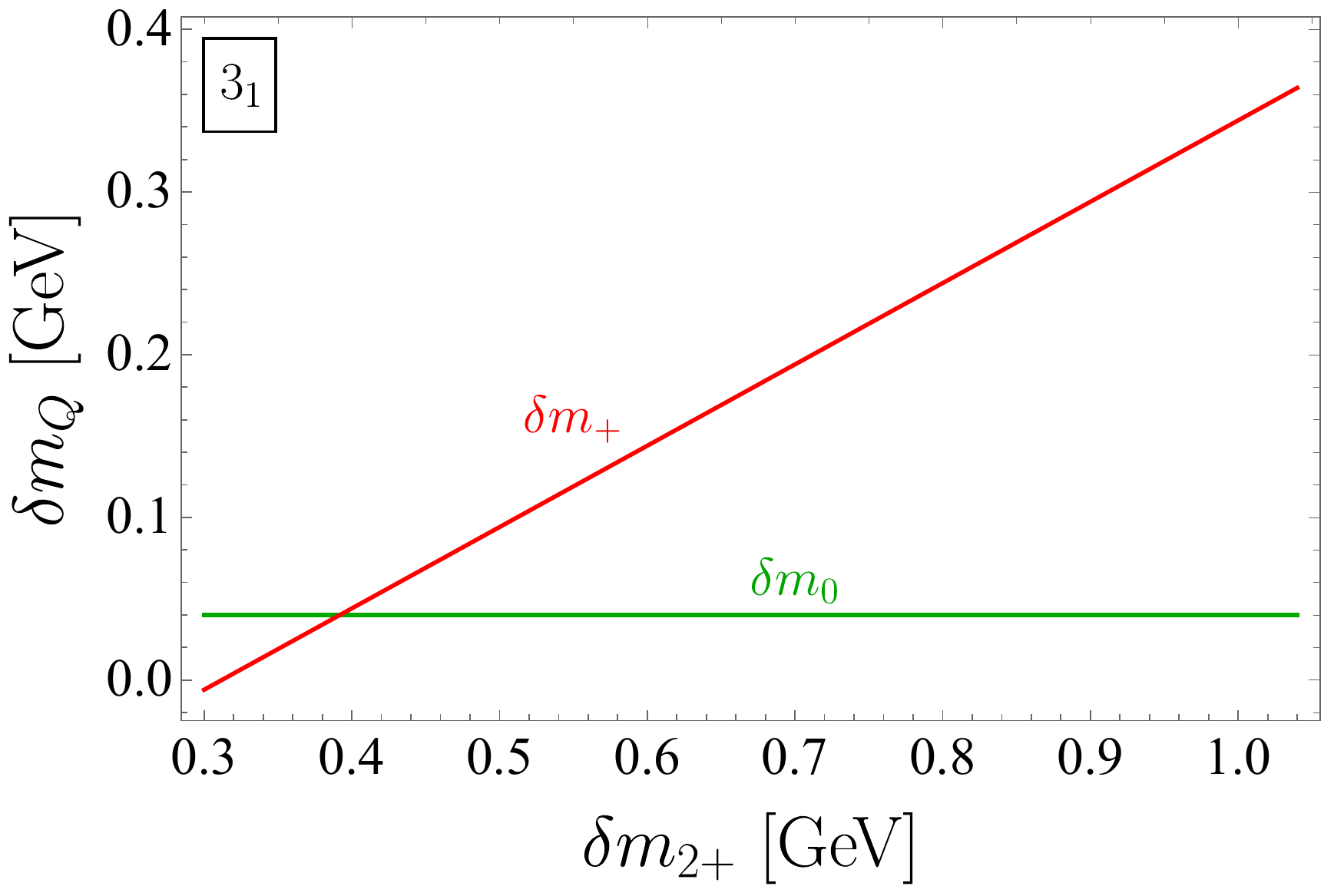}\hfill \includegraphics[width=0.48\textwidth]{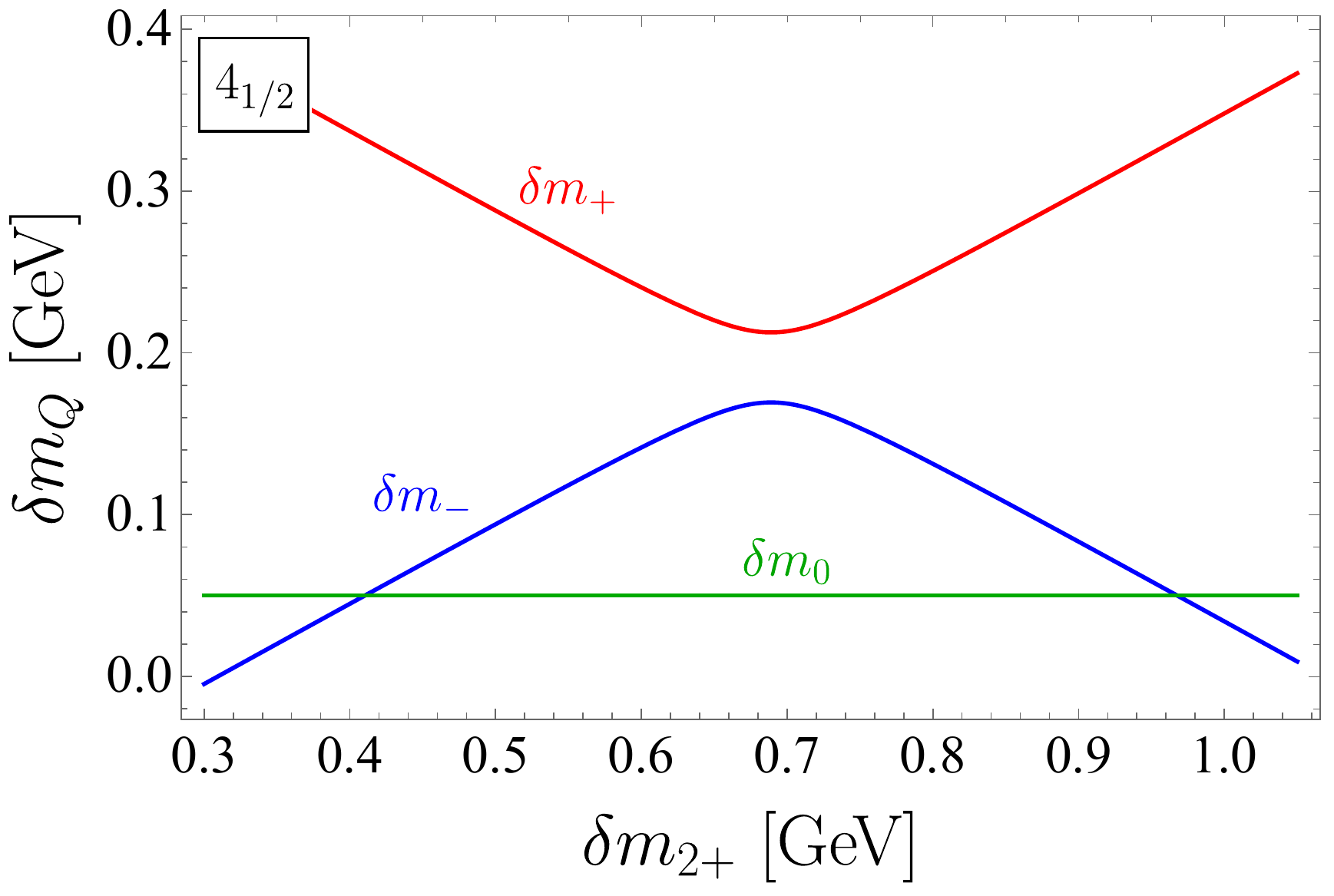}
    \caption{Mass splittings of $3_1$ \textit{(left)} and $4_{1/2}$ \textit{(right)} as a function of $\delta m_{2+}$. In the $3_1$ case, no mixing between the charged components of the multiplet can occur and $\delta m_+$ is a monotonic function of $\delta m_{2+}$. In the $4_{1/2}$ case, instead, because of the mixing induced by $\omix$, the positively charged mass eigenstate $\chi^+$ is always heavier than $\chi^-$. The splitting between $\chi^+$ and $\chi^-$ has a minimum of order $\delta m_0$, which was taken 50~MeV in this plot for display purposes. }
    \label{fig:splitting}
    \end{center}
\end{figure*}

\subsection{Charged splitting}\label{sec:chargedsplit}

The operator $\ochgd$ in \eqref{eq:ren_lag} is necessary to make the DM the lightest state in the EW multiplet for all the $n$-plets where the hypercharge is not maximal. Indeed, EW interactions induce at 1-loop mass splittings between the charged and the neutral components of the EW multiplet which in the limit $m_W\ll M_\chi$ are~\cite{Cheng:1998hc,Feng:1999fu,Gherghetta:1999sw}
\begin{equation}
\label{eq:charged_split_gauge}
    \Delta M_Q^{\text{EW}}= \delta_g \left(Q^2+\frac{2YQ}{\cos\theta_W}\right)\ ,
\end{equation}
where  $\delta_g=(167\pm 4)\mev$ and $Q=T_3+Y$. This implies that negatively charged states with $Q=-Y$ are pushed to be lighter than the neutral ones by EW interactions. Notable exceptions are odd-$n$ multiplets with $Y=0$ and all the multiplets with maximal hypercharge $|Y_{\max}|=(n-1)/2$ where negatively charged states are not present. For these multiplets, having $y_+=0$ would be the minimal and phenomenologically viable choice.  

%The new tree-level masses are given by%\footnote{Including $\delta\mathcal{L}_\text{ch}=y_5\delta\ochgd/\Leff$, we would get: $M_Q=\mdm\sqrt{(1-\epsilon y_+)^2+\epsilon^2y_5^2}$, where $\epsilon=(Q-Y)v^2/(4\mdm\Leff)$. Hence, we are neglecting $\mathcal{O}(\epsilon^2)$ corrections to \eqref{eq:mass_charged}.}
%\begin{equation}
%\label{eq:mass_charged}
%M_Q=M_\chi\left(1-y_+\frac{(Q-Y)v^2}{4 M_\chi\Leff} \right)\, ,
%\end{equation}
%\begin{widetext}
%\begin{equation}
%\label{eq:mass_charged}
%M_Q=M_\chi+\frac{y_+Yv^2}{4 \Leff} +\delta_g Q^2+\text{sgn}(Q)\sqrt{\left(\frac{2Y\delta_g}{\cos\theta_W}-\frac{y_+v^2}{4 \Leff}\right)^2Q^2+\frac{\delta m_0^2}{4}\frac{c_{nYQ}^2}{c_{nY0}^2}} \, ,
%\end{equation}
%\end{widetext}
 Including the contribution of $\ochgd$, the final splittings $\delta m_{Q}= M_Q-\mdm$ between the DM and the charged components read 
\begin{widetext}
\begin{equation}
\label{eq:deltaM}
\delta m_{Q}=\frac{\delta m_{0}}{2}+\delta_g Q^2+\text{sgn}(Q)\sqrt{\left(\frac{2Y\delta_g}{\cos\theta_W}-\frac{y_+v^2}{4 \Leff}\right)^2Q^2+\frac{\delta m_0^2}{4}\frac{c_{nYQ}^2}{c_{nY0}^2}}\ ,
\end{equation}
\end{widetext}
where $\text{sgn}(Q)$ in \eqref{eq:deltaM} accounts for the presence of opposite charge states that are not related by charge conjugation, as implied by to the non-zero hypercharge of our WIMPs. The second term inside the square root comes from the mixing between the charged gauge eigenstates $\chi_Q$ and $\chi_{-Q}^c$ induced by $\omix$. This obviosuly vanishes for $Q>(n-1)/2-Y$. 

The different charged-neutral mass splittings can all be written in terms of two independent splittings, which we choose to be $\delta m_0$ and $\delta m_{\Qbar}$, where $\Qbar\equiv Y+(n-1)/2$ is the largest electric charge in the multiplet. Since $c_{nY\Qbar}=0$, $\delta m_{\Qbar}$ is a monotonic function of $y_+$ and \eqref{eq:deltaM} can be inverted. In Fig. \ref{fig:splitting} we show as an example the mass splittings of $3_1$ and $4_{1/2}$ as a function of $\delta m_{2+}$. In the former case, no mixing occur within the components of the multiplet and $\delta m_+$ is a monotonic function of $\delta m_{2+}$. For the $4_{1/2}$, instead, the mixing induced by $\omix$ between the components with $Q=\pm 1$ makes the positively charged mass eigenstate $\chi^+$  heavier than $\chi^-$. 

In Sec.~\ref{sec:nonminimal splitting} we will explore the parameter space spanned by $\delta m_0$ and  $\delta m_{\Qbar}$, fixing the thermal DM mass of every EW multiplet as shown in Table~\ref{tab:TableI}. Crucially, the operators inducing the splitting in \eqref{eq:ren_lag} also generate new higgs-exchange contributions to the spin-independent scattering cross-section of DM on nucleons. Therefore, the current best upper limit on the DM elestic cross section onto nucleons set by PandaX-4T~\cite{PandaX:2021osp} translates into upper bounds on the neutral and charged splittings as reported in Table~\ref{tab:TableI}. 

Charged-neutral splittings smaller than the EW one in \eqref{eq:charged_split_gauge} require a certain amount of fine-tuning between UV operators and the EW contribution. To quantify this we define the Fine Tuning (F.T.) 
\begin{equation}
    \text{F.T.}\equiv\max_I\left[\frac{\mathrm{d}\log\delta m_Q}{\mathrm{d}\log\delta m_I}\right]\ , \label{eq:FT}
\end{equation}
 
where the index $I$ runs over the three contributions in the definition of $\delta m_Q$ in \eqref{eq:deltaM}. 
Large values of F.T. imply a significant amount of cancellation between two or more parameters.

%%%%%%%%%%%%%%%%%%%%%%%%%%%%%%%%%%%%%%%%%%%%%%%%%%%%%%%%%%%%%%%%%%%%%%%%%%
\subsection{Viable complex WIMPs}
\label{subsec:neutral_splitting}

The EFT approach used to write \eqref{eq:ren_lag} is meaningful only if the UV physics generating $\ochgd$ and $\omix$ is sufficiently decoupled from DM. When $\Lambda_{\text{UV}}$ approaches the DM mass the cosmological evolution of the DM multiplet cannot be studied in isolation, since the heavy degrees of freedom populates the thermal bath at $T\simeq M_{\text{DM}}$ and are likely to modify our freeze-out predictions. To avoid these difficulties we restrict ourselves to $\Lambda_{\text{UV}}\geq 10\mdm$.

This condition, together with the required inelastic splittings in Fig.~\ref{fig:bbn}, can be used to select the viable complex WIMPs. Starting from \eqref{eq:neutral_splitting} and imposing $\delta m_0>\delta m_0^{\min}$, we  derive the viable window for $\Leff$:
\begin{equation}
\label{eq:UV_dm0}
10\mdm<\Leff\leq \left(\frac{4y_0c_{nY0} v^{4Y}}{2^{2Y}\delta m_0^{\min}}\right)^{\frac{1}{4Y-1}}.
\end{equation}
We are now interested in estimating for which multiplets the viable window shrinks to zero. Setting $y_0=(4\pi)^{4Y}$ in \eqref{eq:UV_dm0} that is the largest value allowed by Naive Dimensional Analysis (NDA) we derive the values of $n$ and $Y$ having a non zero cutoff window in \eqref{eq:UV_dm0}. These are for both scalar and fermionic WIMPs $n_{1/2}$ multiplets with $n\leq12$ together with the $3_1$ and the $5_1$ mupltiplets. 

This result can be understood as follows. The upper bounds on $n$ for $Y=1/2$ multiplets come from the perturbative unitarity of the annihilation cross section as discussed in Appendix~\ref{sec:unitaritybound}. The  maximal cutoff required to obtain the phenomenologically viable splittings is of order $\sim 10^7\text{ TeV}$ as shown in the fifth column of  Table~\ref{tab:TableI}. This is many orders of magnitude larger than the DM masses allowed by freeze-out so that \eqref{eq:UV_dm0} results in a wide range of allowed $\Lambda_{\text{UV}}$.

For $Y=1$ multiplets the maximal required cutoff is of order $\sim 10^2$ TeV so that the $n$-dependence of the allowed window in \eqref{eq:UV_dm0} becomes relevant. Given that the DM mass grows with $n$ as $\mdm\sim n^{5/2}$ and the required cutoff stay approximately constant, we expect the allowed window to shrink to zero for large $n$. Numerically we find that the last allowed multiplet has $n=5$. 

We refer to Appendix~\ref{sec:scalarWIMP} for a similar argument for scalar WIMPs. These have a slightly different parametric which however results in the same viable EW multiplets of the fermionic case. 
%\vspace{0.5cm}

%%%%%%%%%%%%%%%%%%%%%%%%%%%%%%%%%%%%%%%%%%%%%%%%%%%%%%%%%%%%%%%%%%%%%%%%%%%%
\begin{figure*}[htp!]
 \centering
\includegraphics[width=0.7\textwidth]{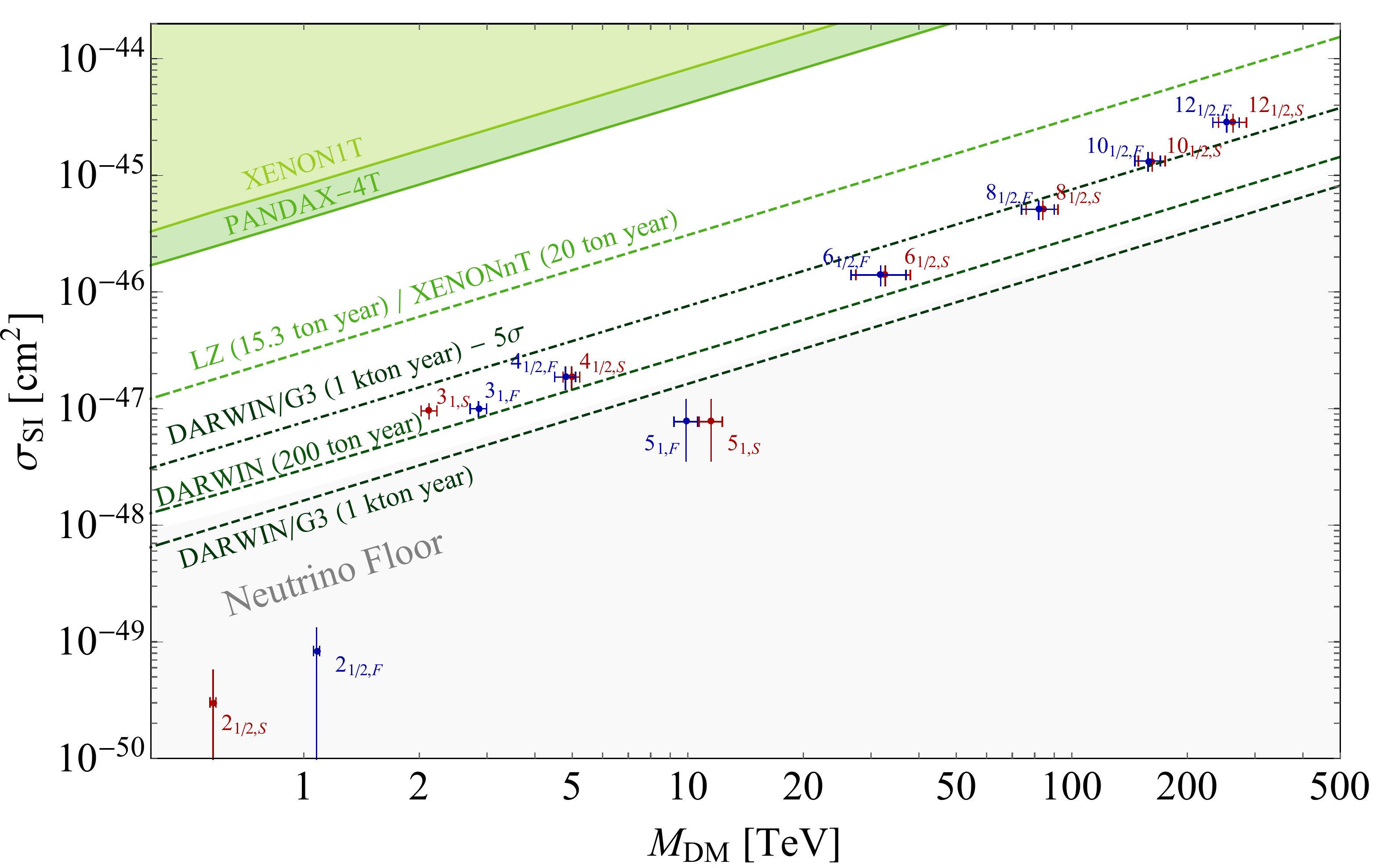}
 \centering
\caption{Expected SI cross-sections for different complex WIMPs for minimal splitting as defined in Sec.~\ref{sec:minimal splitting}. The {\bf blue dots} correspond to Dirac WIMPs and the {\bf red dots} to complex scalar WIMPs. The vertical error bands correspond to the propagation of LQCD uncertainties on the elastic cross-section (\eqref{elasticSIDD}), while the horizontal error band comes from the uncertainty in the theory determination of the WIMP freeze out mass in Table~\ref{tab:TableI}. The {\bf light green} shaded region is excluded by the present experimental contraints from XENON-1T~\cite{Aprile:2018dbl} and  PandaX-4T~\cite{PandaX:2021osp}, the {\bf green dashed} lines shows the expected 95\% CL reach of LZ/Xenon-nT~\cite{Mount:2017qzi,XENON:2020kmp} and DARWIN ~\cite{Aalbers:2016jon,Aalbers:2022dzr}.
\hspace*{\fill} \small}
\label{fig:DD_loop}
\end{figure*}

%%%%%%%%%%%%%%%%%%%%%%%%%%%%%%%%%%%%%%%%%%%%%%%%%%%%%%%%%%%%%%%%%%%%%%%%%%

\subsection{DM stability}\label{sec:stability}

The Lagrangian in \eqref{eq:ren_lag} preserves the DM number and as a consequence the DM is automatically stable. Gauge invariant interactions beyond those of \eqref{eq:ren_lag} can however induce fast DM decay. Here we discuss whether the effect of these operators can be small enough to allow the DM to be accidentally stable. Throughout this discussion we not only require the DM lifetime to be long enough to circumvent cosmological bounds~\cite{Audren:2014bca,Aubourg:2014yra} ($\tau_{\text{DM}}\gtrsim10^{19} \text{ sec}$) but also to satisfy the stronger astrophysical bounds on decaying DM~\cite{Cohen:2016uyg,Ando:2015qda,Cirelli:2012ut} ($\tau_{\text{DM}}\gtrsim10^{28}\text{ sec}$).

For $2_{1/2}$ and $3_1$ we can write renormalizable interactions $\overline{\chi^c}He_R$ and $\overline{\chi}L^c H$ that break the DM number and lead to a fast DM decay. These EW multiplets require a DM number symmetry, for example a discrete $\mathbb{Z}_2$-symmetry acting only on the DM field to provide a viable DM candidate.  

For even multiplets with $Y=1/2$ and $n>2$, we can write the following series of higher dimensional operators inducing DM decay 

\begin{widetext}
\begin{equation}
\begin{split}\label{eq:Leven}
    \mathscr{L}_{\text{even}}=&~\frac{C_1}{\Leff^{n-3}}\overline{L^c}\chi(H^\dagger H)^{\frac{n-2}{2}}+\frac{C_2}{\Leff^{n-3}}\overline{L^c}\chi W_{\mu\nu}W^{\mu\nu}(H^\dagger H)^{\frac{n-6}{2}}+\frac{C_3}{\Leff^{n-3}}\overline{L^c}\chi(W_{\mu\nu}W^{\mu\nu})^{\frac{n-2}{4}}\\
    &+\frac{C_4}{\Leff^{n-3}}\overline{L^c}\sigma^{\mu\nu}\chi W_{\mu\nu}(H^\dagger H)^{\frac{n-4}{2}} + \frac{C_5}{\Leff^2} \overline{\chi} \chi^c \overline{L^c}\chi\ ,
\end{split}
\end{equation}
\end{widetext}
where the Higgs bosons are appropriately contracted to make every term an $SU(2)$ singlet. For $4_{1/2}$, the leading contribution to DM decay comes from the operators with coefficients $C_1$ and $C_4$, while for $n\geq 6$ this is given by the operator with coefficient $C_5$ by closing the $\chi$ loop and attaching $(n-2)/2$ $W^{\mu\nu}$ operators, in a similar fashion to real candidates \cite{Bottaro:2021snn}. In all cases, we need at least $\Leff>10^{10}\mdm$ for $\mathcal{O}(1)$ Wilson coefficients to preserve DM stability. This lower bound is incompatible with the upper bound on the UV physics scale required to generate the inelastic splitting $\delta m_0$. This result implies that the DM stability depends upon the properties of the UV physics generating the neutral splitting in \eqref{eq:neutral_splitting}. 

For the $5_1$ WIMP the lowest dimensional operators inducing DM decay are
\begin{equation}
    \mathscr{L}_5 = \frac{C_1}{\Leff^2}(\overline{\chi}L^c\! H)(H^\dagger \! H)+\frac{C_2}{\Leff^2}(\overline{\chi}\sigma^{\mu\nu}L^c\! H)W_{\mu\nu}.
\end{equation}
These require $\Leff>10^{10}\mdm$ to ensure DM accidental stability which is again incompatible with the upper bound of $\Leff<20 \mdm$ needed to generate the inelastic splitting.

As a result, none of the complex WIMPs with $Y\neq 0$ can be accidentally stable in the sense of Minimal Dark Matter~\cite{Cirelli:2005uq}. Specific UV completions of the physics generating the inelastic splitting in \eqref{eq:neutral_splitting} might allow for DM accidental stability. This question goes beyond the scope of this study. Conversely, complex WIMPs with $Y=0$ can be accidentally stable thanks to the gauging of the unbroken $U(1)$ flavor symmetry in the DM sector~\cite{DelNobile:2015bqo}. The freeze-out predictions for these millicharged WIMPs are given in Appendix~\ref{app:millichargeWIMPs}.    
%%%%%%%%%%%%%%%%%%%%%%%%%%%%%%%%%%%%%%%%%%%%%%%%%%%%%%%%%%%%%%%%%%%%%%%%%
%%%%%%%%%%%%%%%%%%%%%%%%%%%%%%%%%%%%%%%%%%%%%%%%%%%%%%%%%%%%%%%%%%%%%%%%%

\section{Minimal Splitting}\label{sec:minimal splitting}

In this section we discuss the DD signal of Complex WIMPs in the minimal splitting scenario, that is when the mass splittings are chosen to be the smallest possible allowed by the requirements of the previous section. For fermions, we set $\delta m_0= 220\text{ keV}$ for $n_{1/2}$ WIMPs with $n\leq4$ as well as for the allowed $n_1$ WIMPs. For $n_{1/2}$ WIMPs with $n>4$ the BBN bound in Fig.~\ref{fig:bbn} gives the minimal $\delta m_0$. For scalars instead, we set $\delta m_0=4.9$ MeV for $n_{1/2}$ and $\delta m_0=3.7$ MeV for $n_1$ WIMPs, both coming from BBN.

In this minimal setup, $2_{1/2}$ and $3_1$ stand out as the only two multiplets where the DM is automatically the lightest state, with a splitting with the $Q=1$ state given by the pure EW splitting in \eqref{eq:charged_split_gauge} that is $354\text{ MeV}$ for $2_{1/2}$ and $542\text{ MeV}$ for $3_1$. For all the other WIMPs a UV generated splitting is needed to make the DM lighter than the negatively charged states. We fix this splitting to the smallest possible value that gives the DM as the lightest state. As can be seen from \eqref{eq:deltaM} this requires a UV splitting which is of the order of the EW one. We want now to study the phenomenology in direct detection in Sec.~\ref{sec:directminimal} and at a future muon collider in Sec.~\ref{sec:muoncollmin}. 

\subsection{Direct Detection prospects}\label{sec:directminimal}
The spin independent scattering cross-section $\sigma_\text{SI}$ of DM on nuclei receives two  contributions: i) from purely EW loop diagrams ii) from Higgs mediated tree-level diagrams generated by both $\omix$ and $\ochgd$. For minimal splitting Higgs mediated scattering is subdominant and  $\sigma_{\text{SI}}$ can be computed by considering only EW loop diagrams.

Following \cite{Hisano:2011cs, DelNobile:2013sia}, the Lagrangian describing the spin-independent (SI) DM interactions with quarks and gluons is 
\begin{equation}
\label{eq:nogamma}
\mathscr{L}_{\text{eff}}^{\rm{SI}} = f_q m_q \bar{\chi} \chi \bar{q} q + \frac{g_q}{\mdm} \bar{\chi} i \partial^{\mu} \gamma^{\nu} \chi \mathcal{O}^q_{\mu\nu} + f_G \bar{\chi} \chi G_{\mu\nu}G^{\mu\nu},
\end{equation}
where $\mathcal{O}^q_{\mu\nu} \equiv \frac{i}{2} \bar{q} \left( D_{\mu} \gamma_{\nu} + D_{\nu} \gamma_{\mu} - \frac{1}{2} g_{\mu\nu} \slashed{D}   \right)q$ is the quark twist-2 operator. The Wilson coefficients are given by~\cite{Hisano:2011cs}
\begin{comment}
\begin{widetext}
\begin{equation}
\label{eq:EW_loops}
\begin{aligned}
&\hskip 2truecm f_q^{\rm{EW}} = \frac{\alpha_2^2}{4 m_h^2} \left[ \frac{n^2-(4 Y^2+1)}{8 m_W} g_H(w) + \frac{Y^2}{4 m_Z c_w^4} g_H(z) \right] + \frac{((a_q^V)^2-(a_q^A)^2) Y^2}{c_w^4} \frac{\alpha_2^2}{m_Z^3} g_S(z),\\
&\hskip 3.2truecm g_q^{\rm{EW}} =  \frac{n^2-(4 Y^2+1)}{8}  \frac{\alpha_2^2}{m_W^3} g_{T1}(w)+\frac{2 ((a_q^V)^2+(a_q^A)^2) Y^2}{c_w^4} \frac{\alpha_2^2}{m_Z^3} g_{T1}(z),\\
&\hskip-0.6truecm f_G^{\rm{EW}} = - \frac{\alpha_s}{12 \pi} \frac{\alpha_2^2}{4 m_h^2} \left( \sum_{q=c,b,t}  \kappa_q \right) \left[\frac{n^2-(4 Y^2+1)}{8 m_W} g_H(w) + \frac{Y^2}{4 m_Z c_w^4} g_H(z) \right]+ \frac{\alpha_s \alpha_2^2}{4 \pi} \left[\frac{n^2-(4 Y^2+1)}{8 m_W^3} g_W(w,y) + \frac{Y^2}{4 m_Z^3 c_w^4} g_Z(z,y) \right].
\end{aligned}
\end{equation}
\end{widetext}
\end{comment}

\begin{widetext}
\begin{equation}
\label{eq:EW_loops}
\begin{aligned}
& f_q^{\rm{EW}} \simeq -\frac{\pi\alpha_2^2}{16 m_h^2 m_W} \left[ n^2-1-\left(1.03+22 (a_q^{V,2}-a_q^{A,2})\right)Y^2 \right],\\
& g_q^{\rm{EW}} \simeq \frac{\pi\alpha_2^2}{24 m_W^3}\left[ n^2-1-\left(4-18.2 (a_q^{V,2}+a_q^{A,2})\right)Y^2 \right],\\
& f_G^{\rm{EW}} \simeq  \frac{\alpha_s\alpha_2^2}{192 m_h^2 m_W} \left[\left( \sum_{q=c,b,t}  \kappa_q +2.6\right)(n^2-1)-\left( 1.03\sum_{q=c,b,t}  \kappa_q -7.5\right) Y^2 \right]\, ,
\end{aligned}
\end{equation}
\end{widetext}

where $m_h$ is the mass of the Higgs and $\kappa_c = 1.32,\,\kappa_b=1.19,\,\kappa_t=1$. 
\begin{comment}
The loop functions $g_{\{H,S,T1,W,Z\}}$ have been evaluated in what follows in the limits
\begin{equation*}
w \equiv \frac{m_W^2}{M_{DM}^2} \to 0,\,\,\,\,\,z \equiv \frac{m_Z^2}{M_{DM}^2} \to 0,\,\,\,\,\,y \equiv \frac{m_t^2}{M_{DM}^2} \to 0.
\end{equation*}
\end{comment}
Furthermore we have defined $a_q^V = T_{3q}/2 - Q_q s_w^2$, $a_q^A=-T_{3q}/2$ with $c_w,\,s_w$ being the cosine and the sine of the Weinberg angle, respectively. The terms proportional to $Y$ correspond to the exchange of $Z$ bosons inside the EW loops.

After the IR matching of these interactions at the nucleon scale \cite{DelNobile:2013sia}, we can express $\sigma_\text{SI}$ per nucleon (for $\mdm \gg m_N$) as
\begin{equation}
\label{elasticSIDD}
\sigma_{\text{SI}} \simeq \frac{4}{\pi} m_N^4 \vert k_N^{\rm{EW}} \vert^2,
\end{equation}
where $m_N$ is the mass of the nucleon and 
\begin{equation*}
\label{eq:kEW}
k_N^{\rm{EW}} =\!\!\!\!  \sum_{q=u,d,s} f_q^{\rm{EW}} f_{Tq} + \frac{3}{4} (q(2) + \bar{q}(2)) g_q^{\rm{EW}} - \frac{8 \pi}{9 \alpha_s} f_{TG} f_G^{\rm{EW}}\, ,
\end{equation*}
where the nucleon form factors are defined as $ f_{Tq} = \langle N|  m_q \bar{q} q |N\rangle  / m_N$, $f_{TG} = 1-\sum_{q=u,d,s} f_{Tq}$, and $\langle N(p)| \mathcal{O}^q_{\mu\nu}|N(p)\rangle = (p_{\mu} p_{\nu}-\frac{1}{4}m_N^2 g_{\mu\nu})(q(2) + \bar{q}(2))/m_N$, and $q(2)$, $\bar{q}(2)$ are the second moments of the parton distribution functions for a quark or antiquark inside the nucleon \cite{Hisano:2011cs}. The values of these form factors are taken from the results of direct computation on the lattice, as reported by the FLAG Collaboration \cite{Aoki:2021kgd} in the case of $N_f=2+1+1$ dynamical quarks  \cite{Alexandrou:2014sha, Freeman:2012ry}.\footnote{Notice that since $Y \neq 0$, the up and down quarks can give different contributions to the SI cross section \eqref{elasticSIDD}. The ETM Collaboration \cite{Alexandrou:2014sha} has computed the form factors in the case of degenerate light quarks so we take $f_{Tu}=f_{Td}$, ignoring possible differences between these two form factors.} 

The direct detection reach on the different complex multiplets for minimal splitting is summarized in Fig.~\ref{fig:DD_loop}. As can be seen by the figure, most of the WIMPs can be probed for minimal splitting by future large exposure direct detection experiments like DARWIN with the notable exception of the $2_{1/2}$ and $5_1$ that we discuss below.

Combining \eqref{eq:EW_loops} with \eqref{elasticSIDD} the parametric expression for $\sigma_{\text{SI}}$ is
\begin{equation}
\label{eq:cancellation}
    \sigma_{\text{SI}} \approx 10^{-49} \text{ cm}^2 (n^2-1-\xi~ Y^2)^2,
\end{equation}
where $\xi=16.6\pm 1.3$ with the error coming from the lattice determination of the nucleon form factors. This formula makes evident that large cancellations with respect to the natural size of the elastic cross section can take place when $Y\simeq\sqrt{(n^2-1)/\xi}$. In particular, for $n=2$ the exact cancellation takes place at $Y\simeq 0.44\pm 0.02$, which almost matches the exact hypercharge of the doublet leading to the large uncertainty in Fig.~\ref{fig:DD_loop}. Similarly, the cancellation happens at $Y\simeq 1.2$ for $n=5$, which explains why in this case the signal entirely lies below the neutrino floor, while it is within DARWIN reach for their real or millicharged counterparts as shown in Ref.~\cite{Bottaro:2021snn} and Fig.~\ref{fig:DD_loop_milli} respectively. 

Spin dependent (SD) interactions of DM with the nuclei are also induced by EW loops and can lead to a larger cross section compared to the SI one~\cite{Hisano:2011cs}. Unfortunately, the predicted SD cross section for all the complex WIMPs lies always well below the neutrino floor and it will be impossible to test even at future direct detection experiments.

\subsection{$2_{1/2}$ and $3_{1}$ at muon colliders}\label{sec:muoncollmin}
Conversely to the case of real WIMPs studied in Ref.~\cite{Bottaro:2021snn}, large exposure experiments like DARWIN will not be able to entirely close the window of Complex WIMP candidates. This further motivates us to study the expected reach on these DM candidates at a high energy lepton collider such as a muon collider running well beyond TeV center of mass energy. 

\begin{figure*}
    \centering
    \includegraphics[width=0.4\textwidth]{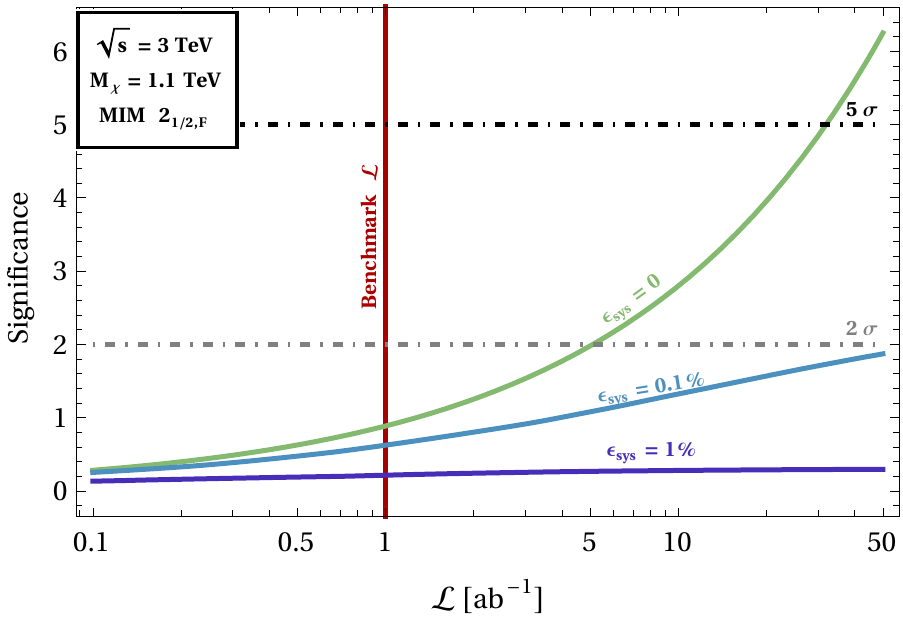}\hfill\includegraphics[width=0.4\textwidth]{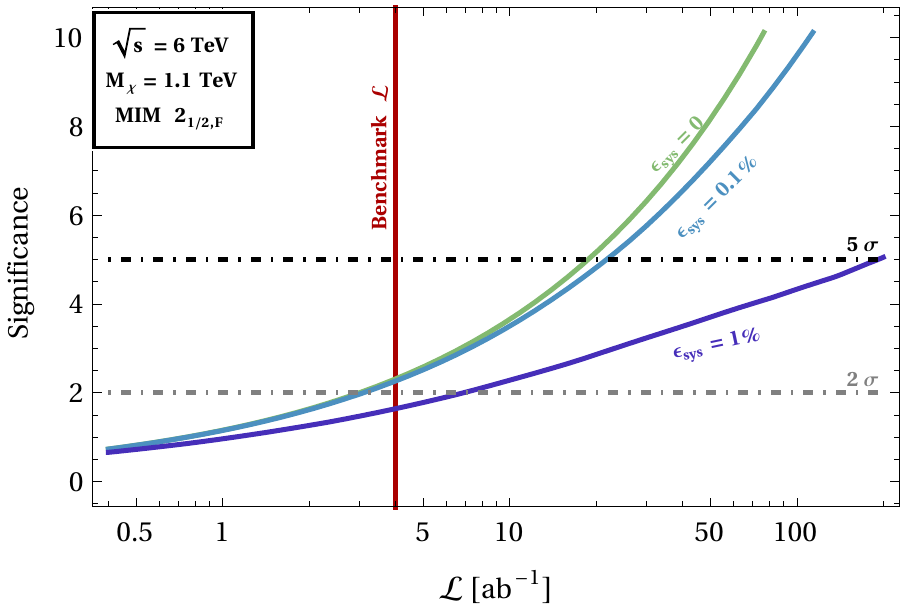}
    \includegraphics[width=0.4\textwidth]{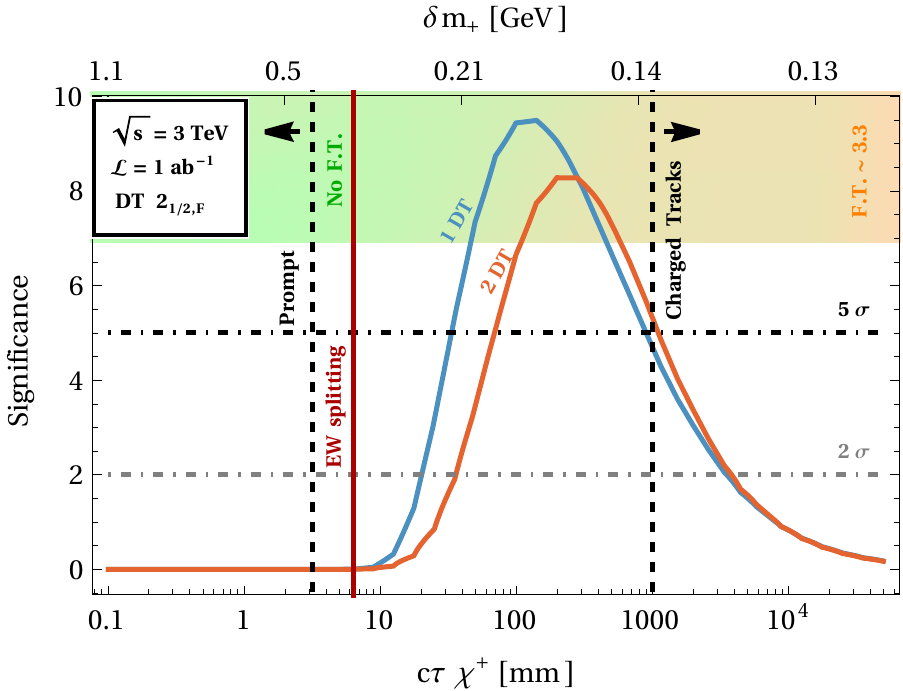}\hfill\includegraphics[width=0.4\textwidth]{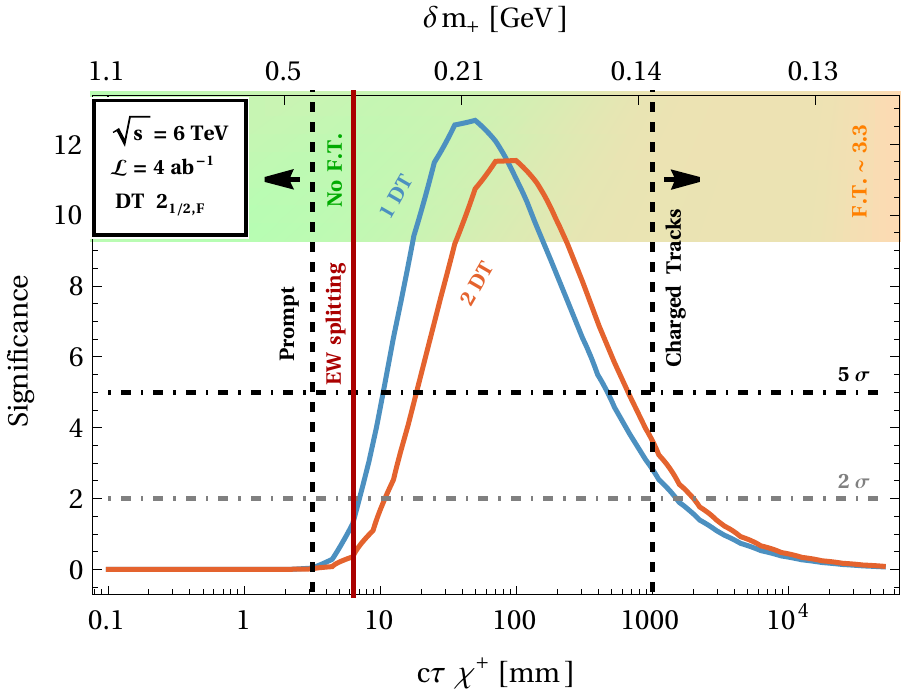}
    \includegraphics[width=0.4\textwidth]{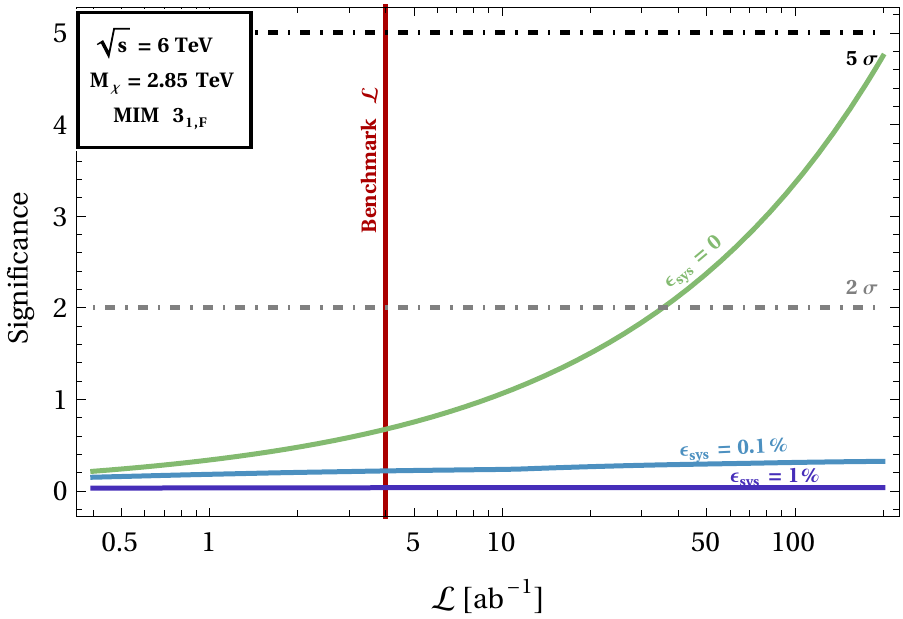}\hfill\includegraphics[width=0.4\textwidth]{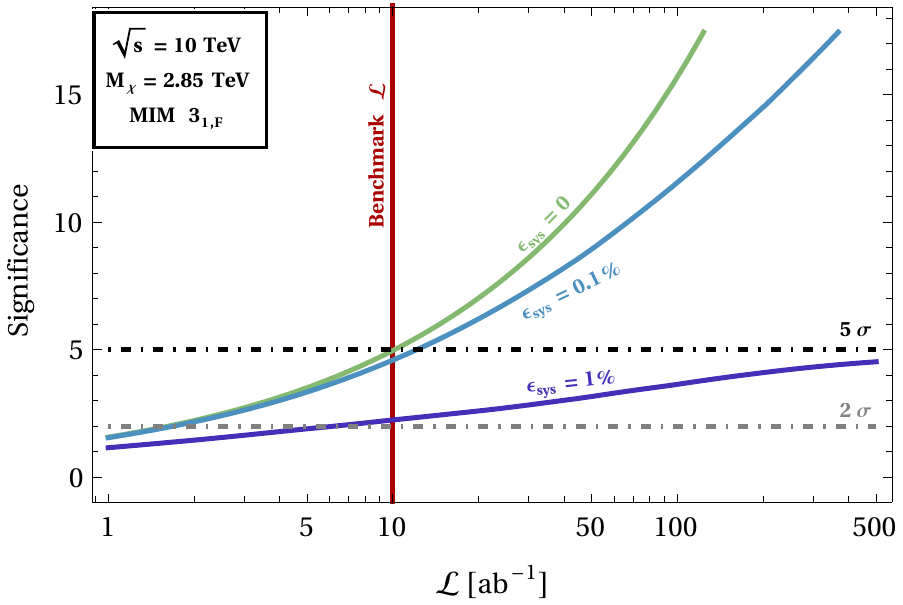}
    \includegraphics[width=0.4\textwidth]{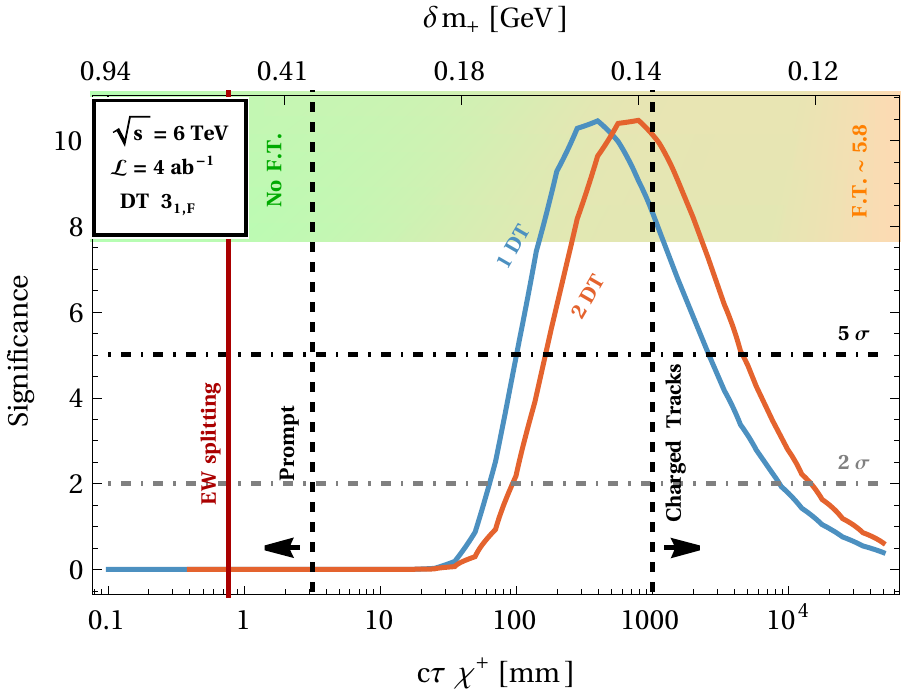}\hfill\includegraphics[width=0.4\textwidth]{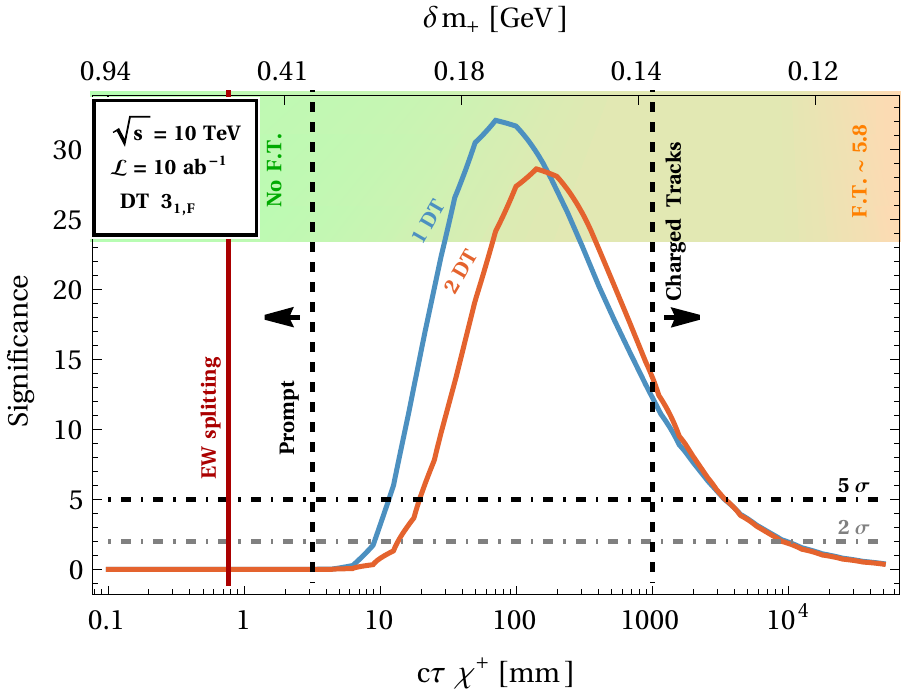}
    \caption{%
    {\bf Top two rows}: Collider reach for the $2_{1/2}$ of mass $M_\chi=1.1$ TeV and $\sqrt{s}=3$ TeV  {\it (left)} and $\sqrt{s}=6$ TeV {\it(right)}. 
    {Upper row:} Combined reach for MIM (mono-$\gamma$+mono-$W$) as a function of the luminosity $\mathcal{L}$ for different value of the expected systematic uncertainties parametrized by $\epsilon_{\text{sys}}$: $1\%$ {\bf (purple)}, $0.1\%$ {\bf (cyan)}, 0 {\bf (green)}. The {\bf red} lines show the benchmark luminosity following \eqref{eq:bench}. %
    {Lower row: } Reach of DT for varying lifetime of the charged track. The {\bf red} vertical line shows the benchmark lifetime for EW splitting. The {\bf blue} and {\bf orange} curves correspond to the reach with 1 and 2 disappearing tracks.  Fine-Tuning values computed on $\delta m_+$ following \eqref{eq:FT} are displayed as color code from green (no fine-tuning) to red (higher fine-tuning). 
    %We may also drop it completely in this plot as the FT, after all, changes only by a factor 10 and it is roughy given by the size of the EW mixing divided by the vertical axis value. %
    %
    {\bf Bottom two rows}: same as for the top rows, but for the case  of the $3_{1}$ of mass $M_\chi=2.85$ TeV and $\sqrt{s}=6$ TeV  {\it (left)} and $\sqrt{s}=10$ TeV {\it(right)}.  %{\bf Top:} Combined reach for MIM (mono-$\gamma$+mono-W) as a function of the luminosity $\mathcal{L}$ for different value of the expected systematic uncertainties parametrized by $\epsilon_{\text{sys}}$ in the cut-and-count scheme: $1\%$ {\bf (blue)}, $0.1\%$ {\bf (cyan)}, 0 {\bf (green)}. The {\bf red} lines show the benchmark luminosity following \eqref{eq:bench}. \textbf{Bottom: } Reach of DT for varying lifetime of the charged track. The {\bf red} line shows the benchmark lifetime for EW splitting. The {\bf blue} and {\bf orange} curves correspond to the reach with 1 and 2 disappearing tracks.
    }
    \label{fig:loglumicombined} %\label{fig:loglumicombined}
\end{figure*}

In this Section we fix the luminosity as a function of the center of mass energy as~\cite{Delahaye:2019omf}
\begin{equation}
\mathcal{L}=10 \text{ ab}^{-1}\cdot\left(\frac{\sqrt{s}}{10\text{ TeV}}\right)^2\,,\label{eq:bench}
\end{equation}
 and we concentrate on machines running at $\sqrt{s}=3,6,10\text{ TeV}$~\cite{Palmer_2014,2203.07224v1}. In Appendix~\ref{app:collider} we provide results for higher center of mass energies, as well as exhaustive summary tables and plots for all the dark matter candidates at center of mass energies up to 30~TeV, including estimates for scenarios in which the luminosity is a free parameter not fixed by \eqref{eq:bench}.

We focus on the collider reach for the complex doublet ($2_{1/2}$) and the complex triplet ($3_{1}$), that are the lightest WIMPs and have the greater chance to be discoverable at $\sqrt{s}\leq 10~{\rm TeV}$. Theoretically, these candidates are the most minimal complex WIMPs since they have maximal hypercharge and the neutral component is automatically the lightest one at the renormalizable level. The only required higher dimensional operator is $\mathcal{O}_0$ which generates the inelastic splitting in \eqref{eq:neutral_splitting}. The sensitivity of high energy muon colliders to heavier multiplets will be discussed in detail in Appendix~\ref{app:collider}.

\bigskip

The expected signatures at future colliders depend very much on the lifetime of the charged states  in the EW multiplet. These might decay back to the neutral DM either promptly or with a macroscopic lifetime on detector scales. 

In the prompt case, the decay of the charged state will give rise to soft radiation which would be impossible to disentangle from the SM background and the DM signal will be characterized only by a large missing energy recoiling against EW radiation. This can be further resolved from the SM background at lepton colliders thanks to the knowledge of the center of mass energy by looking for features in the missing mass (MIM) distribution. We systematically studied the different kinds of EW radiation in Ref.~\cite{Bottaro:2021snn}. We perform a similar study here (details are given in Appendix~\ref{app:collider}), finding that the MIM reach is dominated by the mono-$W$ and the mono-$\gamma$ channels.
%, the latter being comparatively more sensitive for small EW representations. 
Despite the non-zero hypercharge enhancing the coupling to the $Z$, the mono-$Z$ channel has a lower reach, similarly to what was found in Ref.~\cite{Bottaro:2021snn}. 
%This is due to the less distinct signal with respect to the SM background. 

We estimate the reach of MIM searches from cut-and-count experiments using the following estimate for the statistical significance
\begin{equation}
\text{significance}=\frac{S}{\sqrt{B+S+\epsilon_{\text{sys}}^2 (B+S)^2}}\ ,
\end{equation}
where $\epsilon_{\text{sys}}$ parametrizes the expected systematic uncertainties on the measurement. 

In the long-lived case, depending on the decay length of the charged state it might be beneficial to look at disappearing or ``stub'' tracks (DT) as studied in Ref.~\cite{Capdevilla:2021fmj}, or at long-lived charged tracks that could be distinguished from the SM backgrounds by their energy losses in the material and by the measurement of their time of flight (see for example analogous searches performed by ATLAS and CMS at the LHC~\cite{CMS:2016kce,ATLAS:2019gqq}). 

Here we focus on the expected reach from DTs and MIM searches which are relevant for the $2_{1/2}$ and $3_1$, fixing the splitting of the $Q=1$ state with respect to the neutral one to be the one generated purely by EW corrections. In this minimal case the rest-frame lifetime of the charged state decaying back to the DM is $c\tau_+\approx 6.6\text{ mm}$ for the $2_{1/2}$ and $c\tau_+\approx 0.77\text{ mm}$ for the $3_1$.  

Our results for the combination of the MIM searches and the DT tracks search are given in Fig.~\ref{fig:loglumicombined} for fixed thermal mass of the $2_{1/2}$, and similarly for the $3_1$. 
We provide details on the estimate of the MIM reach, as well as for the DT signature, in App.~\ref{app:collider}, also giving results as a function of the mass $M_\chi$ of the candidate.

As can be seen from Fig.~\ref{fig:loglumicombined},  the MIM search at $\sqrt{s}=3$ TeV with the benchmark luminosity $\mathcal{L}=1\; \mathrm{ab}^{-1}$ is not sensitive to the $2_{{1/2}}$. A $\sqrt{s}=6$ TeV collider with benchmark luminosity $\mathcal{L}=4\; \mathrm{ab}^{-1}$, instead,  is able to probe the doublet WIMP at $2\sigma$ C.L. In general the mono-$W$ and mono-$\gamma$ channels give comparable mass reach for the benchmark luminosity adopted here, full detail is given in Tab.~\ref{tab:dirac_ditri_cuts}. The sensitivity of each channel has a specific behavior as a function of the luminosity and for different assumed systematics. The overall combination as function of the total luminosity for fixed thermal mass is shown in Fig.~\ref{fig:loglumicombined}. We remark that the reach deviates from a pure rescaling by $\sqrt{\mathcal{L}}$ because the selections, hence the result, have been optimized as  a function of $\mathcal{L}$ when dealing with  $\epsilon_{\text{sys}}\neq 0$. 

In Fig.~\ref{fig:loglumicombined} we  also display results for the DT search as a function of the lifetime of the charge +1 state, that is in  a 1-to-1 relation with the mass splitting $\delta m_+$.  We see that a collider of $\sqrt{s}=6\text{ TeV}$ can probe the $2_{{1/2}}$ for charge-neutral splitting generated purely by EW interactions. A collider of $\sqrt{s}=3\text{ TeV}$ can  probe a large portion of the allowed lifetimes for the charged tracks corresponding to non-zero UV contributions in \eqref{eq:deltaM}. 

In Fig.~\ref{fig:loglumicombined} we show similar results for the $3_1$ at its thermal mass, which entails interesting results at 6 and 10~TeV center of mass energy machines. For the 10~TeV collider we find that MIM searches are effective probes of this WIMP  candidate and can establish a bound at 95\% CL with a small luminosity or give a discovery with the nominal luminosity. Mono-$W$ and mono-$\gamma$ perform similarly well and their combination is worth being done. The DT search cannot probe the EW splitting for the $3_1$, however it can cover a large portion of allowed lifetimes  in the non-minimal case in which UV physics is contributing to the charged-neutral splitting. 

Finally, it is important to notice that with an $\mathcal{O}(1)$ fine tuning of the UV contribution to the charged neutral splitting against the irreducible EW contribution the charged tracks could be extremely long-lived on detector scale for both the $2_{{1/2}}$ and $3_{1}$. This observation motivates a detailed detector simulation of a muon collider environment to reliably estimate the expected reach in this channel. We will provide some estimates for the signal yield in Fig.~\ref{fig:charged_tracks} in the next section.

\section{Non-minimal splittings}\label{sec:nonminimal splitting}
 
Exploiting the freedom to change the spectrum given by the UV operators in \eqref{eq:ren_lag} one can have significant changes in the phenomenology described in Sec.~\ref{sec:minimal splitting}. We dedicate this section to explore the phenomenological consequences of releasing the minimal splitting assumption.

\subsection{Direct detection}

The Higgs portal operators in \eqref{eq:ren_lag} generate upon EWSB a linear coupling of the DM to the Higgs boson of the form
\begin{equation}
\mathscr{L}_{D,h}=-\frac{\lambda_Dv}{2\Leff}\chidm^2h\, . \label{eq:tree_DMh}
\end{equation}
This coupling mediates tree-level SI scattering processes of DM onto nuclei, therefore it can be constrained by direct detection experiments. As the mass splitting $\delta m_0$ has to be sufficiently large to suppress the scattering mediated by the $Z$ boson, we find   that the allowed parameter space for the non-renormalizable couplings is compact.

Following Ref.~\cite{Arcadi:2019lka} we can integrate out the Higgs boson and write the  couplings of the DM to the SM 
\begin{equation}
    \mathscr{L}_{\text{eff,h}}^{\text{SI}}=\frac{\lambda_D}{2m_h^2\Leff}\chi^2\left(m_q \bar{q}q-\frac{\alpha_s}{4\pi}G^{a}_{\mu\nu}G^{a\mu\nu}
\right),
\end{equation}
so that the matrix elements $f_q$ and $f_G$ required to compute $\sigma_{\text{SI}}$ are simply given by
\begin{equation}
f_q = f_q^{\text{EW}}+\frac{\lambda_D}{2m_h^2\Leff},\, f_G = f_G^{\text{EW}}-\frac{\alpha_s}{8\pi}\frac{\lambda_D}{2m_h^2\Leff},\notag
\end{equation}
where $f_q^{\text{EW}}$ and $f_G^{\text{EW}}$ come from EW loops and are given in \eqref{eq:EW_loops}. We can rewrite the coupling $\lambda_D$ solely in terms of mass splitting as
\begin{equation}
\label{eq:higgs_tree_split}
\frac{\lambda_D}{\Leff}=-\frac{2Y}{v^2}(\delta \mu_{\Qbar}+n\delta \mu_0)\, ,
\end{equation}
where 
\begin{equation}
\label{eq:split_def}
  \!\!\!\! \delta \mu_{\Qbar}\!\equiv\! \frac{2\delta m_{\Qbar}-\delta m_0-2\Delta M_{\Qbar}^{\text{EW}}}{2\Qbar}\, ,\ \delta\mu_0\equiv\frac{\delta m_0}{n}
\end{equation}
and  $\Delta M_{\Qbar}^{\text{EW}}$ is the gauge induced mass splitting in \eqref{eq:charged_split_gauge}.

Replacing \eqref{eq:higgs_tree_split} into \eqref{elasticSIDD} allows us to translate the upper bound on $\sigma_{\text{SI}}$ into an upper bound on $\delta m_0$ and $\delta m_{\Qbar}$. 

\begin{figure}
 \includegraphics[width=1.0\textwidth]{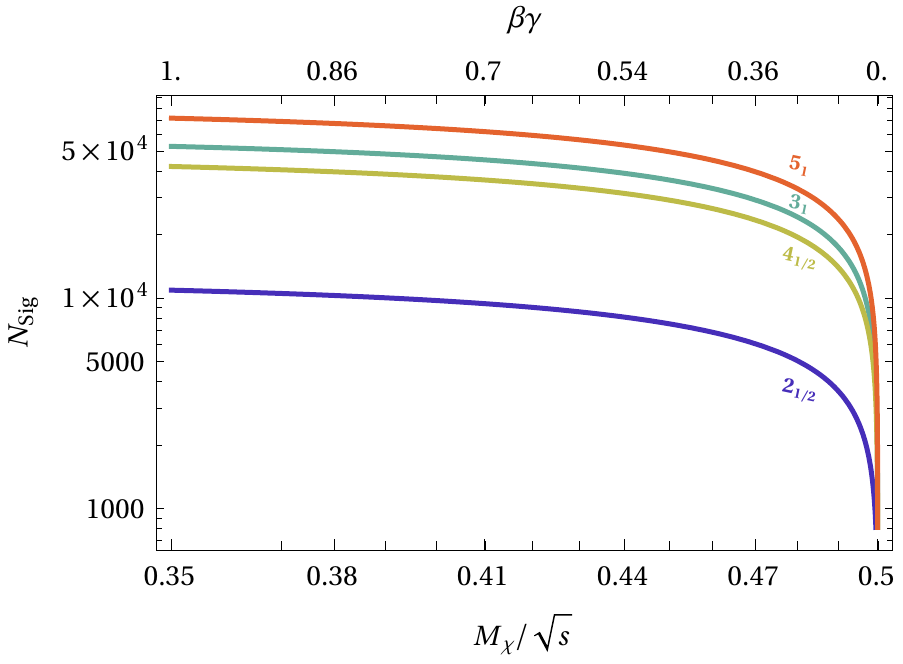}
 \caption{Signal yield of the charged tracks for the $2_{1/2}$, $3_1$, $4_{1/2}$, $5_1$. The plot holds for all $\sqrt{s}$ in the tens of TeV range, assuming the luminosity scales as \eqref{eq:bench}. The cuts applied on the produced charged $\chi$'s are $p_T>200$ GeV, $|\eta_\chi|<2$.}
\label{fig:charged_tracks}
\end{figure}

\begin{figure*}[htbp]
\begin{center}
 \includegraphics[width=0.475 \linewidth]{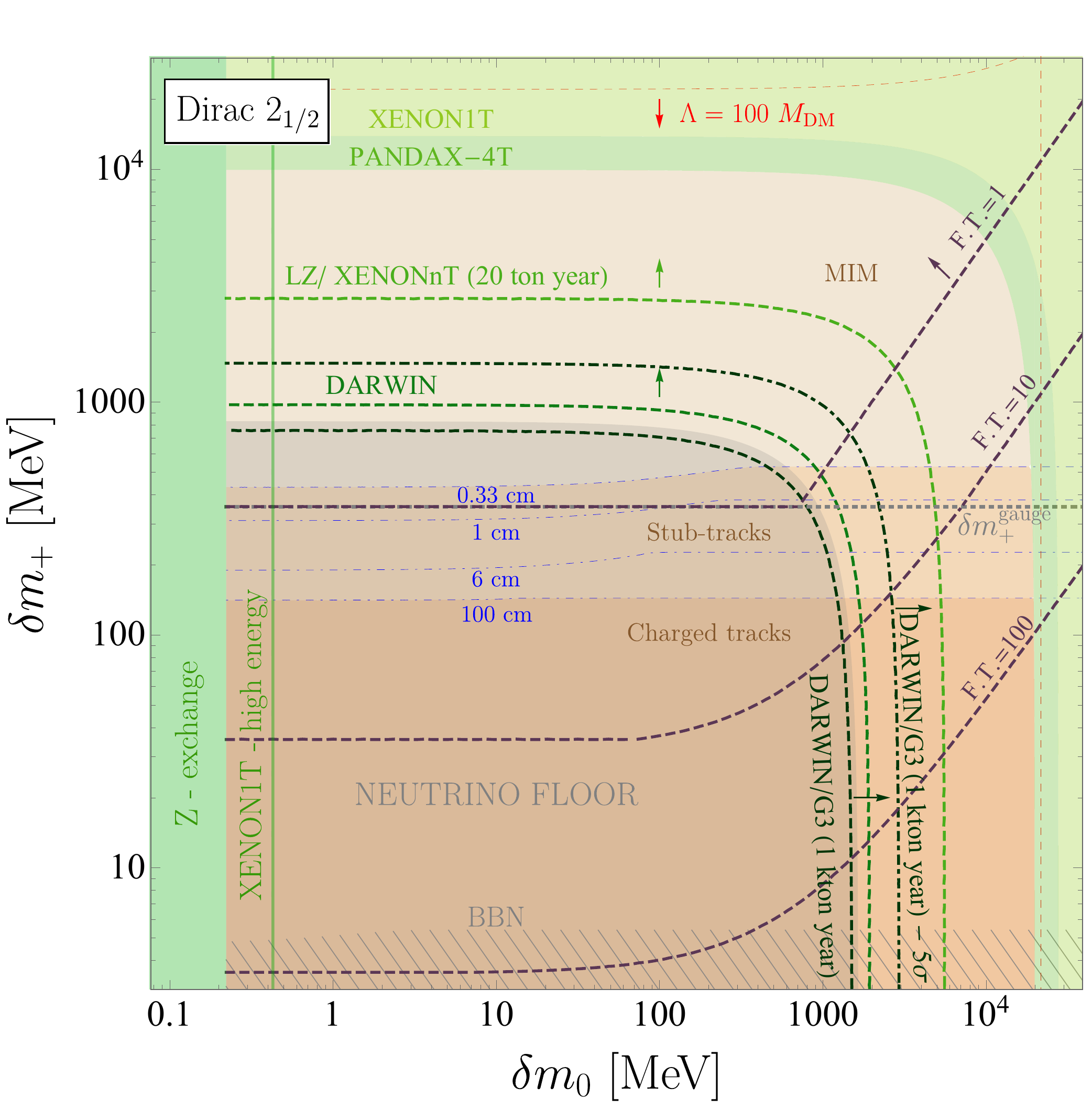}\hfill
     \includegraphics[width=0.5 \linewidth]{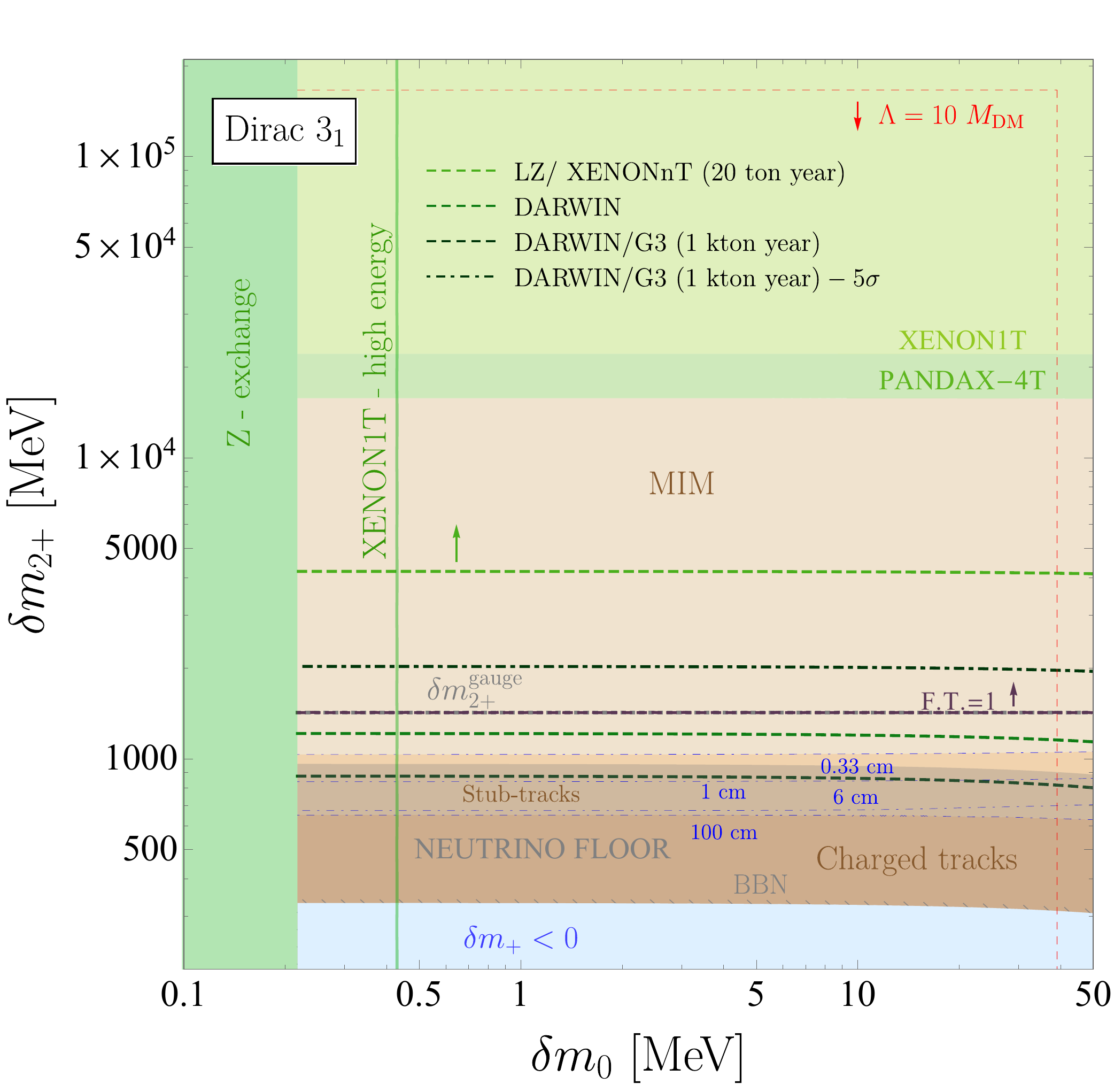}\\
     \includegraphics[scale=0.385]{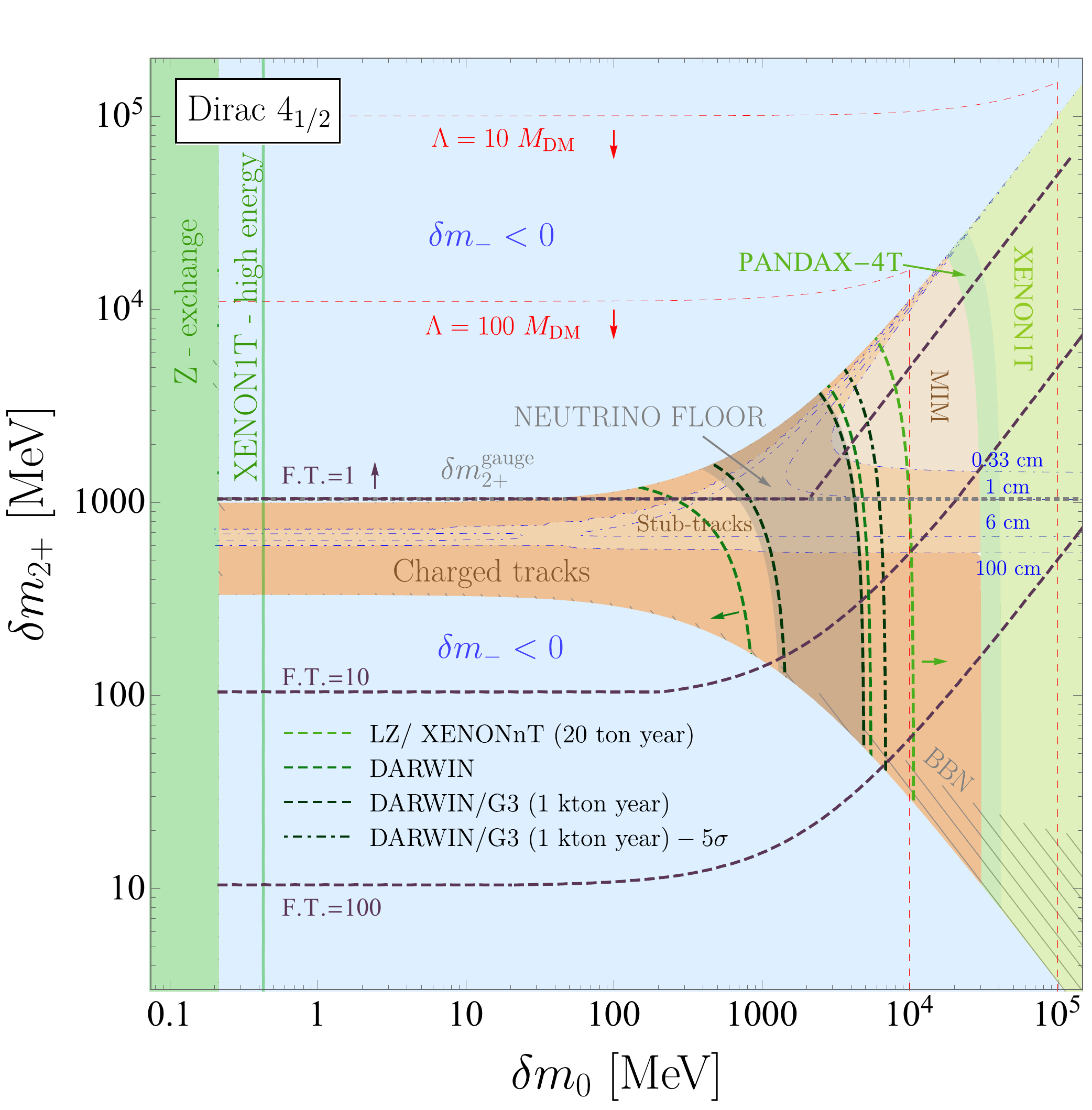}\hfill\includegraphics[scale=0.385]{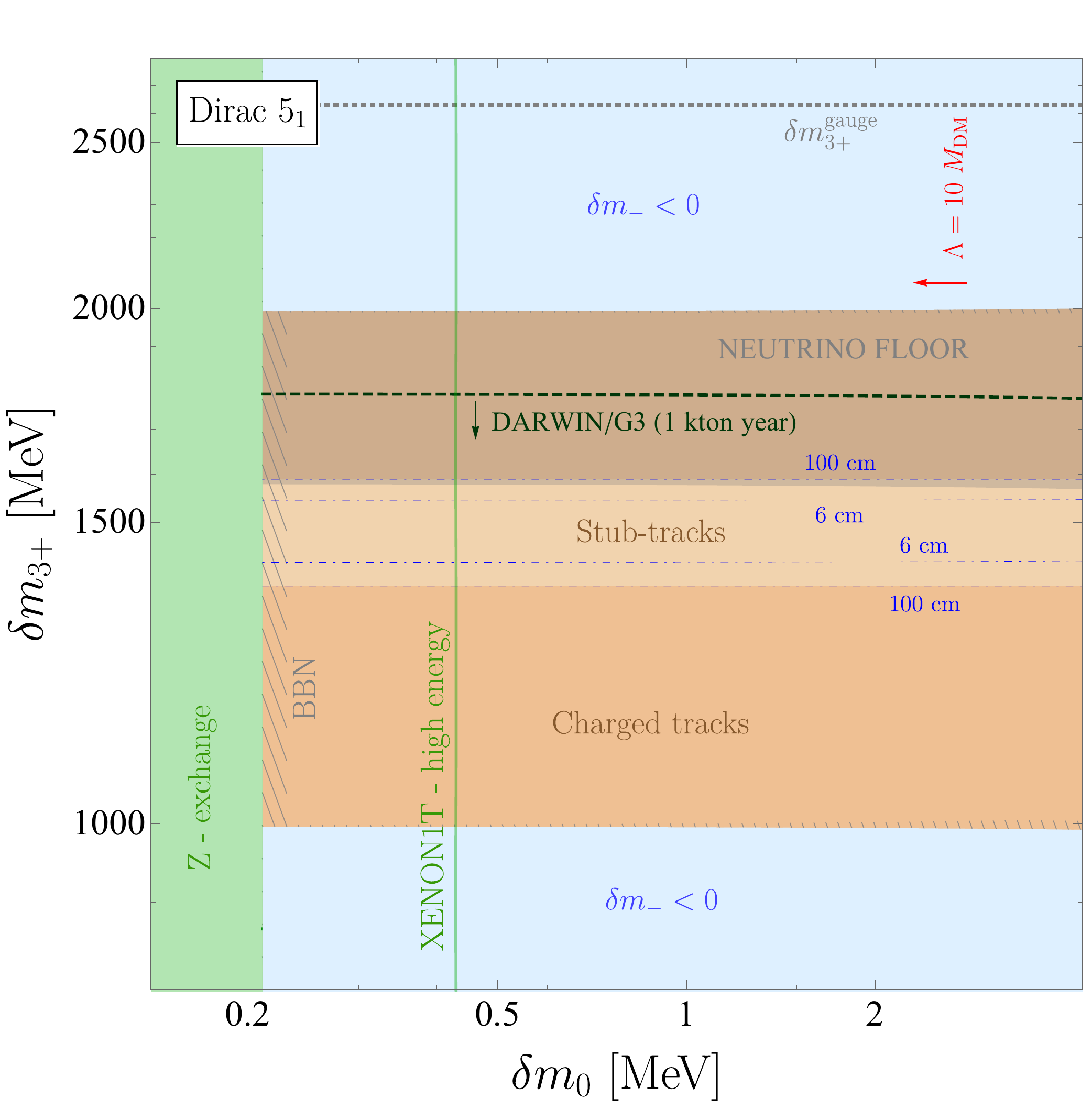}
     
\caption{ Direct Detection and collider signatures  in the plane $\delta m_0$ vs. $\delta m_{\Qbar}$ from the $2_{1/2}$ (\textit{upper left}),  $3_1$ (\textit{upper right}) $4_{1/2}$ (\textit{lower left}) and $5_1$ (\textit{lower right}). The {\bf green} shaded regions are excluded by current DD constraints on inelastic DM at small $\delta m_0$ and on Higgs-exchange elastic scattering at large splittings. {\bf Gray hatched} regions are excluded by BBN constraints on the longest lived unstable particle in the multiplet. {\bf Red} dashed lines delimit the range of perturbative mass splitting at fixed $\Leff/\mdm$ ratio. {\bf Light blue} shaded patches are not viable because the lightest WIMP in the $n$-plet is not the neutral one. {\bf Dashed green} lines show prospects from future high exposure xenon experiments: LZ, DARWIN and DARWIN/G3 (arrows pointing to the direction of the expected probed region). In the {\bf gray shaded} region the DD signal falls below the neutrino floor of xenon experiments~\cite{OHare:2021utq,Aalbers:2022dzr}. The vertical Xenon1T-high energy line show the ultime reach of xenon experiments on inelastic DM (see Fig.~\ref{fig:bbn}). Xenon {\bf Blue} and {\bf black} dashed lines corresponds to different expected lengths of charged tracks. Accordingly, different hues of {\bf brown} distinguish regions where different signatures at future colliders are expected. The {\bf dashed gray} line shows the EW value of $\delta m_{\Qbar}$. \textbf{Purple} dashed lines show the contours of the fine-tuning among the different mass splittings as defined in \eqref{eq:FT}. Mass splittings above the F.T.=1 line are not fine tuned.
\label{fig:doublet-triplet-fourplet-fiveplet} }
\end{center}
\end{figure*}

\subsection{Collider searches}
The splittings $\delta m_0, \delta m_{\Qbar}$ determine the lifetime of the charged components of the EW multiplet, which are pivotal to understand the viable collider signatures. In what follows we take as reference  the detector geometry proposed in \cite{Capdevilla:2021fmj} and we classify the parameter space $(\delta m_0, \delta m_{\Qbar})$  in various regions according to the lifetime $\taulcp$ of the Longest Lived Charged Particle (LCP) of a given multiplet.

For    $c\taulcp>1\;\mathrm{m}$ the LCP gives long charged tracks (CT)  with an average length roughly corresponding to the middle layer of the outer tracker. 
The SM background processes for this ``long'' track with anomalous properties strongly depends on the properties of the detector, therefore its study is outside the scope of this work. We limit ourselves to estimate the number of expected signal events of charged $\chi$ pair production at a muon collider, as shown in Fig.~\ref{fig:charged_tracks}.  We highlight that this results approximately hold for generic $\sqrt{s}$ in the domain of tens of TeV, as long as  \eqref{eq:bench} for the luminosity holds. In Fig.~\ref{fig:charged_tracks} we show the $\beta\gamma$ of the produced charged particle, which plays a crucial role to disentangle these special tracks from the SM background. 

Our estimates for the LCP signature is based on the counting the number of LCP produced. For the LCP we require $p_T>200$~GeV, $|\eta_\chi|<2$, inspired by LHC searches \cite{ATLAS:2019gqq}. For the $2_{1/2}$ and $3_1$ we include in our counting all charged particles production, assuming that $\chi^{2+}$  promptly decay to the long lived $\chi^+$. For the $4_{1/2}$ and $5_1$, which have more complicated spectra, we stick to the minimal splitting scenario for the estimate of the charged tracks yield, which corresponds to the minimal $y_{+}$ necessary to lift $\chi^{-}$ above $\chi_{\text{DM}}$. In this spectrum configuration $\chi^{-}$ is the most natural and only candidate to make long charged tracks, therefore we only consider this contribution in our result.\footnote{We neglect any possible contribution coming from $\chi^+$ decaying into $\chi^{-,c}$, since in the region where mixing is relevant, the charged particles lifetimes are too short and are in the ``stub tracks region'', as shown in Fig.~\ref{fig:doublet-triplet-fourplet-fiveplet}.
}

 For $1\;  \mathrm{m}>c\taulcp >0.33\;\mathrm{cm}$ the lightest charged WIMP gives a disappearing track signal. This search has been introduced in Sec.~\ref{sec:minimal splitting}  and it is further described in Appendix~\ref{app:collider}, in particular in Fig.~\ref{fig:dt_2dplot} where the reach in the plane mass-lifetime is displayed for judiciously chosen center of mass energies for every WIMP up to the $5_1$.

For $c\taulcp<0.33\;\mathrm{cm}$ the lightest charged  WIMP decays promptly on collider scales and gives missing energy signatures such as the mono-$W$ and mono-$\gamma$ discussed in Sec.~\ref{sec:minimal splitting}. The choice for the threshold value to consider the signal as prompt corresponds to $c \tau/2$ of the charged track for the $2_{{1/2}}$ WIMP in the minimal splitting scenario,  which is a known benchmark where DT reconstruction starts to become challenging. Full details on MIM searches are given in Appendix~\ref{app:collider} and in particular in Fig.~\ref{fig:fermionic_barchart}.

\subsection{Parameter space for complex WIMPs}

In the rest of this section, we examine in greater detail the parameter space spanned by $\delta m_0$ and $\delta m_{\Qbar}$  specifically looking at each of the lightest multiplets of Tab.~\ref{tab:complex} up to the $5_1$ WIMP.  
%\rf{Why not say something for $n>5$ and direct detection?}
 This will allow us to discuss the salient phenomenological features of the complex WIMPs.

\subsubsection{The Dirac $2_{1/2}$ and $3_1$}

The $2_{1/2}$ and $3_1$ are special WIMPs because they are the only multiplets with maximal hypercharge compatible with our assumptions. In particular, for $2_{1/2}$ requiring $\delta m_{0,+}>0$ automatically implies the neutral WIMP candidate is the lightest one.  Perturbativity requires $\delta m_0<40\mev$ for the $n=3_1$ WIMP, because of the strong suppression of the 7 dimensional operator generating the neutral splitting. The narrower range of $\delta m_0$ for the $3_{1}$ with respect to $2_{1/2}$ was expected from the  higher dimensionality of $\omix$ for $Y=1$ as compared to $Y=1/2$.

In Fig.~\ref{fig:doublet-triplet-fourplet-fiveplet} we show the constraints on the Dirac $2_{1/2}$ and $3_1$ in the parameter space spanned by $\delta m_0$ and $\delta m_{\Qbar}$ coming from present DD experiments like Xenon1T and PANDAX-4T, as well as prospect from future high exposure Xenon experiments, $i.e.$ LZ, XENONnT, DARWIN and DARWIN/G3. As we can see from \eqref{eq:higgs_tree_split}, these bounds depend solely on the combination 
\begin{equation}
\delta \mu\equiv 2\delta m_{\Qbar}+(2\Qbar-1)\delta m_0\ . 
\end{equation}
In particular we find
\begin{equation}
\label{eq:tree_vs_loops_2-3}
    \sigma_\text{SI}\approx 10^{-48}\text{cm}^2\left\{
    \begin{aligned}
    &\left(0.3-\frac{\delta\mu}{1 \gev}\right)^2,\  [2_{1/2}]\\
    &\left(0.2+\frac{\delta\mu}{1 \gev}\right)^2, \  [3_{1}]
    \end{aligned}\right.\, .
\end{equation}

In the region of low mass splitting for both $2_{1/2}$ and $3_1$ the direct detection cross section lies below the neutrino floor. For $2_{1/2}$, \eqref{eq:tree_vs_loops_2-3} shows that large cancellations between EW loops and tree-level Higgs exchange occur around $\delta\mu\simeq 300\mev$, while for $3_1$ the minimum $\sigma_\text{SI}$ is obtained for $\delta\mu=0$ and falls below the neutrino floor. To produce a direct detection cross section above the neutrino floor for the $2_{1/2}$ $(3_{1})$ we need $\delta \mu>1.6$ GeV ($\delta \mu> 2.0$ GeV).  PANDAX-4T already excludes mass splitting $\delta\mu<20$ GeV ($\delta \mu<30$ GeV) for the $2_{1/2}$ $(3_1)$. 

All in all, DD still leaves a large portion of parameter space unconstrained between the neutrino floor and the PANDAX-4T constraints. Remarkably, this region corresponds to mass splittings that do not require tuned adjustments of the three contributions to $\delta m_0$ and $\delta m_+$. The region of large splittings down to $\delta \mu\sim 1\text{ GeV}$ can be covered by large exposure Xenon experiments while the large portion of the parameter space lying below the neutrino floor should be taken as a major motivation for a future muon collider. 

For $2_{1/2}$ WIMP the neutrino floor region can be fully probed  only via a combination of charged tracks, DT and  MIM searches. The different search strategies become relevant depending on the $\delta m_+$ value, as shown in Fig.~\ref{fig:doublet-triplet-fourplet-fiveplet}. For the $3_1$ WIMP stub and charged tracks can exclude the entire neutrino floor region, while MIM and DT searches can be complementary to large exposure DD experiments to probe the rest of the parameter space.

While the plot in Fig.~\ref{fig:doublet-triplet-fourplet-fiveplet} show the possible regions in which DT can be reconstructed efficiently, they do not show if the reach is large enough to guarantee exclusion (or discovery) of the WIMP candidate. For these two WIMP candidates a detailed discussion was given in Sec.~\ref{sec:muoncollmin}. The general message is that for $\sqrt{s}$ sufficiently larger than twice the thermal DM mass DT searches will be powerful enough to probe the parameter space.

\subsubsection{The Dirac $4_{1/2}$ and $5_1$}

Being the first multiplet with non-maximal hypercharge, the constraints coming from DM stability shape the allowed parameter space of the $4_{1/2}$. In particular, we must require $\delta m_{-}$ to be positive. This constraint excludes the light blue shaded region in Fig.~\ref{fig:doublet-triplet-fourplet-fiveplet} bottom left.

 The region where the direct detection cross section lies below the neutrino floor is reduced to a tiny band for $4\gev<\delta \mu< 15\gev$. This region can be probed by direct searches at a future high energy muon collider with $\sqrt{s}>10 \text{ TeV}$ because of the thermal mass of the $4_{1/2}$ lies around 4.8 TeV. In particular, stub and charged tracks searches could cover almost the entire neutrino floor region, except for a small portion accessible only to MIM searches. The rest of the viable parameter space for the $4_{1/2}$ can be in principle probed by large exposure direct detection experiments.

We show the parameter space of the $5_1$ WIMP in Fig.~\ref{fig:doublet-triplet-fourplet-fiveplet} bottom right. The dominant constraint comes from the perturbativity of the operator generating the neutral splitting, which requires $\delta m_0<3\mev$. As a consequence, testing larger splittings for inelastic DM at Xenon1T can already probe large portions of the allowed parameter space of the $5_1$. DM stability requires $1\gev\lesssim\delta m_{3+}\lesssim 2\gev$. The parameter space can be fully excluded by charged and stub track searches if a muon collider of $\sqrt{s}>20$ TeV will be constructed. Besides, all the allowed mass splittings except a small window $1.8\gev\lesssim \delta m_{3+}\lesssim 2.0\gev$ can be probed by DARWIN/G3.

\section{Indirect probes of complex WIMPs}\label{sec:indirect}
In this section we briefly comment on indirect probes of complex WIMPs, focusing in particular on the distinctive imprints of their Higgs interactions on precision electroweak observables (Sec.~\ref{sec:EWPT}) and on the electron dipole moment (Sec.~\ref{sec:EDM}).

\subsection{Electroweak precision observables}\label{sec:EWPT}
The addition of new EW multiplets to the SM leads to deviations in EW observables which could be tested with LEP data, at the LHC and at future colliders~\cite{Farina:2016rws,DiLuzio:2018jwd}. Since the presence of new EW multiplets affects mainly gauge bosons self energies, their indirect effects can be encoded in the oblique parameters~\cite{Peskin:1991sw,Barbieri:2004qk}. Here we briefly summarize the main features of these contributions following the notation of Ref.~\cite{Barbieri:2004qk}. The most relevant oblique parameters can be related to the SM gauge boson vacuum polarizations 
\begin{equation}
\begin{aligned}
&\yhat = \frac{g_Y^2m_W^2}{2}\Pi''_{BB}(0)\quad ,\quad \what = \frac{g_2^2 m_W^2}{2}\Pi''_{33}(0)\ ,\\
&\shat = \Pi_{3B}'(0)\quad ,\quad \that = \frac{\Pi_{33}(0)-\Pi_{WW}(0)}{m_W^2}\ ,
\end{aligned}
\end{equation} 
where $\Pi_{ij}$ appear in the kinetic terms of the EFT describing the SM vectors interactions at energies smaller than the DM mass
\begin{equation}
    \mathscr{L}_\text{oblique}=-\frac{1}{2}V_i^\mu\Pi_{ij}V_{j,\mu}\ .
\end{equation}

In particular, all the EW multiplets (both real and complex) give a universal contribution to the $\hat{W}$, $\hat{Y}$ as previously found in \cite{Cirelli:2005uq}  
\begin{widetext}
\begin{align}
    &\hat{W}=\frac{\alpha_\text{em}}{180\pi}\frac{m_Z^2}{\mdm^2}\kappa n(n^2-1)\simeq 8\kappa\times 10^{-7}\left(\frac{1\tev}{\mdm}\right)^2\!\!\left(\frac{n}{2}\right)^3\ ,\label{eq:WWIMP}\\
    &\hat{Y}=\frac{\alpha_\text{em}}{15\pi}\frac{m_Z^2}{\mdm^2}\kappa nY^2\!\simeq 5\kappa\times 10^{-7} \left(\frac{Y}{1/2}\right)^2\left(\frac{1\tev}{\mdm}\right)^2\left(\frac{n}{2}\right) \ ,\label{eq:YWIMP}
\end{align}
\end{widetext}
where $\kappa= 1,1/2,1/8,1/16$ for Dirac fermions, real fermions, complex scalars and real scalars respectively. In the numerical estimates we normalized the expected contributions to the fermionic $2_{1/2}$. The rough expectations for heavier WIMPs can be obtained from the equations above by remembering that $M_{\text{DM}}\sim n^{5/2}$. The  contribution to $\hat{W}$ scale like as $\sim 1/n^2$ while the one to $\hat{Y}$ as $\sim 1/n^4$.

$\hat{W}$ and $\hat{Y}$ induce effects in SM observables that grow with energy so that LHC searches have already ameliorated the sensitivity on these operators compared to LEP. In particular $\hat{W}$ has been recently bounded by CMS~\cite{CMS:2022yjm} and $\hat{Y}$ is also set to be tested with a similar precision~\cite{Strumia:2022qkt}. Even if the current precision is not sufficient to probe the WIMP thermal masses, further improvements are expected in a future high energy muon collider. This could provide interesting indirect tests of EW WIMPs~\cite{DiLuzio:2018jwd,Chen:2022msz,RFXZ}. The size of the expected WIMP contributions in \eqref{eq:WWIMP} and \eqref{eq:YWIMP}  the results of Ref.~\cite{RFXZ} indicates that a muon collider of $\sqrt{s}=30\text{ TeV}$ would be needed to observed these deviations.

As extensively discussed in this paper, complex WIMPs require contact interactions with the SM Higgs to be phenomenologically viable. These interactions give an additional contributions to the $\hat{S}$ parameter. Moreover, the mass splittings inside the EW multiplet induced by $\omix$ and $\ochgd$ break the custodial symmetry in the Higgs sector. As a consequence, complex WIMPs give irreducible contributions also to the $\hat{T}$ parameter. The explicit formulas for fermionic WIMPs are 
\begin{widetext}
\begin{align}
    &\hat{S}_F=-\frac{\alpha_{\text{em}}Yn^3}{9\pi s_{2W}\mdm}\delta\mu_{\Qbar}\simeq 1.2 \times 10^{-5}\left(\frac{n}{2}\right)^3\left(\frac{1 \tev}{\mdm}\right)\left(\frac{\delta\mu_{\Qbar}}{10\gev}\right)\, ,\label{eq:SWIMP}\\
    &\hat{T}_F=-\frac{\alpha_{\text{em}}}{ 18\pi s_{2W}^2}n^3\frac{\delta\mu_{Q_M}^2+\delta\mu_{0}^2}{m_Z^2}\simeq 1.6  \times 10^{-5}\left(\frac{n}{2}\right)^3\frac{\delta\mu_{\Qbar}^2+\delta\mu_0^2}{100 \gev^2}\, ,
\end{align}
\end{widetext}
where $s^2_{2W}=\sin^22\theta_W\simeq0.83$ while $\delta\mu_{\Qbar}$ and $\delta\mu_0$ are defined in \eqref{eq:split_def}. %$\delta\mu_{\Qbar}=y_+v^2/4\Leff$ and $\delta\mu_0=\delta m_0/n$. 
The corresponding formulas for scalar WIMPs are easily obtained from the fermionic ones by replacing $\shat_S=\frac{1}{4}\shat_F$ and $\that_S=-\that_F$. Our result for $\that_S$ is consistent with the result in Ref.~\cite{DiLuzio:2015oha}, first computed in \cite{Lavoura:1993nq}.

Both $\hat{S}$ and $\hat{T}$ are currently constrained at the level of $3\times 10^{-3}$  by LEP data (where the precise value will of course depend on the correlation between these two parameters~\cite{Barbieri:2004qk}). High luminosity colliders further exploring the $Z$ pole like FCC-ee will improve this precision typically by a factor of 10~\cite{Fan:2014vta} with the ultimate goal of pushing the precision of EW observables to the level of $10^{-5}$~\cite{Buttazzo:2020uzc}. 

Concerning $\hat{S}$ we conclude that the deviations induced by complex WIMPs are unlikely to be visible unless for light multiplets with the ultimate EW precision. Conversely, at fixed large splitting and increasing size of the multiplet, the deviations on $\hat{T}$ are enhanced to the point that a $12_{1/2}$ multiplet is giving $\hat{T}\simeq 3\times 10^{-3}$, which is within the reach of current LEP data and could help reducing the tension the $m_W$ mass extracted from the EW fit and the one measured directly by the CDF collaboration~\cite{CDF:2022hxs}. Of course this extreme scenario requires very large couplings at the boundary of perturbativity for the contact operators inducing the splitting in \eqref{eq:ren_lag}. 

\subsection{Electric dipole moment}\label{sec:EDM}
As mentioned in Sec.~\ref{sec:WIMP} and detailed in Appendix~\ref{sec:operators}, the operators generating the splittings for fermionic WIMPs have in general two different chiral structures and different relative phases. This generically induces operators contributing to the electron EDM at 1-loop~\cite{Panico:2018hal,Cesarotti:2018huy}. Following the notation of Ref.~\cite{Panico:2018hal}, the leading operator generated by WIMPs loops is 
\begin{equation}
    \mathscr{L}_{\text{EDM}}\supset \frac{c_{WW}}{\Lambda^2_{\text{UV}}} \vert H\vert^2 W\tilde{W}\ ,\label{eq:edmleading}
\end{equation}
where we neglect for this discussion operators such as $\vert H\vert^2 B\tilde{B}$ and $H^{\dagger}\sigma^a H W\tilde{B}$ that would be suppressed by powers of $Y/n$ with respect to the one with two $SU(2)$ field strength insertions. 

The Wilson coefficient in \eqref{eq:edmleading} is constrained by the recent measurement of the ACME collaboration~\cite{ACME:2018yjb} to be 
\begin{equation}
c_{WW}< 4\times 10^{-3}g_2^2\left( \frac{ \Lambda_{\text{UV}}}{10\text{ TeV}}\right)^2\frac{d_e}{1.1\times 10^{-29}\text{e\,cm}}\ .\label{eq:EDMbound}
\end{equation}

The effective operator in \eqref{eq:edmleading}, is generated at $4Y$-loops from the $\mathcal{O}_0$ operator and its axial partner and at 2-loop from the $\mathcal{O}_+$ operator and its axial partner. For example the 2-loops Wilson coefficients can be estimated as 
\begin{equation}
 c_{WW}\vert_{2-\text{loop}}\simeq \frac{g_2^2 n (n^2-1)^2\mathcal{I}[y_I y_{I,5}^{\ast}]}{32 (16\pi^2)^2\Lambda_{
\text{UV}}^2}\ ,
\end{equation}
where the index $y_I$ controls the contribution from $\mathcal{O}_0$ or $\mathcal{O}_+$ and $y_{I,5}$ the ones from their axial counterpart. If only  $\mathcal{O}_0$ is present, like it would be for instance in the minimal splitting case for the $2_{1/2}$ multiplet, the bound in \eqref{eq:EDMbound} can be rewritten as an upper bound on the neutral splitting 
\begin{equation}
\delta m_0<16\text{ GeV}\left(\frac{2}{n}\right)^{\frac{3}{2}}\left|\theta_{\text{CP}}\right|^{\frac{1}{2}}\sqrt{\frac{|d_e|}{1.1 \times 10^{-29}\text{e\,cm}}}\ ,
\end{equation}
where we defined $\theta_{\text{CP}}=y_I^2/\mathcal{I}[y_I y_{I,5}^{\ast}]$. A similar bound is found for $\delta m_{\Qbar}$ in the minimal splitting case, where $\omix$ can be neglected. Hence, for $\mathcal{O}(1)$ couplings and CP-violating phases, the EDM already gives competitive if not stronger bounds with respect to those shown in Table~\ref{tab:TableI}. Interestingly, future experimental upgrades are expected to be sensitive to electron EDM as small as $d_e\simeq 10^{-34}\text{e\, cm}$~\cite{Alarcon:2022ero} which would provide sensitivity to the WIMP parameter space down to splittings of order $\sim 50\text{ MeV}$ for $\mathcal{O}(1)$ couplings and CP-violating phases. 

It would be interesting to further explore this direction in models where the $\mathcal{O}(1)$ CP-violating phase is motivated by a mechanism producing the observed baryon asymmetry in the Universe such as WIMP baryogenesis~\cite{Cui:2011ab,Cui:2015eba}.    
 
\section{Conclusions}
In this study we derived the first full classification of WIMP in complex representations of the EW group. Our only assumption is that below a certain UV scale the SM is extended by the addition of a single EW multiplet containing the DM candidate.  Our main results are contained in Table~\ref{tab:TableI}, where we give all the freeze-out predictions for the WIMP thermal masses, the ranges of phenomenologically viable splittings and the maximal scale separation between the DM mass and the UV physics generating the splittings. 

We then inspect the phenomenology of complex WIMPs with the goal of understanding what would be required experimentally to provide the ultimate test on the EW nature of DM. We focus on the interplay between future large exposure direct detection experiments and a future high energy lepton collider, e.g. a muon collider. The reach from the current and forthcoming indirect detection experiments of pure electroweak multiplets are very promising (see e.g.~\cite{Cirelli:2015bda, HESS:2018kom} for current reaches and~\cite{Lefranc:2016fgn, Rinchiuso:2020skh} for prospects). Neverthless a robust assessments of the sensitivities for all the complex WIMPs classified here and the real WIMPs classified in \cite{Bottaro:2021snn} would require a careful evaluation of the expected annihilation cross sections and an assessment of the astrophyisical uncertanties~\cite{Lefranc:2016dgx}. We leave this work for a future investigation. Similarly, the prospects of discovering WIMPs via kinetic heating of compact astrophysical objects, like neutron stars~\cite{Baryakhtar:2017dbj, Bell:2020jou} or white dwarfs~\cite{Bell:2021fye}, should be further assessed by the future James Webb Space Telescope infrared surveys~\cite{Gardner:2006ky}. 

Our main result is that a kiloton exposure Xenon experiment would be able to probe most of the complex and the real WIMPs, with the notable exceptions of the complex doublet with $Y=1/2$ whose cross section lies naturally well below the neutrino floor. This result for the complex doublet was of course well known from the many previous studies on the SUSY Higgsino~\cite{Krall:2017xij}, but it gets further substantiated in the context of our  WIMP classification. Our findings  constitute a major motivation to push forward the research and development of a future high energy lepton collider. Specifically, we find that a future muon collider with $\sqrt{s}=6 \text{ TeV}$ would be able to fully probe the existence of a fermionic complex WIMP doublet at its thermal mass around 1.1~TeV. 

We further study the parameter space of the different complex WIMPs and find regions where the direct detection cross section can drop below the neutrino floor because of accidental cancellations of the EW elastic scattering (like for the $5_1$) or at small (mildly tuned) values of the charged-neutral mass splittings (like for the $3_1$) or for the destructive interference of EW and Higgs induced scattering (like for the $4_{1/2}$). In all these cases a high energy lepton collider might be again the only way of discovering these WIMPs. 

Interestingly, the list of WIMP candidates we furnished provides a series of targets that can be probed at successive stages of a future machine. Moreover, our analysis shows that future searches for long-lived charged tracks will be crucial to fully probe the WIMP parameter space. This strongly motivates a detailed collider study assessing the expected sensitivity of these searches in a realistic muon collider environment. We hope to come back to this issue in the near future. 

\section*{Acknowledgments}
We thank Ranny Budnik and Itay Bloch for useful insights on Xenon1T, Scott Haselschwardt for enlightening discussions about the future of Xenon experiments, Rodolfo Capdevilla, Federico Meloni, Rosa Simoniello and Jose Zurita for comparison on the disappearing track prospects. Simon Knapen, Ennio Salvioni and Andrea Tesi for discussions.

\appendix

\section{Complex WIMP classification}
In this appendix we give further details on our complex WIMP classification. In Sec.~\ref{sec:operators} we explicitly classify the possible UV operators, substantiating the result of Sec.~\ref{sec:WIMP}. In Sec.~\ref{sec:scalarWIMP} we give the results for scalar WIMPs. 
\subsection{UV Operators}\label{sec:operators}
In Sec.~\ref{sec:WIMP} we showed that in order to make complex WIMPs with $Y\neq 0$ viable, new UV sources of splitting between the charged and neutral components and among the neutral components are necessary. In \eqref{eq:ren_lag} we showed how the neutral splitting can be generated by $\mathcal{O}_{0}$ and the charged-neutral splitting by $\mathcal{O}_{+}$. Here we take a step back and investigate the generality of this choice. 

The general form of an operator responsible for a Majorana mass term for $\chi_N$ after EWSB is
\begin{equation}
\label{eq:omix_form}
\overline{\chi}(y_0+y_{0,5}\gamma_5)\mathcal{I}\chi^c H^{4Y}\ ,
\end{equation}
where the $H^{4Y}$ is necessary to match the hypercharge of the $\chi^2$ piece, while $\mathcal{I}$ is a SU(2)$_L$ tensor. We are crucially assuming that the Higgs is the only scalar picking a VEV after EWSB. Under this assumption, since the Higgs is a boson, the only  surviving SU(2)$_L$ structure is the totally symmetric combination. This can be seen as the symmetric combination of $2Y$ Higgs pairs in the isotriplet representation (the isosinglet is antisymmetric and vanishes identically), so that we are left with $\omix$ defined in \eqref{eq:ren_lag} and its axial counterpart with the $\gamma_5$ insertion. The latter can be shown to give only subleading contributions to the mass splitting between the neutral components of $\chi$. Assuming $y$ to be real, the shift on $\delta m_0$ induced by  $y_{0,5}\neq0$ with respect to its expression in \eqref{eq:neutral_splitting} is
\begin{equation}
    \delta m_0 \to \delta m_0\sqrt{ 1 + 4\Re[y_{0,5}]^2\frac{\Leff^2}{\mdm^2}\left(\frac{v}{\sqrt{2}\Leff}\right)^{8Y}}\, ,
\end{equation}
which is highly suppressed for $\Leff>\mdm>v$. The operator $\omix$ in \eqref{eq:ren_lag} is then the dominant contribution to the inelastic splitting among the neutral components.

The isospin structure and field content of $\ochgd$ is already the minimal required to generate additional splitting between the charged components, the only possibility is again to change its chiral structure writing the general operator inducing charged-neutral splitting as
\begin{equation}
   \overline{\chi}T^a(y_+ + y_{+,5}\gamma_5)\chi H^\dagger\sigma^a H\ .
\end{equation}
Similarly to the neutral case, the $\gamma_5$ insertion leads to a subleading shift in the mass splittings with respect to the value of \eqref{eq:deltaM}
\begin{equation}
M_Q^2\to M_Q^2 +y_{+,5}^2\frac{v^4}{16\mdm^2\Lambda^2}(Q-Y)^2 \ .
\end{equation}

Finally, we comment on the implications on DD signals. Both operators with the $\gamma_5$ insertions give additional contributions to the SI cross-section and match to the effective operators:
\begin{equation*}
\begin{split}
\mathscr{L}_{\text{eff}}^{\rm{SI}} = &\tilde{f}_q m_q \bar{\chi} i\gamma_5\chi \bar{q} q + \frac{\tilde{g}_q}{\mdm} \bar{\chi} i \partial^{\mu} \gamma^{\nu} i\gamma_5\chi \mathcal{O}^q_{\mu\nu}\\
&+ \tilde{f}_G \bar{\chi} i\gamma_5\chi G_{\mu\nu}G^{\mu\nu}.
\end{split}
\end{equation*}
However, all these operators are strongly momentum-suppressed~\cite{Cirelli:2013ufw, Freytsis:2010ne}, so that their corrections to $\sigma_{\text{SI}}$ are expected to be $(q/m_N)^2 \sim 10^{-6}$ smaller than those from \eqref{eq:nogamma}, and thus negligible.

%\DR{add the most general direct detection cross section at least for maximal Y multiplets}\SK{Done}

\subsection{Complex Scalar WIMPs}~\label{sec:scalarWIMP}
Following the discussion in Sec.~\ref{sec:WIMP} for the fermions, supported by the previous Appendix~\ref{sec:operators}, the minimal Lagrangian for a scalar complex WIMP is
\begin{align}
&\mathscr{L}_{\text{S}} =|D_\mu\chi|^2-M_\chi|\chi|^2 +\frac{y_0}{\Leff^{4Y-2}}\mathcal{O}_{0}^S + y_+\mathcal{O}_{+}^S+\hc\, ,\notag\\
&\omix^S=\frac{1}{2(4Y)!}\left(\chi^\dagger(T^a)^{2Y}\chi^c\right)\left[H^{c\dagger}\frac{\sigma^a}{2} H\right]^{2Y}\, , \label{eq:ren_lag_scal}\\
&\ochgd^S=-\chi^\dagger T^a\chi H^\dagger\frac{\sigma^a}{2} H\, .\notag
\end{align}

Conversely to the fermionic case, no additional operators can be written to generate the fundamental mass splitting. The neutral and charged \textit{squared} mass splitting, $\mu_Q^2=M_Q^2-\mdm^2$, can be written as
\begin{widetext}
\begin{equation}
\begin{split}
    &\mu_0^2 =4y_0 c_{nY0}\Leff^2\left(\frac{v}{\Leff\sqrt{2}}\right)^{4Y}\, ,\\
    &\mu_{Q}^2=\frac{\mu_0^2}{2}+2\mdm\delta_gQ^2+\text{sgn}(Q)\sqrt{\left(\frac{4Y\delta_g\mdm}{\cos\theta_W}-\frac{y_+v^2}{4}\right)^2Q^2+\frac{\mu_0^2}{4}\frac{c_{nYQ}^2}{c_{nY0}^2}}.
\end{split}
\end{equation}
\end{widetext}
The linear mass splittings, $\delta m_Q=M_Q-\mdm$, are then given by $\delta m_Q=\mu^2_Q/(2\mdm)$. Notice that for $Y=1/2$ all the operators are renormalizable and any dependence on the cutoff disappears.\\
The lower bound on $\delta m_0$ from DD is identical to that for fermions, since $\sigma_\text{SI}$ does not change. Instead, the BBN bound differs since the one-loop decay channel $\chi_0\rightarrow\chidm \gamma$ now is now heavily suppressed with respect to the three-body decays. As a consequence, the BBN condition becomes
\begin{equation}
    \Gamma_{\chi_0}\equiv \Gamma_{\bar{\nu}\nu}+\Gamma_{\bar{e}e}>6.58 \times 10^{-25}\text{ GeV}\, ,
\end{equation}
which explains why the bound on $\delta m_0^{\text{min}}\simeq 4-5\mev$ for scalars is so much stronger than the one for fermions, as shown in Fig.~\ref{fig:bbn}. For $Y\neq 1/2$, the lower bound on $\delta m_0$ sets the upper bound of the allowed window for $\Leff$ 
\begin{equation}
10\, M_{\text{DM}}<\Leff\leq \left(\frac{2y_0c_{nY} v^{4Y}}{2^{2Y}\delta m_0^{\min}\mdm}\right)^{\frac{1}{4Y-2}}\, , \label{eq:scalarwindow}
\end{equation}
which is the analogous of \eqref{eq:UV_dm0} for scalar WIMP. Once set $y_0=(4\pi)^{4Y}$ to its NDA maximal value and given the scaling $\mdm\sim  n^{5/2}$ \eqref{eq:scalarwindow} can be used to determine the viable EW multiplets. It turns out that the allowed multiplet are the same as for fermions, $i.e.$ all even $n_{1/2}$ plus $3_1$ and $5_1$. 

Finally, we discuss DM stabiity for scalars. All the $Y=1/2$ scalars are never accidentally stable, since DM decay can be induced by the renormalizable operator $\chi^2\chi^\dagger H^c$. Similarly, for $3_1$ we can write $\chi^\dagger H^2$, while for $5_1$ the lowest dimensional operators are:

\begin{equation}
    \mathscr{L}=\frac{C_1}{\Leff}\chi^\dagger H^2 (H^\dagger H) + \frac{C_2}{\Leff}(\chi^\dagger\chi)(\chi^\dagger H^2)\, ,
\end{equation}
which ensures stability for $\Leff>10^{21}\mdm$, way larger than $\Leff<20\mdm$ from $\delta m_0$ perturbativity. 

Results for the collider reach on the scalars are given in Fig.~\ref{fig:scalar_barchart} in Appendix~\ref{app:collider}.

\section{Complex WIMP cosmology}\label{app:boundstates}

In this Section, we briefly describe the cosmological evolution of the Complex WIMP candidates. In particular, we  highlight the main differences with respect to real WIMPs and refer the reader to \cite{Bottaro:2021snn,Mitridate:2017izz,Harz:2018csl} for further details.

\subsection{Sommerfeld Enhancement and Bound State Formation}

In the non-relativistic regime, long range interactions in the form of Yukawa or Coulomb potentials significantly affect the WIMP annihilation dynamics. In particular, these interactions can distort the wave function of the incoming particles leading to Sommerfeld Enhancement (SE) or lead to WIMP Bound State Formation (BSF). For WIMPs, the potential induced by EW interactions is attractive for all isospins $I\lesssim \sqrt{2}n$, $I$ denoting the dimension of the isospin representation of the pair of particles. In these channels, the SE increases the annihilation cross-section by a factor which, for small velocities, can be approximated as
\begin{equation}
\label{eq:som}
S_E^I \approx \frac{2\pi \alphaeff}{\vrel}\ ,
\end{equation}
where $\alphaeff = \alpha_2(I^2+1-2n^2)/8  + Y^2 \alpha_Y\equiv \alpha_{2,\text{eff}} + Y^2 \alpha_Y$. Hence, the annihilation cross-section reads
\begin{equation}
\langle\sigma_{\text{ann}}^\text{SE}\vrel\rangle=\sum_I \langle S_E^I\sigma_{\text{ann}}^I\vrel\rangle\ ,
\end{equation}
where $\sigma_{\text{ann}}^I$ is the perturbative, hard annihilation cross-section in the isospin channel $I$. The $SU(2)_L$ symmetric limit approximate better the exact result the more we increase the dimensionality of the EW multiplet. This is because most of the freeze-out dynamics for large $n$-plets occurs before the EWSB. For smaller multiplets the symmetric approximation generically fails because EWSB become important. In our computation, we always assume the symmetric limit for $n\geq 6$, while we compute the SE in the broken phase for smaller multiplets. 

Besides, the mass splittings generated by $\omix$ and $\ochgd$ can alter the pattern of resonances in the SE, so that the thermal mass becomes a function of $\delta m_0$ and $\delta m_+$ as well. In Table \ref{tab:TableI} we assume the minimal $\delta m_0$, where the effect of the splitting is safely negligible, and account for the dependence on $\delta m_+$ in the theoretical uncertainty. More details about the effect of these UV operators are discussed in Sec.~\ref{sec:splitting}. Another source of uncertainty in the computation of $\sigma_{\text{ann}}^\text{SE}$ comes from neglecting partial waves higher than the s-wave. We estimate the correction to $\sigma_{\text{ann}}^\text{SE}$ as

\begin{equation}
    \frac{\sigma_{\text{ann},\text{p-wave}}^\text{SE}}{\sigma_{\text{ann},\text{s-wave}}^\text{SE}}\simeq \vrel^2+\alphaeff^2\ .
\end{equation}
\\

In addition to the SE, attractive isospin channels lead to BSF through the emission of one (or more) EW gauge boson in the final state $\chi_i + \overline{\chi}_j \rightarrow \text{BS}_{i'j'} +V^a$, where $V_a$ can be either a weak or hypercharge gauge boson. In the non-relativistic limit, and for single gauge boson emission, the  BSF process is induced by an electric dipole moment emission whose dynamics is encoded in the effective Hamiltonian described in Ref.~\cite{Mitridate:2017izz,Harz:2018csl}. The main difference with respect to real WIMPs are the different selection rules selecting the bound states allowed by symmetries. In the real case, the bound states are of the form $\chi\chi$ and the (anti-)symmetry of the wave function allows only for those BS satisfying $P_\text{BS}\equiv (-1)^{L+S+\frac{I-1}{2}}=1$ , while for complex WIMPs the bound states are of the form $\overline{\chi}\chi$ and the above selection rule no longer applies. \\
The BSF cross-section can be computed in the $SU(2)_L$ symmetric limit with good approximation and scales as:
\begin{equation}
\label{eq:BSF}
\sigma_{B_I}\vrel\simeq \frac{E_{B_I}a_B}{\mdm n^2}(\alpha_{2,\text{eff}}S_E^{I\pm 2}+Y^2\alpha_Y S_E^{I})\ ,
\end{equation}

where the first and second term account for weak and hypercharge vector boson emmission, respectively,  $E_{B_I}\approx \alphaeff^2\mdm/4n_B^2$ - $n_B$ being the energy level - is the binding energy and $a_B=1/\alphaeff \mdm$ the Bohr radius.\\
Once the BS is formed, it can annihilate into SM, decay to lower lying BS, both with rate $\Gamma\approx \alpha_{2,\text{eff}}^2\alphaeff^3\mdm$ or $Y^2\alpha_{Y}^2\alphaeff^3\mdm$, depending on the mediator, or be broken by the interactions with the plasma (ionization). The ionization rate limits the efficiency of DM annihilation through BSF and it is Boltzmann suppressed as $\sim e^{E_{B_I}}/T$. The Boltzmann suppression gets bigger as we increase the dimensionality of the multiplet, due to the larger binding energies, so that neglecting BS ionizations is a good approximation for large EW multiplets. Under this approximation, the overall DM effective annihilation cross-section reads
\begin{equation}
\label{eq:effxsec}
\langle\sigma_{\text{eff}}\vrel\rangle\equiv \langle\sigma_{\text{ann}}^\text{SE}\vrel\rangle+\sum_{B_J}\langle \sigma_{B_J}\vrel\rangle,
\end{equation}
that is, once a BS is formed it eventually annihilates to SM without being destroyed. In \cite{Bottaro:2021snn} we checked the validity of this approximation for $n\leq 7$ and found that it overestimates the thermal mass of about 5 TeV. We assume this holds also for complex WIMPs and include this uncertainty in the estimate of our theoretical uncertainty. In the symmetric limit the effect of the mass splittings generated after EWSB are not taken into account. We expect these to make the heavier components of the multiplet decouple earlier than the lighter ones, thus reducing the cross-section. In order to estimate the error due to this approximation, we compute the thermal mass first setting to zero $\sigma_{B_I}$ for $T<\max\delta m_Q$ and then including BSF until the DM abundance saturates. These two effects are the dominant sources of error for $n\lesssim 8$. 

For larger multiplets, the effective cross-section in \eqref{eq:effxsec} suffers from additional theoretical uncertainties. Estimating from the results of Ref.~\cite{Urban:2021cdu} the NLO correction to the potentials controlling the SE we get
\begin{equation}
\label{eq:1loop}
\frac{\Delta V_{\text{NLO}}}{V_{\text{LO}}}\sim \beta_2^{\text{SM}}\frac{\alpha_2}{4\pi }\log\left(\frac{m_W}{\mdm}\right)\, ,
\end{equation}
 where $\beta_2^{\text{SM}}=-19/6$ is the SM contribution to the SU(2)$_L$ beta function. \eqref{eq:1loop} results in a correction to the Sommerfeld factor $S_E$, which affects both $\sigma_{\text{ann}}^\text{SE}\vrel$ and $\sigma_{B_J}\vrel$ as introduced in \eqref{eq:BSF}. NLO corrections start dominating the theoretical uncertainty for $n\gtrsim 8$.

\subsection{Perturbative Unitarity Bound}\label{sec:unitaritybound}
 
The perturbative unitarity of the S-matrix sets an upper bound on the size of each partial wave contribution to the total annihilation cross-section \cite{Cabibbo:1961sz,Cabibbo:1974ke, Griest:1989wd}:

\begin{equation}
(\sigma_\text{eff} \vrel)^{J}\leq \frac{4\pi(2J+1)}{\mdm^2 \vrel}\ ,
\end{equation}

where $\vec{J} = \vec{L} + \vec{S}$ is the total angular momentum. The stronger inequality comes from the $s$-wave channel (i.e. $J=0$) which can be written as
\begin{equation}
\label{eq:unit_bound}
(\sigma_\text{ann} \vrel)+\sum_{B_J} f_{B_J}^0 (\sigma_{B_J} \vrel)\leq \frac{4\pi}{\mdm^2 \vrel}\ ,
\end{equation}
where $f_{B_i}^0$ selects the BS contributions that can be formed by $J=0$ initial wave. As we noticed in the previous section, the BSF contribution is larger for complex WIMPs compared to the real case due to the larger multiplicity of bound states that can be formed. While for scalar WIMPs only p-orbitals bound states contribute to the s-wave unitarity, for fermionic WIMPs additional contributions arise from selecting the projection onto the $J=0$ partial wave of the BSF cross-section of $ns_I^{S=1}$ states.

\subsection{Impact of mass splittings}\label{sec:splitting}

Here we discuss under which conditions the effect of the UV splittings can be neglected in the prediction of the thermal mass.  The UV splittings can affect the freeze-out computation in two ways: i) they directly contribute to the DM annihilation cross-section into Higgs bosons, ii) after the EWSB, the interaction \eqref{eq:tree_DMh} generates a Higgs-mediated Yukawa potential, thus affecting the Sommerfeld enhancement. For definiteness, we focus on fermionic WIMPs and on the $\mathcal{O}_0$ contributions. We checked that similar conclusions hold for scalar WIMPs and for the contrbutions from  $\mathcal{O}_+$. 

For $Y=1/2$ WIMPs, the hard annihilation cross-section into higgses induced by $\omix$ can be estimated as
\begin{equation}
    \sigma_{2H} \vrel \simeq \frac{3y_{0,+}^2}{32n^2\pi\Leff^2}T_R\, ,
\end{equation}
where $T_R=n(n^2-1)/16$ is the Dynkin index of the DM $SU(2)_L$ representation. This should be compared with the typical EW hard cross-section, which is
\begin{equation}
    \sigma_{\text{EW}} v_\text{rel}\simeq \frac{\pi\alpha_2^2(n^2-1)^2}{32n\mdm^2}\, .
\end{equation}
The condition $\sigma_{2H}/\sigma_{\text{EW}}<1$ can be translated into an upper bound on the mass splittings. For instance for the inelastic mass splitting we get
\begin{equation}
\label{eq:up_delta0_higgs}
    \delta m_0< 1.6\gev \left(\frac{1\tev}{\mdm}\right) n^2\ .
\end{equation}
Since $\mdm\sim n^{5/2}$ we get stronger upper bound on the splitting for large multiplets. All in all, we do not expect significant changes in the picture presented in Sec.~\ref{sec:nonminimal splitting} even though for large splittings (above 1 GeV) one can get $\mathcal{O}(1)$ effects on the thermal freeze-out predictions for small multiplets. 

For $Y=1$ the contribution from $\omix$, which now is a 7D operator which controls the 4-body process $\overline{\chi}\chi\rightarrow 4H$ whose cross section can be estimated as
\begin{equation}
    \sigma_{4H} v_\text{rel}\approx\frac{\pi y_0^2\mdm^4}{(16\pi^2)^3\Leff^6}\frac{(n^2-1)^2}{64n}\, .
\end{equation}
The condition $ \sigma_{4H}/\sigma_{\text{EW}}<1$ translates into
\begin{equation}
    \delta m_0<0.45\gev \left(\frac{1\tev}{\mdm}\right)^3 n^2,
\end{equation}
which, for $n=3,5$ is much looser than the upper bound on $\delta m_0$. As a consequence, the UV splittings do not impact the freeze-out prediction for $Y=1$ WIMPs.

Finally, concerning the contribution to the Sommerfeld enhancement, the potential arising from the Yukawa interaction in \eqref{eq:tree_DMh} is:

\begin{equation}
    V_H(r)=-\left(\frac{\lambda_D v}{8\pi\Leff}\right)^2\frac{e^{-m_h r}}{r}.
\end{equation}

Such potential is shorter range with respect to typical EW Yukawa potentials, due to the larger Higgs mass as compared to weak boson masses. We find that $V_H(r)$ is negligible compared to the size of the EW potential $\alpha_2/r$ for the mass splittings of our interest. 

\begin{table*}[t]
\renewcommand{\arraystretch}{1.3}
\begin{center}
 \begin{tabular}{ c | c |  c | c | c  }
            DM spin             &$n_\epsilon$ & $\mdm$ (TeV)& $\Lambda_{\text{Landau}}/M_{\text{DM}}$ & $(\sigma v)^{J=0}_{\text{tot}}/(\sigma v)^{J=0}_{\text{max}}$ \\ \hline
  \multirow{6}{*}{Complex scalar} %& &&&&\\[-4pt]
                                                & $3$ & $1.60 \pm 0.01 - 2.4^*$ & $> M_{\text{Pl}}$ & -   \\
                                               &  $5$ & $11.3 \pm 0.6$ &  $> M_{\text{Pl}}$  & 0.003 \\ 
                                               &  $7$ & $47\pm 3$& $2\times 10^{6}$ &0.02\\
                                               &  $9$ & $118\pm 9$& $110$ &0.09 \\
                                               &  $11$ &  $217\pm 17$                    &  7 &    0.25                                            \\
                                               &  $13$ & $ 352 \pm 30$ & 3 &0.6 \\ \hline
     \multirow{6}{*}{Dirac fermion}%& &&&&\\[-4pt]
      & $3$ & $2.0 \pm 0.1 - 2.4^*$& $> M_{\text{Pl}}$ & -  \\
                                                          & $5$ & $9.1\pm0.5$ & $4\times10^{6}$&0.002    \\
                                                          & $7$ & $45\pm 3$ & 80&0.02 \\
                                                           &$9$ & $115\pm 9$ &6&0.09 \\
                                                           &$11$ & $211\pm 16$ &2.4&0.3 \\
                                                           &$13$ & $340\pm 27$ &1.6&0.7 \\
  \end{tabular}
 \caption{Freeze-out mass predictions for millicharged WIMP DM. The annihilation cross-section includes both the contribution of SE and BSF. For the triplets, a second prediction, denoted with a *, for the thermal mass is present due to the emergence of a large resonance in the Sommerfeld enhancement for $\mdm\approx 2.4$ TeV. We provide a measure of how close the DM annihilation cross-section is  to the unitarity bound for $s$-wave annihilation $(\sigma v)^{J=0}_{\text{max}}=4\pi/M_{\text{DM}}^2v$. We derive the scale where EW gauge coupling will develop a Landau pole by integrating-in the WIMP multiplet at its freeze-out mass. \label{table:summarymilli}}
 \label{tab:complex}
\end{center}
\end{table*}

\section{Millicharged WIMPs}\label{app:millichargeWIMPs}
Complex WIMPs with $Y=0$ have an unbroken $U(1)$ flavor symmetry which can be gauged by a new dark photon. Generically the dark photon would mix with the visible one through a kinetic mixing operator $\epsilon F F'$ and the complex WIMP would acquire a EM charge $\epsilon$. In this scenario the dark gauge symmetry makes the DM accidentally stable as noticed in Ref.~\cite{DelNobile:2015bqo} at the price of giving up charge quantization. Here we want to summarize the freeze-out predictions and the basic phenomenology of millicharged WIMPs in the limit of very small $\epsilon$ (i.e. $\epsilon<10^{-10}$) when their phenomenology resemble the one of the real WIMPs discussed in Ref.~\cite{Bottaro:2021snn}.

Concerning the freeze-out dynamics, the only difference between real and millicharged WIMPs is in the existence of BS with $P_\text{BS}= (-1)^{L+S+\frac{I-1}{2}}=-1$ formed by $\bar{\chi}\chi$ pairs of millicharged WIMPs. These are forbidden by the spin-statistic properties of the $\chi\chi$ wave function for real WIMPs. Since $P_\text{BS}$ is preserved by dipole interactions for $Y=0$, to leading order no transitions can occur between states with opposite $\pbs$ and excited BS with $\pbs=-1$ will dominantly decay to $1s$ and $2s$ states with the same $\pbs$. The latter have small decay widths with respect to their $\pbs=1$ counterparts. $ns_1^{S=1}$ and $ns_5^{S=1}$ annihilate into four vectors with a rate
\begin{equation}
    \Gamma(ns_{1,5}^{S=1}\rightarrow VVVV)\simeq \frac{\alphaeff^4}{16\pi^2 \mdm^2}\frac{|R_{n0}(0)|^2}{\mdm^2}\, ,
\end{equation}
while $ns_3^{S=0}$ annihilates into three vectors with rate:
\begin{equation}
    \Gamma(ns_{3}^{S=0}\rightarrow VVV)\simeq \frac{\alphaeff^3}{4\pi \mdm^2}\frac{|R_{n0}(0)|^2}{\mdm^2}\, .
\end{equation}
where $R_{n0}\sim (\alphaeff \mdm)^{3/2}$ is the radial wave function of the BS at the origin.  As noticed in Ref.~\cite{DelNobile:2015bqo} for the millicharged 3-plet a large resonance in the Sommerfeld enhancement leads to two different freeze out predictions. We summarize our freeze out predictions in Table \ref{tab:complex}. 

\begin{figure*}[htp!]
 \centering
\includegraphics[width=0.7\textwidth]{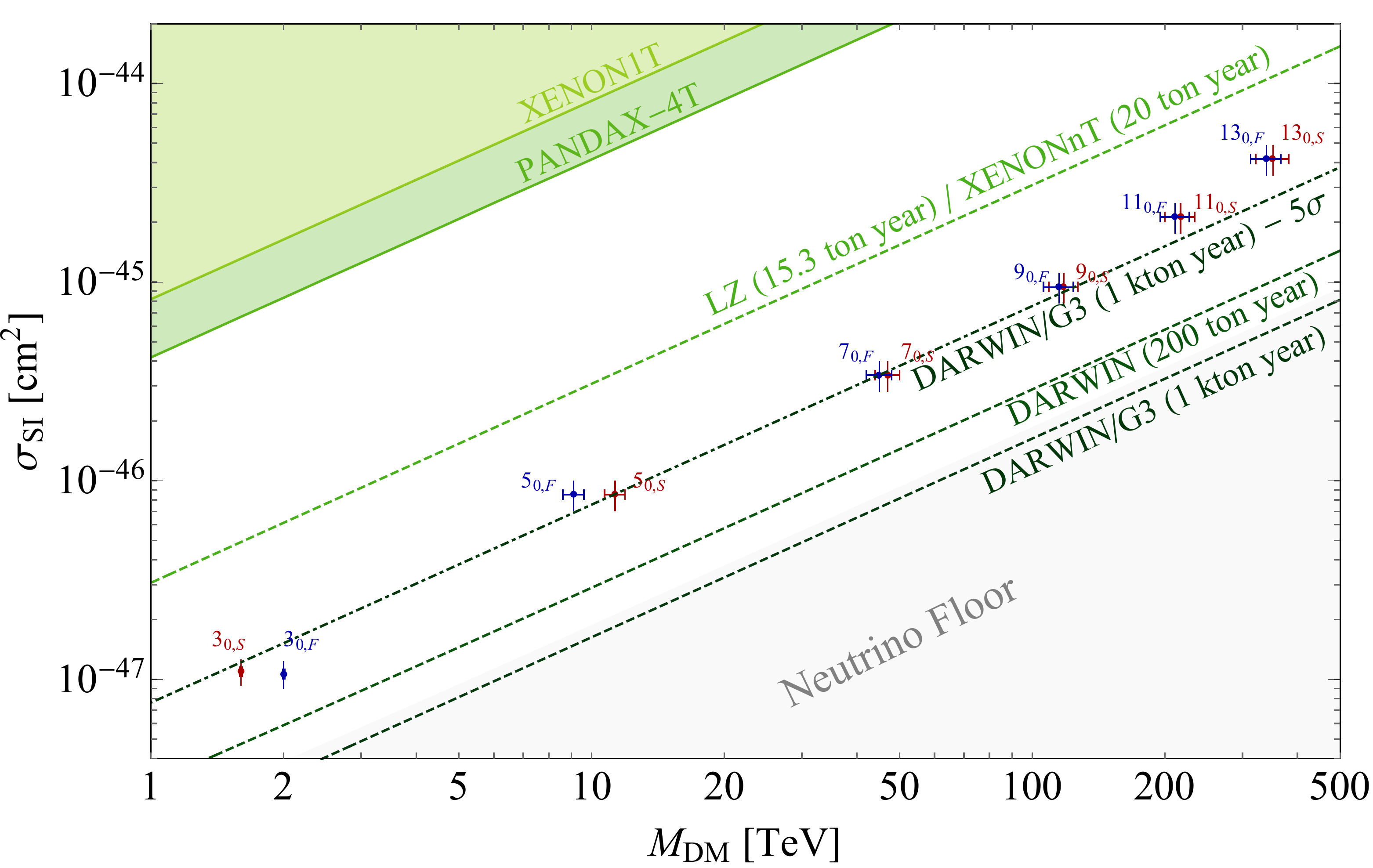}
 \centering
\caption{Expected SI cross-sections for different millicharged complex WIMPs with $\epsilon \lesssim 10^{-10}$. The {\bf blue dots} correspond to Dirac WIMPs and the {\bf red dots} to complex scalar WIMPs. The vertical error bands correspond to the propagation of LQCD uncertainties on the elastic cross-section (\eqref{elasticSIDD}), while the horizontal error band comes from the uncertainty in the theory determination of the WIMP freeze out mass in Table~\ref{table:summarymilli}. The {\bf light green} shaded region is excluded by the present experimental contraints from XENON-1T~\cite{Aprile:2018dbl} and  PandaX-4T~\cite{PandaX:2021osp}, the {\bf green dashed} lines shows the expected 95\% CL reach of LZ/Xenon-nT~\cite{Mount:2017qzi,XENON:2020kmp} and DARWIN ~\cite{Aalbers:2016jon,Aalbers:2022dzr}.
\hspace*{\fill} \small}
\label{fig:DD_loop_milli}
\end{figure*}

For completeness, in Fig.~\ref{fig:DD_loop_milli} we show the SI scattering cross-section of DM on xenon nuclei for the different millicharged candidates. This cross-section is identical to that computed for real candidates with $\epsilon \lesssim 10^{-10}$~\cite{DelNobile:2015bqo, Cirelli:2016rnw}, while larger $\epsilon$ would open up new opportunities for direct detection (see e.g.~Fig~1 of \cite{DelNobile:2015bqo}). From  Fig.~\ref{fig:DD_loop_milli} we see that even in the worst case scenario of very small millicharge large exposure experiments will be able to fully probe millicharged WIMPs with 200 ton-year exposure. The heavier multiplets could be also firmly discovered at DARWIN with kiloton exposure.

%\subsection{The 4-plet in detail}

\section{Details on the Collider reach}\label{app:collider}

The kinematic constraint of pair producing the DM particles makes direct searches at colliders of energies $\sqrt{s}\le 14 $ TeV impossible for $n\ge 5$.
We consider masses $M\gtrsim \sqrt{s}/10$, including only DY production channels. Indeed in this range VBF production is subdominant.\\
For the $n<5$ candidates, the strategies available are the \emph{missing invariant mass} (MIM), \emph{charged tracks} (CT) and \emph{disappearing tracks} (DT) searches. The first is essentially independent on the splitting values: the collider energies of few TeV needed for DM pair production are much larger than the maximum perturbatively generated splitting.
The DT searches in general require, at fixed $\mdm$, a scan in the $(\delta m_0, \delta m_\Qbar)$   to compute the signal, since the spectrum and the decay width of the particle $\chi^Q$ is determined uniquely by the splittings $\delta m_\Qbar$, $\delta m_0$.

\begin{figure*}[htbp]
\begin{center}
 \includegraphics[width=0.46\textwidth]{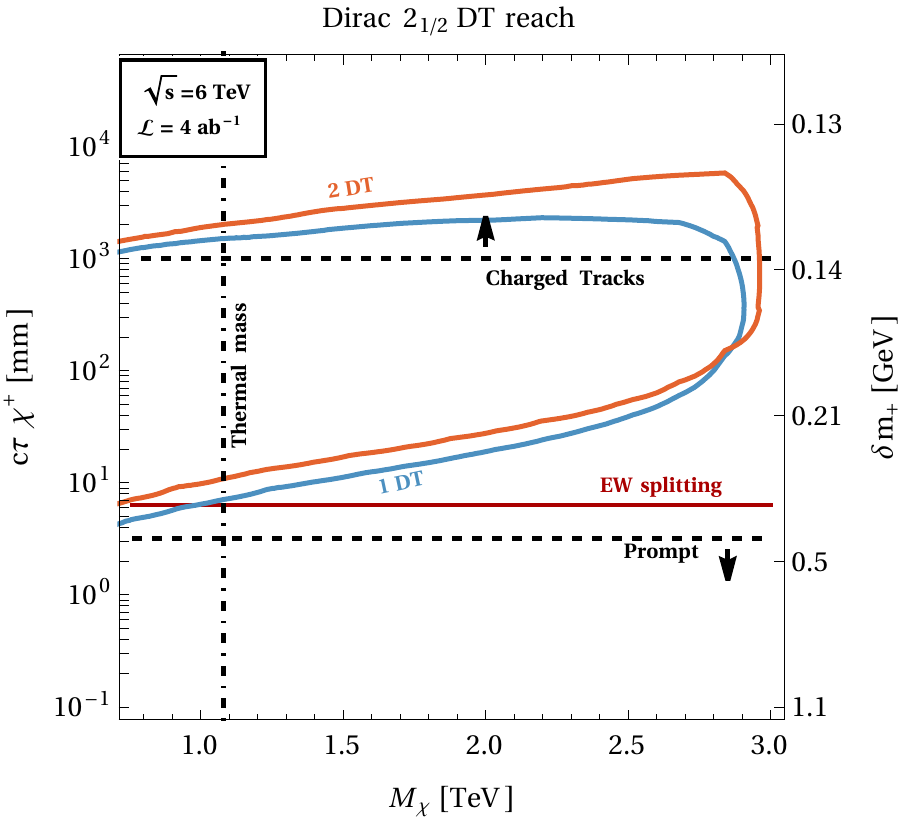}\hfill\includegraphics[width=0.46\textwidth]{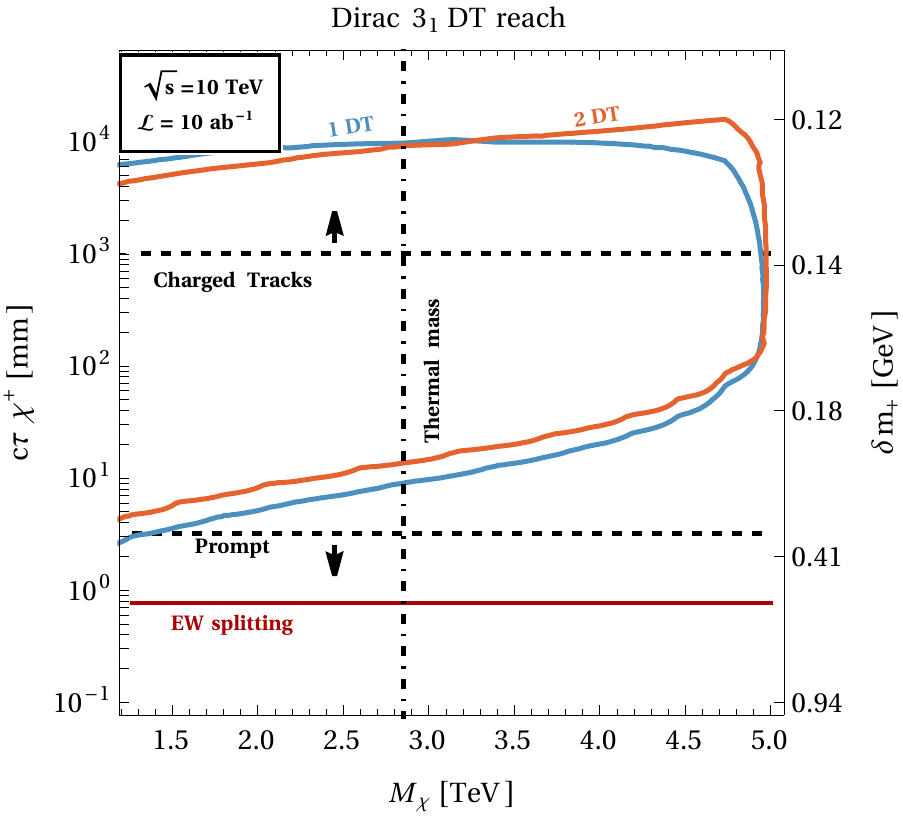}
    \includegraphics[width=0.46\textwidth]{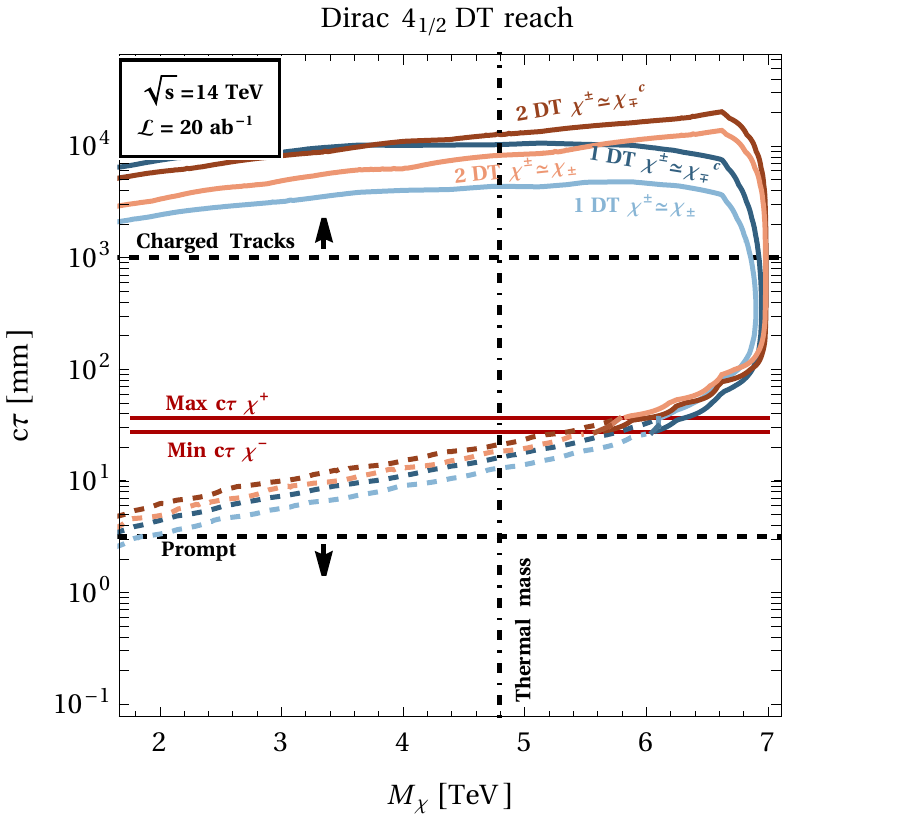}\hfill\includegraphics[width=0.46\textwidth]{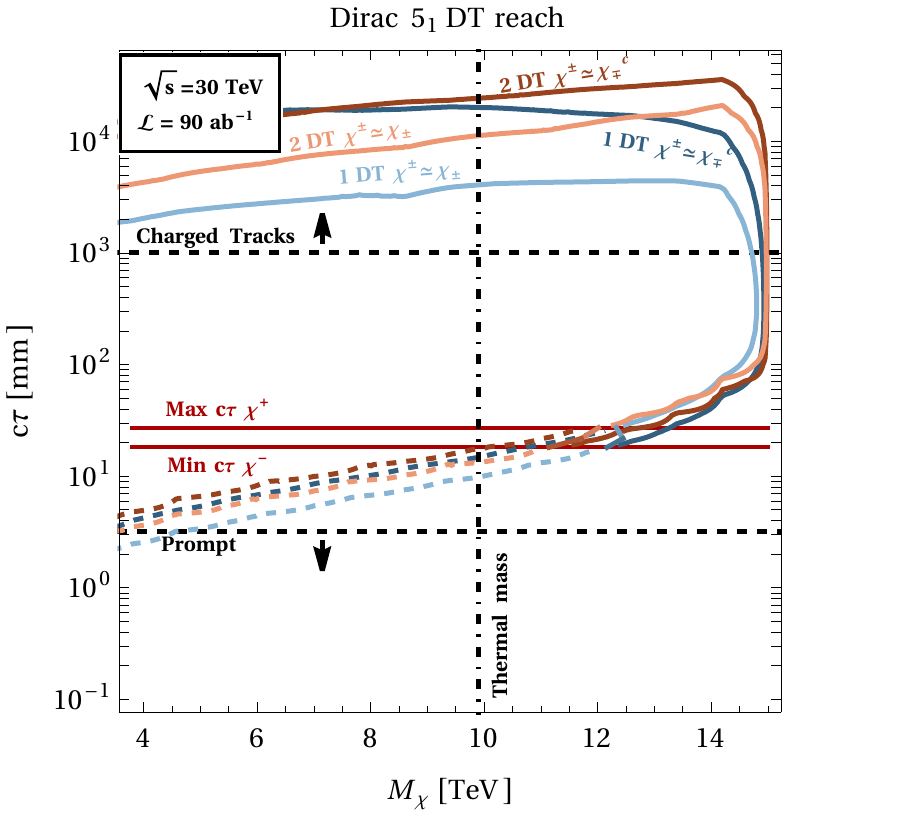}
\caption{Sensitivity at 95\% C.L. from DT searches as a function of $M_\chi$ and the proper lifetime of the charged particle $\chi^+$. The mass splitting $\delta m_+$ is also shown in the extra vertical axis in two upper panels. In the lower panels, the two different contours for each search channel correspond to the different physical situations where the heavy mass eigenstate corresponds to one of the two gauge eigenstates $\chi_+$ or $\chi_{-}^{c}$. \textbf{Top Left: } Dirac $2_{1/2}$ at $\sqrt{s}=6$ TeV, $\mathcal{L}= 4\; \mathrm{ab}^{-1}$. \textbf{Top Right:} Dirac $3_1$ at $\sqrt{s}=10$ TeV, $\mathcal{L}= 10\; \mathrm{ab}^{-1}$.  \textbf{Bottom Left: } Dirac $4_{1/2}$ at $\sqrt{s}=14$ TeV, $\mathcal{L}= 20\; \mathrm{ab}^{-1}$. \textbf{Bottom Right: } Dirac $5_1$ at $\sqrt{s}=30$ TeV, $\mathcal{L}= 90\; \mathrm{ab}^{-1}$.}
 \label{fig:dt_2dplot}
\end{center}
\end{figure*}

%because the signal depends exponentially on the lifetime of the charged particles, which are determined by a combination of both $\delta m_0, \delta m_+$, with a cubic or quintic power depending on the number of final state particles available. 
To get a feel of how powerful DT searches can be, we fix the neutral splitting to a representative  value $\delta m_0 \simeq 200 {\rm KeV }$ for the fermions. Our results are given in Fig.~\ref{fig:dt_2dplot} where we display the region of the plane WIMP mass versus $c\tau$  where experiments can probe the several WIMPs considered in each panel of the figure. For this result we consider only $\chi^+$ and its conjugate as candidate long-lived for the $2_{1/2}$. For the  $3_1$, the $4_{1/2}$ and $5_1$ all states with charge greater than 1 are assumed to decay promptly to the candidate charged long lived states at the bottom the spectrum.\footnote{We keep track in detail of the mixing of the charge $\pm 1$ gauge eigenstates into the suitable charge $\pm 1$ mass eigenstates.}
As we have fixed $\delta m_0$ much smaller than the mass splitting between the neutral and charged states, we can effectively consider decays into both $\chi^0$ and $\chi^{\rm DM}$ as if they were degenerate in mass. Considering the possible decay channels $\chi^-$ into 
$ e^- \nu_e \chi^{0,\mathrm{DM}}\,$, 
$ \mu^- \nu_\mu \chi^{0,\mathrm{DM}}\,$, 
$ \pi^- \chi^{0,\mathrm{DM}}$ and the charge conjugates for $\chi^+$
we get 
\begin{equation}
    \Gamma \left( \chi^\pm\rightarrow \chi^{0,\mathrm{DM}}+\mathrm{SM} \right) = \Gamma_{e^\pm}+\Gamma_{\mu^\pm}+\Gamma_{\pi^\pm}  \;,
\end{equation}
where the RHS indicates the decay width into neutral states plus the SM states indicated by the subscript.
The relevant widths are given by:
\begin{equation}
\begin{split}
    &\Gamma_{e^\pm }= g\left(n,Y, \pm 1 \right)\frac{G_F^2 \delta m_\pm ^5}{60 \pi^3}\;,\\
    &\Gamma_{\mu^\pm}=g\left(n,Y,\pm 1 \right)\frac{G_F^2 \delta m_\pm^5}{2\pi^3} \Phi (\delta m_\pm,m_\mu) \;,\\
    &\Gamma_{\pi^\pm}=g\left(n,Y, \pm 1 \right)\frac{G_F^2 f_\pi^2 |V_{ud}|^2 \delta m_\pm^3}{4\pi}\sqrt{1-\frac{m_\pi^2}{\delta m_\pm^2}} \; ,
\end{split}
\end{equation}
where $g\left(n,Y,Q \right)$ accounts for the different strengths of the $W$ boson coupling to each of the $\chi^{Q}$:
\begin{equation}
    g\left(n,Y,Q \right)=n^2-1-4(Q-Y)(Q-Y-\operatorname{sgn}(Q) )\,,\label{eq:qnYQ}
\end{equation}
with $n$ the dimensionality of the multiplet and $Y$ its hypercharge, $G_F$ is the Fermi constant, $f_\pi=131$ MeV is the pion decay constant.  In the above formula $\Phi$ is the full phase space of the 3-body decay of a massive particle of mass $M$ into a massless lepton (e.g neutrino), a massive lepton with mass $m_l \ll M$ (e.g. muon), and a heavy particle with a mass $M-\delta$, for splitting $\delta\ll M$:
\begin{equation}
\begin{split}
    \Phi(\delta ,m_l)= &\frac{1}{60}\sqrt{\delta^2-m_l^2}\left(2 \delta^4-9\delta^2m_l^2-8m_l^4 \right)+\\
    & +\frac{1}{4}m_l^4\delta \operatorname{ArcCoth}\left( \frac{\delta}{\sqrt{\delta^2-m_l^2}}\right)\;.
\end{split}
\end{equation}
The massless limit was employed for the electron since in the regions relevant for DTs the splittings allow to neglect $m_e$. 
In principle, for large splittings, also kaon and tau channels should be considered. However, for such large splittings the decays into the other SM particles are so fast that tracks will never be reconstructed.

If we move away from the $\delta m_0 \simeq 0$ points, we have to consider cases in which $\chi^+$ is heavier than some of the particles with $Q>1$. 
%
% If this happens the lifetime of the particle with Q>1 typically is very long.
We checked that for the $3_1$, $4_{1/2}$, $5_1$ there are choices of the couplings $y_0$ and $y_+$ for which $M_{\chi^{++}}<M_{\chi^+}$.  
In this case, $\chi^{++}$ cannot decay into $\chi^+$, and is forced to decay to the neutral states by emitting twice an off-shell $W$ boson that gives rise to $l^+ \nu_l$ or $\pi^+$. 
%This decay is very suppressed, and varies steeply with the splitting. This can be seen for example in Fig.~\ref{fig:doublet-triplet-fourplet-fiveplet} \rf{(bottom left for the $4_{1/2}$),} and causes the bending upward of the iso-lifetime lines
%\footnote{This is also seen for the $3_1$, but in a region that is cut out by perturbativity not shown in Fig.~\ref{fig:doublet-triplet-fourplet-fiveplet} (top right).}.
This decay rate has been estimated by summing over all possible final states using NDA.  With respect to the decay with just one off-shell $W$ emission, we get an extra factor $g(n,Y,2)G_F f_\pi \delta/(16\pi^2)$ and $g(n,y,2)G_F \delta^2/(16\pi^2)^2$ for any extra $\pi$ or $l^+ v_l$ pair, respectively. Decays into negatively charged states are even more suppressed, since they need more extra final states. These configurations can give $\chi^{++}$ DT signals as well as CT signals, which are potentially interesting for their peculiar detector response. 

Following \cite{Capdevilla:2021fmj}, we employ two different DT signal regions: i) events with at least one disappearing track in addition to a hard photon with energy $E_\gamma>25 $ GeV; ii) events with two disappearing tracks, one of which is short, and the hard photon.
The former has an estimated background cross section of $\sigma_B=18.8$ ab at the 10~TeV collider, leading to a signal-to-noise ratio of order 1. The latter is background free.
To compute the number of expected signal events, we do a Monte Carlo simulation with MadGraph5\_aMC@NLO~\cite{Alwall:2014hca}. We also got similar results employing the analytic approximation explained in the appendix of \cite{Bottaro:2021snn}.
The sensitivity to disappearing  tracks is shown in the plots of Fig.~\ref{fig:dt_2dplot}, for minimal $\delta m_0$ imposed by DD and BBN, as a function of the dark matter mass and the lifetime $c \tau $ for each of the considered WIMP candidates.

Our results in Fig.~\ref{fig:dt_2dplot} show that the thermal mass can be excluded over the entire range of lifetimes that give rise to the DT signature. For the $2_{1/2}$ this corresponds to the full parameter space with splitting larger than the EW one.
The contours for the $4_{1/2}$ and $5_1$ WIMPs deserve some comment. In this case two singly-charged states are present: at least one state is in the DT region, and DT searches for this long-lived state, shown as solid contours, can probe the thermal mass over the full range of possible lifetimes. We also show the DT reach for the state with shorter lifetime, which could produce a second observable signal in parts of the parameter space. We show with different colors the two different physical solutions where the long-lived mass eigenstate $\chi^+$ corresponds to the gauge eigenstate $\chi_+$ or $\chi_-^c$ (corresponding to the regions of low or high $\delta m_{2+}$ in Fig.~\ref{fig:splitting}), which have slightly different production cross-sections.

\bigskip

For the MIM searches we have proceeded similarly to our earlier work~\cite{Bottaro:2021snn} and we have  optimized the selection for the fermion $2_{1/2}$ and $3_1$ at $\sqrt{s}=3, 6$ TeV and $\sqrt{s}=6, 10$ TeV, respectively. For the other candidates we used the optimal cuts that we derived from our previous results for real candidates with odd $n$~\cite{Bottaro:2021snn}.  For each level of systematics $\epsilon$ we have interpolated our previous results as functions of $n$ and $\sqrt{s}$ and we have derived optimal cuts for the new $n$ plets studied in this work. We remark that this procedure is potentially inaccurate as the real odd $n$-plets contain 2 times fewer degrees of freedom, hence have rates smaller by a factor 2. This can in principle affect the result of the optimization of the selection. We checked that the difference with a dedicated optimization is negligible, thus our event selection should be quite close to the optimal one.

Following the mono-$W$ discussion in \cite{Bottaro:2021snn}, we have computed the background rate with MadGraph5\_aMC@NLO~\cite{Alwall:2014hca} using the full 3-body process for $\eta \lesssim 6$, and matched it to a computation using the Weizs\"acker-Williams approximation for larger values of pseudorapidity.
%with MadGraph5\_aMC@NLO~\cite{Alwall:2014hca}  the $\eta_\mathrm{match}$ of the charged lepton, needed to correctly combine the background computed using the Weizsacker-Williams approximation %and the full 3-body process. The obtained values are: \DB{ometterei questa discussione}\rf{Ometterei di omettere}
%\begin{equation*}
%    \eta_\mathrm{match}=6.2,6.5 \qquad \mathrm{for} \qquad \sqrt{s}=6,10 \; \mathrm{TeV} \;.
%\end{equation*}

\begin{figure*}[htbp]
\begin{center}
    \includegraphics[width=0.45\textwidth]{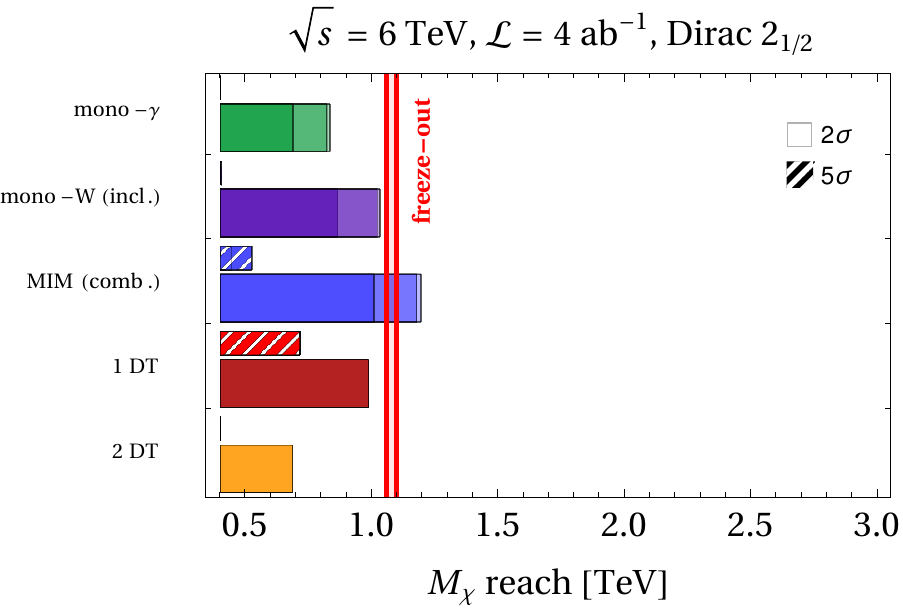}\hfill\includegraphics[width=0.45\textwidth]{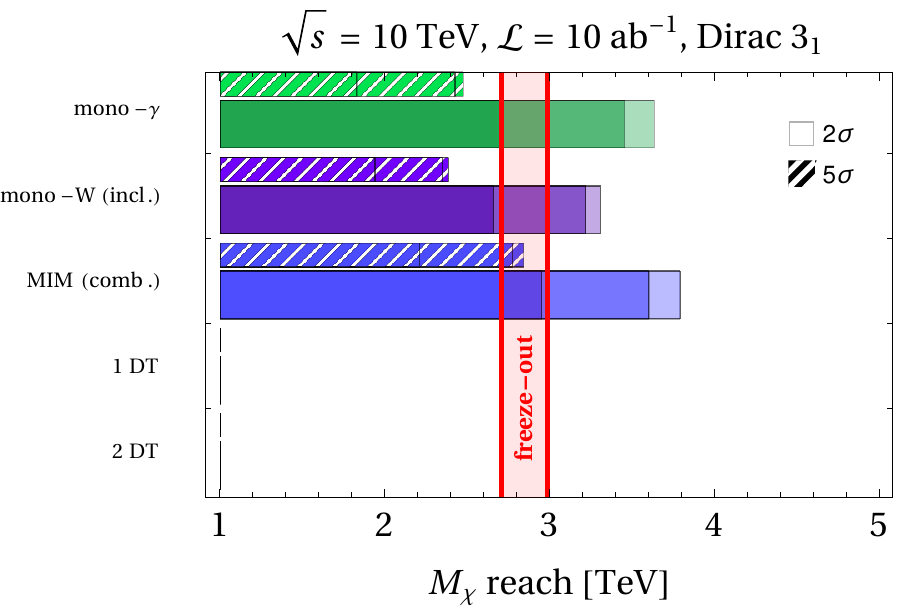}
    \includegraphics[width=0.45\textwidth]{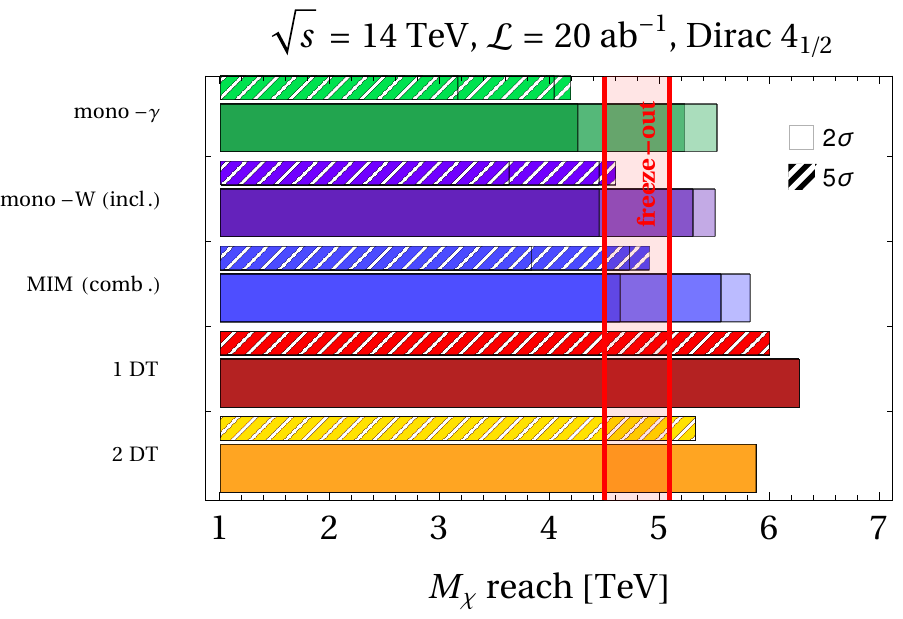}\hfill\includegraphics[width=0.45\textwidth]{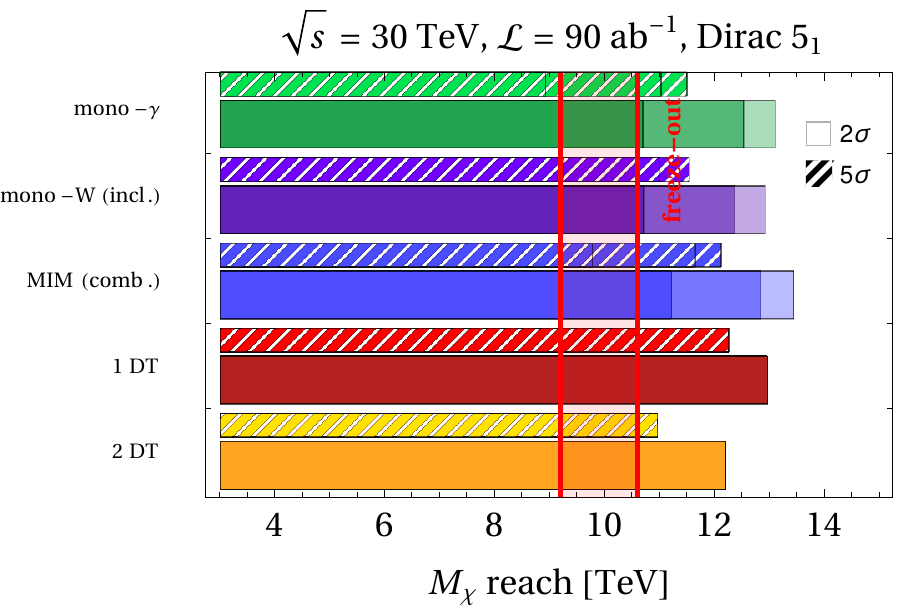}
\caption{Different bars show the 2 $\sigma$(solid wide) and 5$\sigma$ (hatched thin) reach on the fermionic  WIMP mass at a muon collider for different search channels. The first three bars show the channels discussed in Sec.~\ref{sec:nonminimal splitting} where DM would appear as missing invariant mass (MIM) recoiling against one or more SM objects: mono-$\gamma$, inclusive mono-$W$, and the combination of all these MIM channels (blue). The mono-$Z$ channel is not reported since it gives results below the minimum mass shown. The last two bars show the reach of disappearing tracks, requiring at least 1 disappearing track (red), or at least 2 tracks (orange). All the results are shown assuming systematic uncertainties to be 0 (light), 1\textperthousand (medium), or 1\% (dark). The vertical red bands show the freeze-out prediction. Tracks have been computed assuming minimally split neutral states, and a charged-neutral splitting $\delta m_+$ equal to the gauge contribution for the $2_{1/2}$ and $3_1$. For the $4_{1/2}$ and $5_1$ the splitting-inducing couplings $y_0$ and $y_+$ have been tuned so that $\delta m_+ = \delta m_-$ (up to minimal $\delta m_0$ corrections), which represents a least favorable condition for DT  searches. \textbf{Top Left: } Dirac $2_{1/2}$ for $\sqrt{s}=6$ TeV and $\mathcal{L} = 4 \; \mathrm{ab}^{-1} $. \textbf{Top Right: } Dirac $3_{1}$ for $\sqrt{s}=10$ TeV and $\mathcal{L} = 10 \; \mathrm{ab}^{-1}$. \textbf{Bottom Left: } Dirac $4_{1/2}$ for $\sqrt{s}=14$ TeV and $\mathcal{L} = 20 \; \mathrm{ab}^{-1} $. \textbf{Bottom Right: } Dirac $5_{1}$ for $\sqrt{s}=30$ TeV and $\mathcal{L} = 90 \; \mathrm{ab}^{-1}$. }
\label{fig:fermionic_barchart}
\end{center}
\end{figure*}

 We report in Fig.~\ref{fig:fermionic_barchart} the sensitivity for the $2_{1/2}$, $3_1$, $4_{1/2}$, $5_1$, at colliders with suitable $\sqrt{s}$ that can provide 95\% C.L.\ exclusion. Full results for all benchmark colliders at the nominal luminosity of \eqref{eq:bench} are given in Tables~\ref{tab:dirac_ditri_cuts} and \ref{tab:dirac_tetrapenta_cuts}, for the various search channels under consideration. Results for generic center-of-mass energy and integrated luminosity are displayed in Fig.~\ref{fig:energy+lumi+plane} for the mono-$\gamma$ and mono-$W$ channels. The results on this figure have been obtained by rescaling the results at $\sqrt{s}=3,6,10,14,30~$TeV, under the assumption that all cross-sections and kinematical cuts scale trivially with collider energy, and neglecting systematic errors.

\begin{figure*}[htbp]
\begin{center}
    \includegraphics[width=0.45\textwidth]{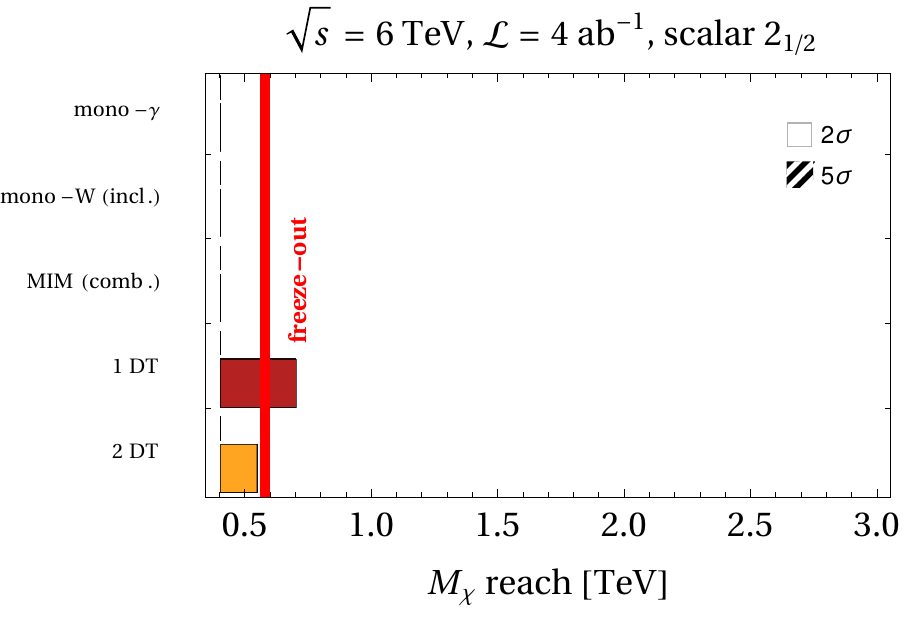}\hfill\includegraphics[width=0.45\textwidth]{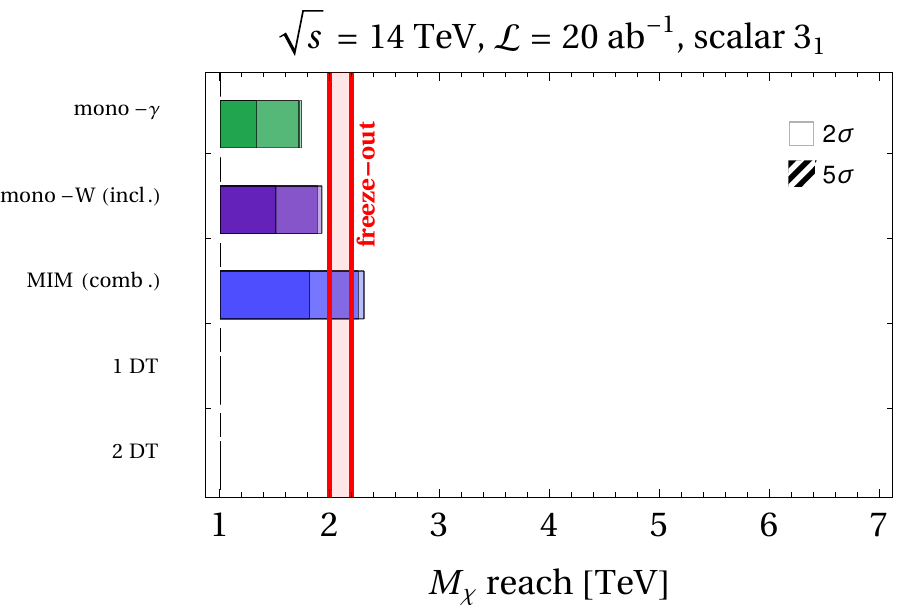}
    \includegraphics[width=0.45\textwidth]{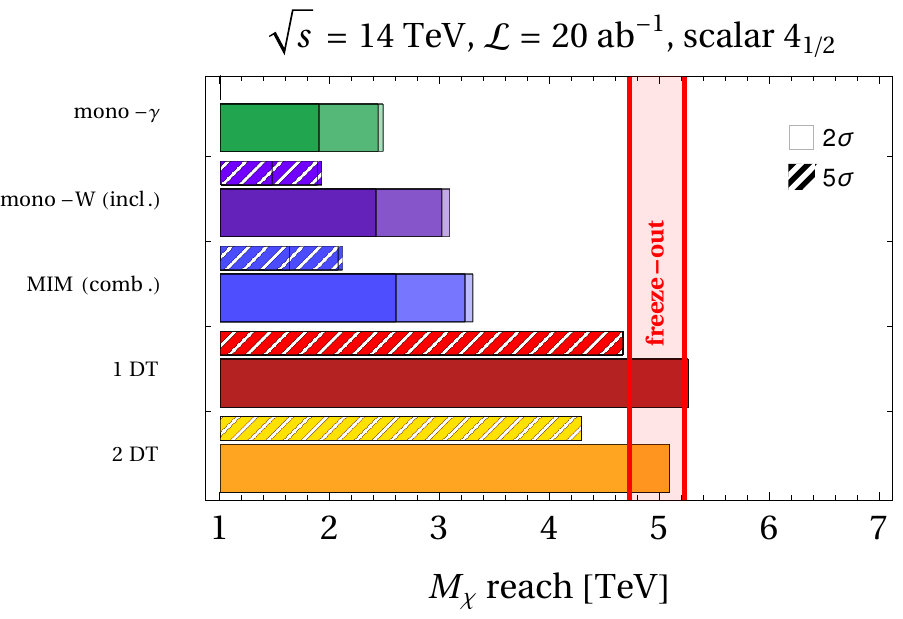}\hfill\includegraphics[width=0.45\textwidth]{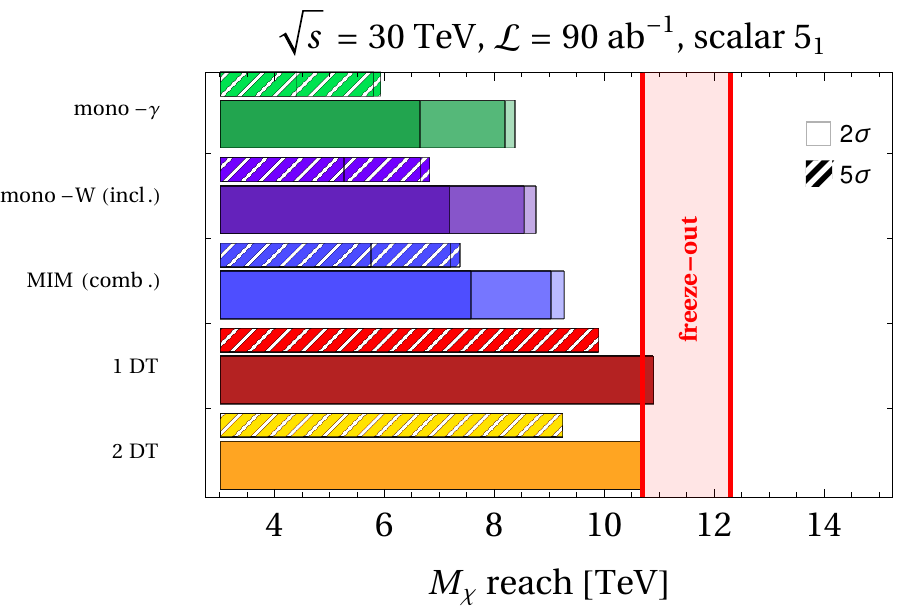}

\caption{%Different bars show the 2 $\sigma$(solid wide) and 5$\sigma$ (hatched thin) reach on the scalar WIMP mass at a muon collider for different search channels. The first three bars show the channels discussed in Section \ref{sec:nonminimal splitting} where DM would appear as missing invariant mass (MIM) recoiling against one or more SM objects: mono-gamma, inclusive mono-W, and the combination of all these MIM channels (blue). The mono-Z channel is not reported since it gives results below the minimum mass shown. The last two bars show the reach of disappearing tracks, requiring at least 1 disappearing track (red), or at least 2 tracks (orange). All the results are shown assuming systematic uncertainties to be 0 (light), 1\textperthousand (medium), or 1\% (dark). The vertical red bands show the freeze-out prediction. Tracks have been computed assuming degenerate neutral states, and a charged-neutral splitting $\delta m_+$ equal to the gauge contribution for the $2_{1/2}$ and $3_1$. For the $4_{1/2}$ and $5_1$ the splitting-inducing couplings $y_0$ and $y_+$ have been tuned so that $\delta m_+ = \delta m_-$, which represents a least favorable condition for DT searches. \textbf{Top Left: } Scalar $2_{1/2}$ for $\sqrt{s}=6$ TeV and $\mathcal{L} = 4 \; \mathrm{ab}^{-1} $. \textbf{Top Right: } Scalar $3_{1}$ for $\sqrt{s}=10$ TeV and $\mathcal{L} = 10 \; \mathrm{ab}^{-1}$. \textbf{Bottom Left: } Scalar $4_{1/2}$ for $\sqrt{s}=14$ TeV and $\mathcal{L} = 20 \; \mathrm{ab}^{-1} $. \textbf{Bottom Right: } Scalar $5_{1}$ for $\sqrt{s}=30$ TeV and $\mathcal{L} = 90 \; \mathrm{ab}^{-1}$. 
Results for scalar WIMPs. Same notation as in Fig.~\ref{fig:scalar_barchart}.}
\label{fig:scalar_barchart}
\end{center}
\end{figure*}

% For the $4_{1/2}$ and $5_1$, the reaches are not computed using the optimized procedure. We used interpolated cuts from the one obtained for the $3_0$ and $5_0$ at different energies in \cite{Bottaro:2021snn}. This procedure has also been repeated for $2_{1/2}$ and $3_1$, and the results obtained with the interpolated cuts match the ones obtained with the optimized procedure.

The DT results shown in Fig.~\ref{fig:fermionic_barchart} are a particular sub-case of those shown in Fig.~\ref{fig:dt_2dplot}, as in Fig.~\ref{fig:fermionic_barchart} we fixed some choice of the mass splitting governing the spectrum specifically for each WIMP as described in the following. For the $2_{1/2}$ the splitting between the charged and neutral components has been set to its gauge contribution $\delta m_+=354$~MeV. For the $3_1$ the splitting has been set to its gauge value of $\delta m_+=540$~MeV. 
For the $4_{1/2}$, the splitting $\delta m_+$ has been taken equal to $\delta m_-$. This choice implies that the $\chi^+$ and $\chi^-$ particles are maximal admixtures of the gauge eigenstates, hence they have same cross-sections and lifetimes. The latter are governed by the mass splitting $\delta m_+$ and $\delta m_-$, which, from \eqref{eq:deltaM} are constrained to satisfy $\delta m_+ + \delta m_- = \delta m_0 +2\delta_g$. This equation implies that for $\delta m_0 \ll \delta_g$ in the $\delta m_+ =\delta m_- $ case we obtain lifetimes that have relatively challenging detector acceptances. Thus we can consider this choice as a representative but somewhat pessimistic estimate of the mass reach.  The $\chi^{++}$ has been assumed to promptly decay equally into  $\chi^+$ and $\chi^{-,c}$. 
%The production cross sections for the mass eigenstates states within the $4_{1/2}$ have been computed separately, as necessary due to different couplings of these states to the weak bosons (see eq.(\ref{eq:qnYQ})).
For the $5_1$ WIMP we proceeded similarly to the $4_{1/2}$ case.

A summary of the capabilities of the several stages of a high energy muon collider  is given in Fig.~\ref{fig:summary_all_colliders} under the assumption of luminosity following the scaling of \eqref{eq:bench}. The upshot of these studies is that, as the center of mass energy of the collider is increased, the higher energy machine gains sensitivity to heavier WIMP candidates. All in all, the list of WIMP candidates that we have described in this work and earlier Ref.~\cite{Bottaro:2021snn} provides a series of targets that can be probed at successive stages of a future high energy muon collider. 

%In Tables~\ref{tab:dirac_ditri_cuts} and \ref{tab:dirac_tetrapenta_cuts} we report the results of the analysis for the Dirac $2_{1/2}$,$3_1$ and for the Dirac $4_{1/2}$, $5_1$ respectively. The cuts for the $2_{1/2}$ and $3_1$ at $\sqrt{s}=3,6$ TeV and $\sqrt{s}=6,10$ TeV respectively have been done by repeating the optimization procedure outlined in \cite{Bottaro:2021snn}, while in all the other cases we used the cuts obtained by interpolating the cuts obtained for the odd $n$ real candidates. The two procedure match well. 
%The cuts for the doublet at $\sqrt{s}=3,6$ TeV has been obtained with the full optimization procedure, while the other cases have used the interpolated cuts. The optimized cuts match the interpolated ones.

\begin{table*}
\footnotesize
\begin{centering}
\renewcommand{\arraystretch}{1.2}
\begin{tabular}{|c|c|c||c|c|c|c|c||c|c|c|c|c|}
\cline{4-13}
\multicolumn{3}{c|}{} & \multicolumn{5}{c||}{Dirac $2_{1/2}$} & \multicolumn{5}{c|}{Dirac $3_1$}\\
\cline{2-13}
\multicolumn{1}{c|}{} & $\sqrt{s}$ & $\epsilon_{\rm sys}$ & $\eta_{X}^{\rm cut}$& $p_{T,X}^{\rm cut}$ {[}TeV{]} & $S_{95\%}$& $S_{95\%}/B$& $M_{95\%}$ {[}TeV{]} & $\eta_{X}^{\rm cut}$& $p_{T,X}^{\rm cut}$ {[}TeV{]} & $S_{95\%}$ & $S_{95\%}/B$& $M_{95\%}$ {[}TeV{]}\\
\cline{2-13}
\multicolumn{13}{c}{}\\[-9pt]
\hline\multirow{15}{*}{\rotatebox{90}{Mono-$\gamma$}} & \multirow{3}{*}{3 TeV} & 0 & 2.4 & 0.24 & 492 & 0.01 & 0.5 & 2.4 & 0.01 & 3181 & 0.001 & 1.3 \\
 &  & 1\permil & 1.6 & 0.54 & 213 & 0.02 & 0.49 & 1.4 & 0.12 & 1217 & 0.004 & 1.2 \\
 &  & 1\% & 1. & 0.9 & 94 & 0.05 & 0.43 & 0.6 & 0.42 & 353 & 0.03 & 0.94 \\
\cline{2-13}
 & \multirow{3}{*}{6 TeV} & 0 & 1.8 & 0.96 & 296 & 0.02 & 0.83 & 2.4 & 0. & 6895 & 0.0006 & 2.5 \\
 &  & 1\permil & 1.4 & 1.4 & 282 & 0.02 & 0.82 & 1.4 & 0.24 & 1453 & 0.004 & 2.2 \\
 &  & 1\% & 1. & 1.9 & 83 & 0.06 & 0.69 & 0.8 & 1.1 & 309 & 0.03 & 1.7 \\
\cline{2-13}
 & \multirow{3}{*}{10 TeV} & 0 & 1.2 & 3. & 101 & 0.05 & 1.1 & 2. & 0.2 & 3159 & 0.001 & 3.6 \\
 &  & 1\permil & 1.2 & 3. & 99 & 0.04 & 1.1 & 1.4 & 0.8 & 867 & 0.006 & 3.5 \\
 &  & 1\% & 0.8 & 3.8 & 56 & 0.08 & 0.98 & 1. & 2. & 348 & 0.03 & 2.7 \\
\cline{2-13}
 & \multirow{3}{*}{14 TeV} & 0 & 1.2 & 4.2 & 107 & 0.04 & 1.5 & 1.8 & 0.56 & 2080 & 0.002 & 4.9 \\
 &  & 1\permil & 0.8 & 4.5 & 110 & 0.04 & 1.5 & 1.6 & 1.1 & 991 & 0.005 & 4.7 \\
 &  & 1\% & 0.6 & 5.9 & 46 & 0.1 & 1.3 & 0.6 & 2.8 & 290 & 0.03 & 3.7 \\
\cline{2-13}
 & \multirow{3}{*}{30 TeV} & 0 & 1. & 9.6 & 107 & 0.04 & 3. & 1.8 & 1.8 & 1423 & 0.003 & 10. \\
 &  & 1\permil & 1. & 9.6 & 107 & 0.04 & 3. & 1.6 & 3. & 911 & 0.005 & 9.6 \\
 &  & 1\% & 0.8 & 10. & 98 & 0.05 & 2.7 & 1. & 6.6 & 309 & 0.03 & 7.6 \\
\hline
\multirow{15}{*}{\rotatebox{90}{Mono-$W$ (inclusive)}} & \multirow{3}{*}{3 TeV} & 0 & 1.4 & 0.6 & 336 & 0.01 & 0.53 & 2. & 0.18 & 1833 & 0.002 & 1.1 \\
 &  & 1\permil & 1.2 & 0.72 & 228 & 0.02 & 0.52 & 1.6 & 0.3 & 1180 & 0.004 & 1. \\
 &  & 1\% & 0.8 & 1. & 92 & 0.06 & 0.47 & 1. & 0.66 & 345 & 0.03 & 0.85 \\
\cline{2-13}
 & \multirow{3}{*}{6 TeV} & 0 & 1.4 & 1.2 & 361 & 0.01 & 1. & 2. & 0.48 & 1572 & 0.003 & 2. \\
 &  & 1\permil & 1.4 & 1.2 & 351 & 0.01 & 1. & 1.4 & 0.72 & 1051 & 0.005 & 2. \\
 &  & 1\% & 1. & 1.8 & 140 & 0.04 & 0.86 & 0.8 & 1.2 & 480 & 0.03 & 1.6 \\
\cline{2-13}
 & \multirow{3}{*}{10 TeV} & 0 & 1.6 & 2. & 336 & 0.01 & 1.5 & 1.6 & 1. & 1424 & 0.003 & 3.3 \\
 &  & 1\permil & 1.6 & 2. & 331 & 0.01 & 1.5 & 1.4 & 1.2 & 1065 & 0.005 & 3.2 \\
 &  & 1\% & 0.8 & 3.4 & 100 & 0.05 & 1.3 & 0.8 & 2. & 429 & 0.03 & 2.7 \\
\cline{2-13}
 & \multirow{3}{*}{14 TeV} & 0 & 1.2 & 3.4 & 298 & 0.01 & 2.1 & 1.6 & 1.4 & 1566 & 0.003 & 4.6 \\
 &  & 1\permil & 1.2 & 3.4 & 322 & 0.01 & 2.1 & 1.4 & 1.7 & 1066 & 0.005 & 4.4 \\
 &  & 1\% & 0.8 & 4.8 & 107 & 0.05 & 1.7 & 0.8 & 3.6 & 225 & 0.03 & 3.6 \\
\cline{2-13}
 & \multirow{3}{*}{30 TeV} & 0 & 1.4 & 6.6 & 344 & 0.01 & 4.1 & 1.8 & 3. & 1433 & 0.003 & 9.6 \\
 &  & 1\permil & 1.4 & 7.2 & 283 & 0.01 & 4.1 & 1.6 & 4.2 & 976 & 0.005 & 9.3 \\
 &  & 1\% & 0.8 & 11. & 75 & 0.07 & 3.4 & 0.8 & 8.4 & 216 & 0.03 & 7.5 \\
\hline
\end{tabular}

\end{centering}

\caption{95\%C.L.\ reach on the mass of the Dirac $2_{1/2}$ and $3_{1}$ from the various mono-X channels. The excluded number of signal events $S_{95\%}$ and the relative precision $S_{95\%}/B$ are also given,together with the values of the optimal event selection cuts on $\eta_X$ and $p_{T,X}$,where $X$ is the single vector boson.The numbers are shown for different collider energies $\sqrt{s}$ and systematic uncertainties $\epsilon_{\rm sys}$. \label{tab:dirac_ditri_cuts}}
\end{table*}

\begin{table*}
\footnotesize
\begin{centering}
\renewcommand{\arraystretch}{1.2}
\begin{tabular}{|c|c|c||c|c|c|c|c||c|c|c|c|c|}
\cline{4-13}
\multicolumn{3}{c|}{} & \multicolumn{5}{c||}{Dirac $4_{1/2}$} & \multicolumn{5}{c|}{Dirac $5_1$}\\
\cline{2-13}
\multicolumn{1}{c|}{} & $\sqrt{s}$ & $\epsilon_{\rm sys}$ & $\eta_{X}^{\rm cut}$& $p_{T,X}^{\rm cut}$ {[}TeV{]} & $S_{95\%}$& $S_{95\%}/B$& $M_{95\%}$ {[}TeV{]} & $\eta_{X}^{\rm cut}$& $p_{T,X}^{\rm cut}$ {[}TeV{]} & $S_{95\%}$ & $S_{95\%}/B$& $M_{95\%}$ {[}TeV{]}\\
\cline{2-13}
\multicolumn{13}{c}{}\\[-9pt]
\hline\multirow{15}{*}{\rotatebox{90}{Mono-$\gamma$}} & \multirow{3}{*}{3 TeV} & 0 & 2.4 & 0.01 & 3026 & 0.001 & 1.4 & 2.4 & 0.01 & 2823 & 0.001 & 1.5 \\
 &  & 1\permil & 1.4 & 0.01 & 4092 & 0.003 & 1.3 & 1.6 & 0.01 & 3906 & 0.003 & 1.4 \\
 &  & 1\% & 0.6 & 0.3 & 517 & 0.02 & 1.1 & 0.4 & 0.12 & 1245 & 0.02 & 1.2 \\
\cline{2-13}
 & \multirow{3}{*}{6 TeV} & 0 & 2.4 & 0.01 & 6776 & 0.0006 & 2.7 & 2.4 & 0.01 & 6400 & 0.0006 & 2.9 \\
 &  & 1\permil & 1.2 & 0.24 & 1732 & 0.004 & 2.4 & 1.2 & 0.01 & 6084 & 0.003 & 2.7 \\
 &  & 1\% & 0.8 & 0.96 & 478 & 0.02 & 2. & 0.6 & 0.72 & 697 & 0.02 & 2.3 \\
\cline{2-13}
 & \multirow{3}{*}{10 TeV} & 0 & 2.4 & 0.01 & 7217 & 0.0008 & 4.1 & 2.4 & 0.01 & 10886 & 0.0004 & 4.7 \\
 &  & 1\permil & 1.2 & 0.6 & 1264 & 0.004 & 3.8 & 1.2 & 0.2 & 3073 & 0.003 & 4.4 \\
 &  & 1\% & 0.8 & 1.6 & 424 & 0.03 & 3.1 & 0.4 & 1.2 & 531 & 0.02 & 3.7 \\
\cline{2-13}
 & \multirow{3}{*}{14 TeV} & 0 & 2.2 & 0.28 & 3174 & 0.001 & 5.5 & 2.4 & 0.01 & 15838 & 0.0003 & 6.4 \\
 &  & 1\permil & 1.2 & 1.1 & 936 & 0.005 & 5.2 & 1. & 0.56 & 1967 & 0.003 & 6. \\
 &  & 1\% & 0.6 & 2.8 & 276 & 0.03 & 4.3 & 0.4 & 2. & 483 & 0.02 & 5.1 \\
\cline{2-13}
 & \multirow{3}{*}{30 TeV} & 0 & 1.6 & 1.2 & 2374 & 0.002 & 11. & 1.6 & 0.6 & 4145 & 0.001 & 13. \\
 &  & 1\permil & 1.2 & 2.4 & 1134 & 0.004 & 11. & 0.8 & 1.2 & 2078 & 0.003 & 13. \\
 &  & 1\% & 0.8 & 6. & 356 & 0.03 & 8.7 & 0.6 & 4.2 & 671 & 0.02 & 11. \\
\hline
\multirow{15}{*}{\rotatebox{90}{Mono-$W$ (inclusive)}} & \multirow{3}{*}{3 TeV} & 0 & 2.4 & 0.06 & 5317 & 0.0008 & 1.2 & 2.4 & 0. & 6314 & 0.0006 & 1.3 \\
 &  & 1\permil & 1.2 & 0.24 & 1409 & 0.004 & 1.2 & 1. & 0.18 & 1705 & 0.004 & 1.3 \\
 &  & 1\% & 0.8 & 0.54 & 460 & 0.02 & 1. & 0.6 & 0.36 & 1255 & 0.02 & 1.1 \\
\cline{2-13}
 & \multirow{3}{*}{6 TeV} & 0 & 2.2 & 0.12 & 8969 & 0.0004 & 2.4 & 2.4 & 0. & 16721 & 0.0002 & 2.6 \\
 &  & 1\permil & 1.2 & 0.48 & 1855 & 0.003 & 2.3 & 1. & 0.36 & 2629 & 0.003 & 2.5 \\
 &  & 1\% & 0.8 & 1.1 & 467 & 0.03 & 2. & 0.6 & 0.72 & 1010 & 0.02 & 2.2 \\
\cline{2-13}
 & \multirow{3}{*}{10 TeV} & 0 & 2. & 0.4 & 4213 & 0.001 & 4. & 2.2 & 0.2 & 9661 & 0.0004 & 4.4 \\
 &  & 1\permil & 1.4 & 1. & 1202 & 0.004 & 3.8 & 1.2 & 0.6 & 2786 & 0.003 & 4.2 \\
 &  & 1\% & 0.6 & 2. & 441 & 0.03 & 3.2 & 0.4 & 1.6 & 622 & 0.02 & 3.7 \\
\cline{2-13}
 & \multirow{3}{*}{14 TeV} & 0 & 1.8 & 0.56 & 4093 & 0.001 & 5.5 & 2.2 & 0.28 & 10846 & 0.0004 & 6.1 \\
 &  & 1\permil & 1.2 & 1.4 & 1215 & 0.004 & 5.3 & 1.2 & 0.84 & 2859 & 0.003 & 5.8 \\
 &  & 1\% & 1. & 2.8 & 486 & 0.02 & 4.4 & 0.2 & 2.5 & 315 & 0.03 & 5. \\
\cline{2-13}
 & \multirow{3}{*}{30 TeV} & 0 & 1.6 & 2.4 & 2888 & 0.002 & 12. & 2.2 & 0.6 & 8056 & 0.0006 & 13. \\
 &  & 1\permil & 1.2 & 2.4 & 1922 & 0.003 & 11. & 1.2 & 2.4 & 1829 & 0.003 & 12. \\
 &  & 1\% & 1. & 6. & 627 & 0.02 & 9.3 & 0.4 & 5.4 & 395 & 0.03 & 11. \\
\hline
\end{tabular}

\end{centering}

\caption{95\%C.L.\ reach on the mass of the Dirac $4_{1/2}$ and $5_{1}$ from the various mono-X channels. The excluded number of signal events $S_{95\%}$ and the relative precision $S_{95\%}/B$ are also given,together with the values of the optimal event selection cuts on $\eta_X$ and $p_{T,X}$,where $X$ is the single vector boson.The numbers are shown for different collider energies $\sqrt{s}$ and systematic uncertainties $\epsilon_{\rm sys}$. \label{tab:dirac_tetrapenta_cuts}}
\end{table*}

\begin{figure*}
    \centering
    \includegraphics[width=0.44\textwidth]{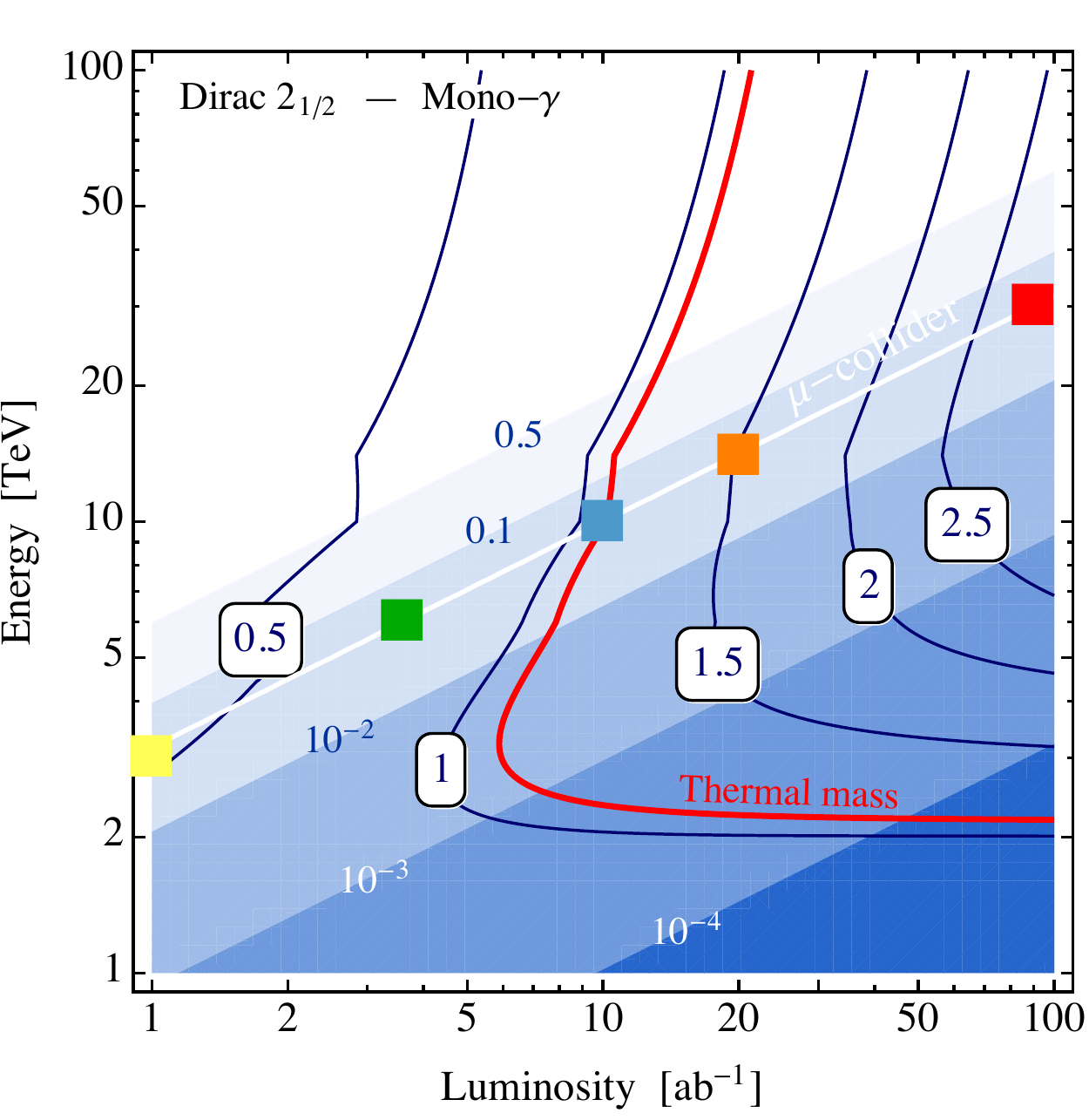}\hfill%
    \includegraphics[width=0.44\textwidth]{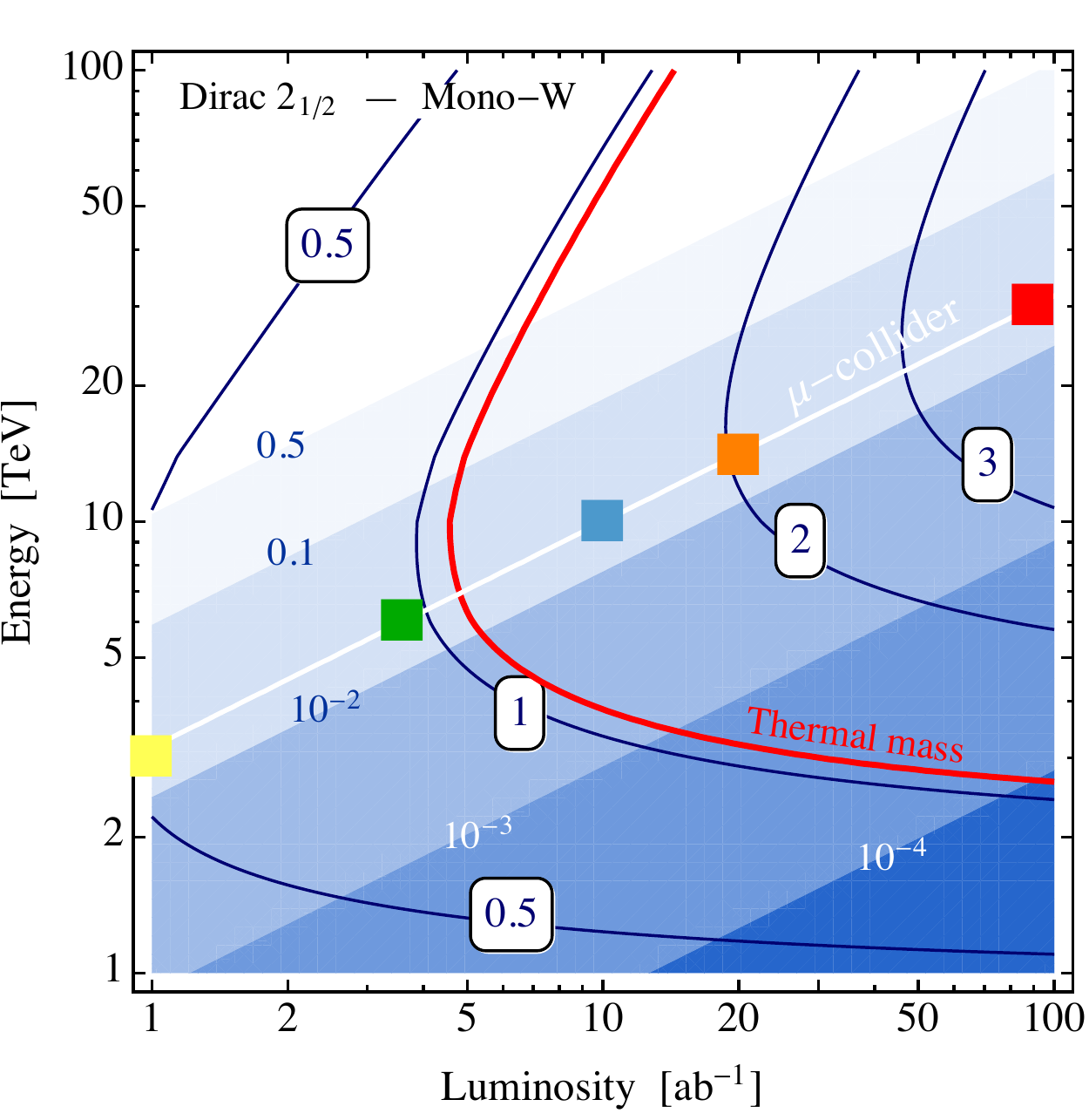}\\
    \includegraphics[width=0.44\textwidth]{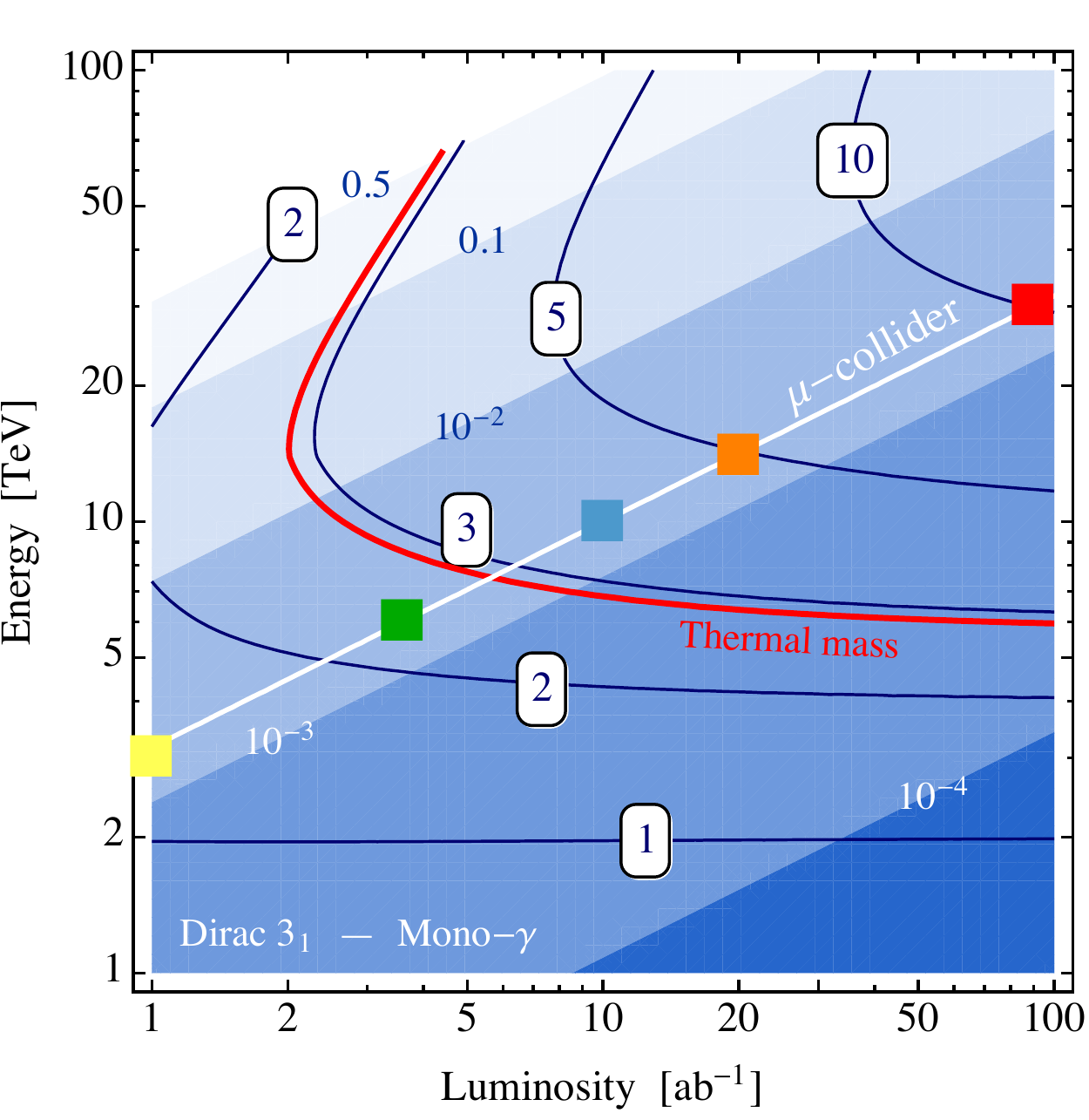}\hfill%
    \includegraphics[width=0.44\textwidth]{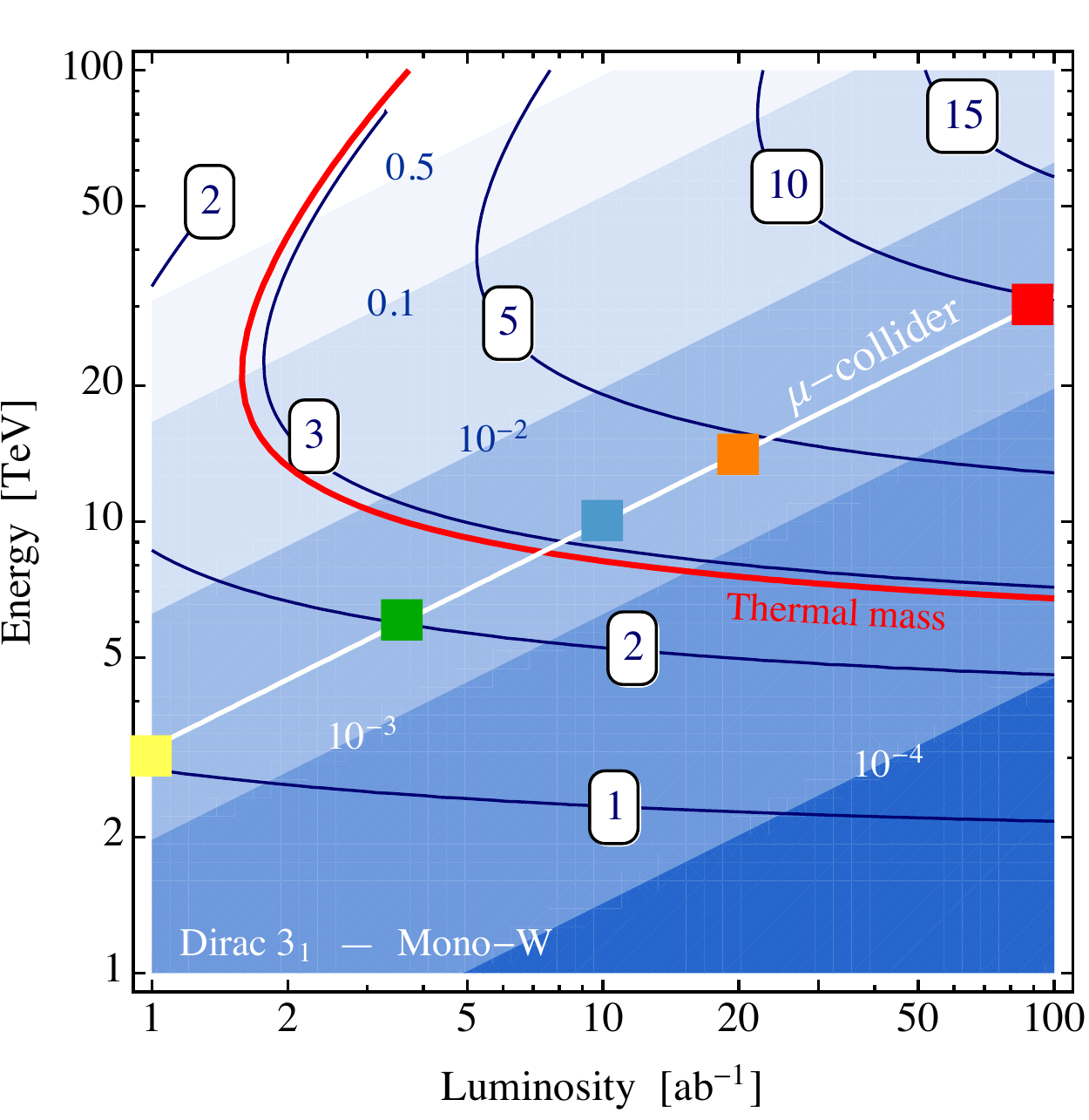}\\
    \includegraphics[width=0.44\textwidth]{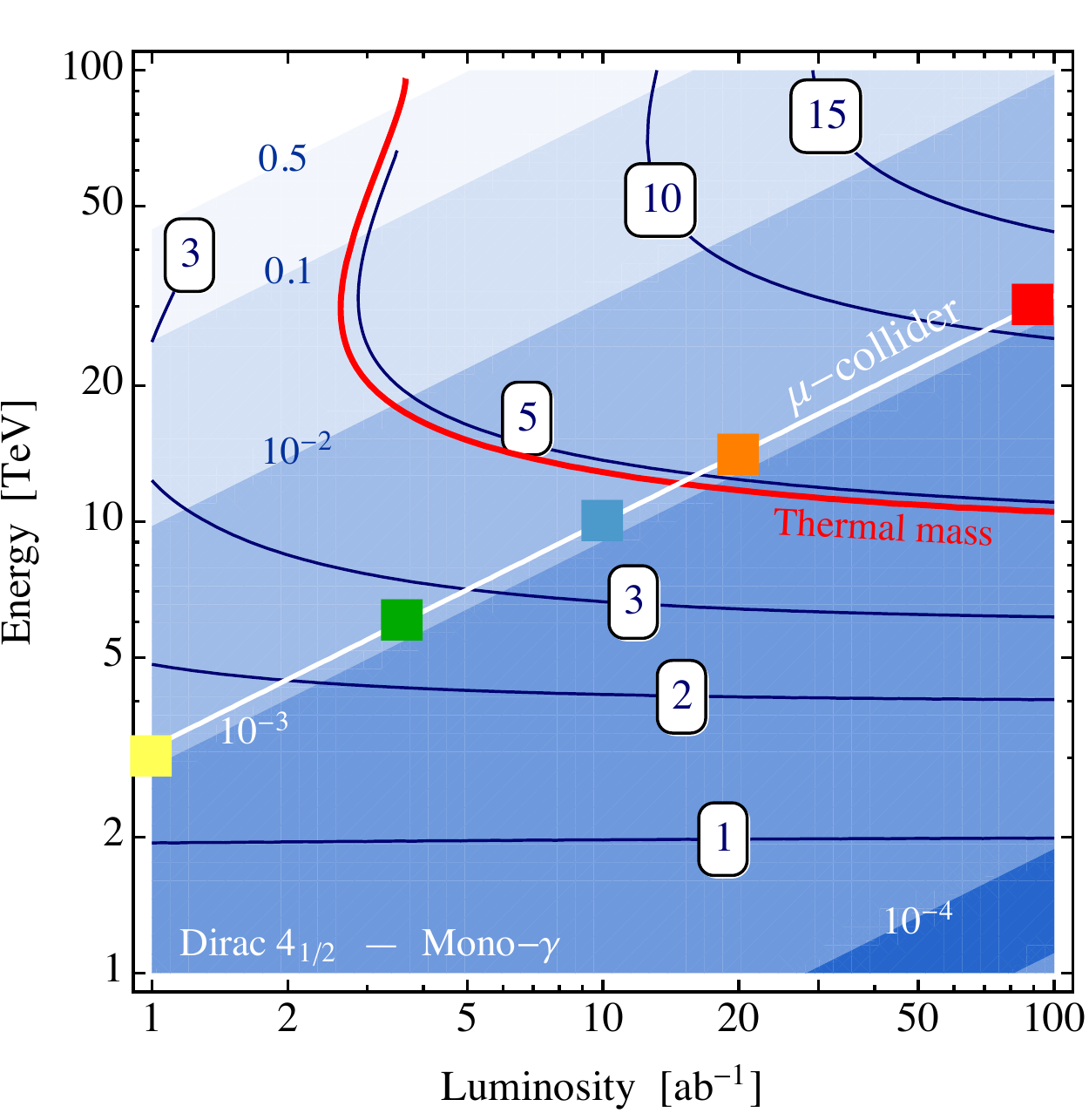}\hfill%
    \includegraphics[width=0.44\textwidth]{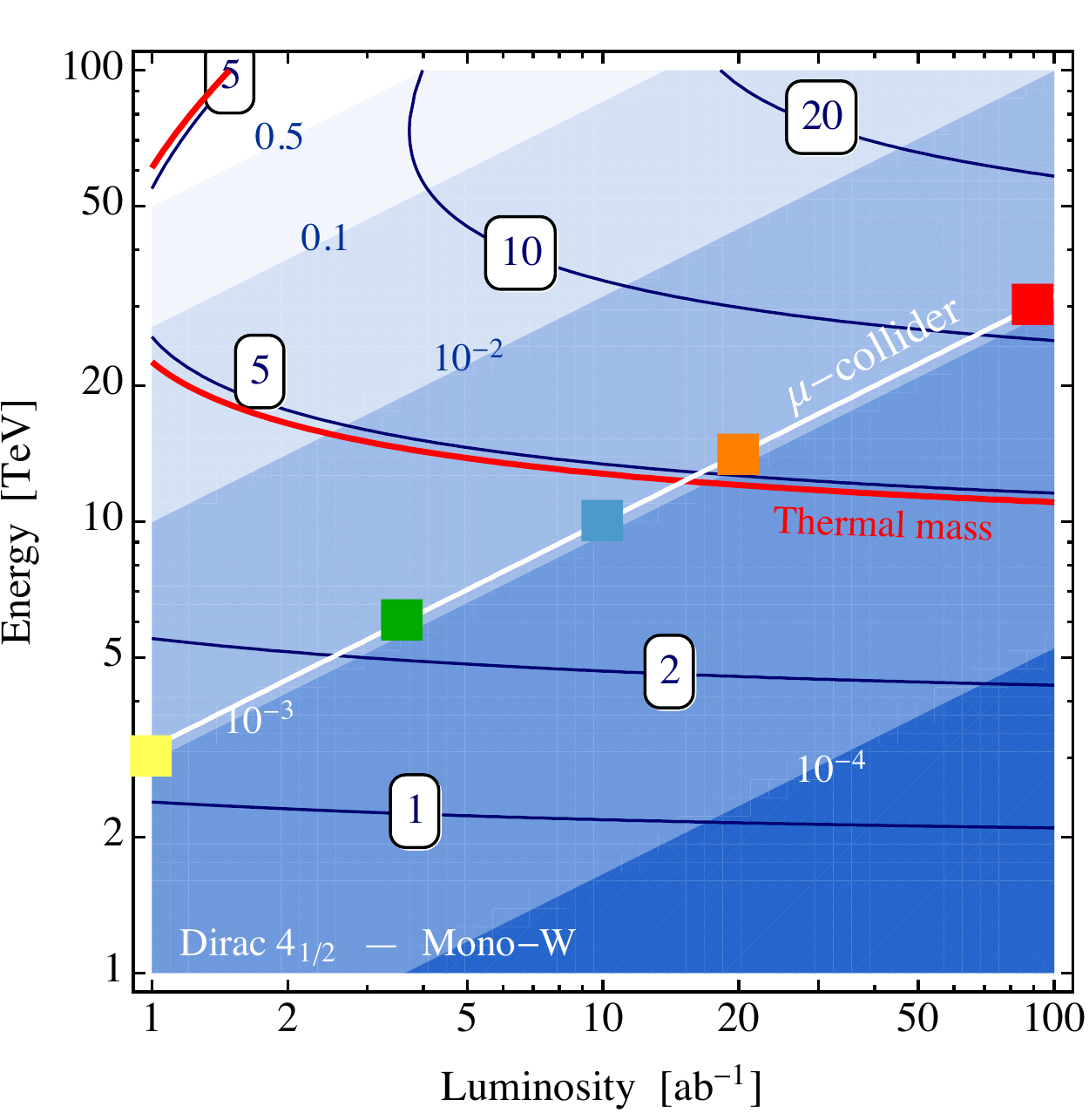}\\
    \caption{Mass reach (in TeV) in the mono-$\gamma$ and mono-$W$ channels as a function of collider center-of-mass energy and luminosity, for Dirac fermion doublets, triplets, and 4-plets (blue lines). The thermal freeze-out mass is shown in red. Blue shades show the expected values of the ratio of the signal rate over background. Systematic uncertainties are set to zero. The muon collider luminosity \eqref{eq:bench} is shown as a white line, with the benchmark values of $\sqrt{s}$ highlighted by the colored squares.}
    \label{fig:energy+lumi+plane}
\end{figure*}

\begin{figure*}
    \centering
    \includegraphics[width=0.46\textwidth]{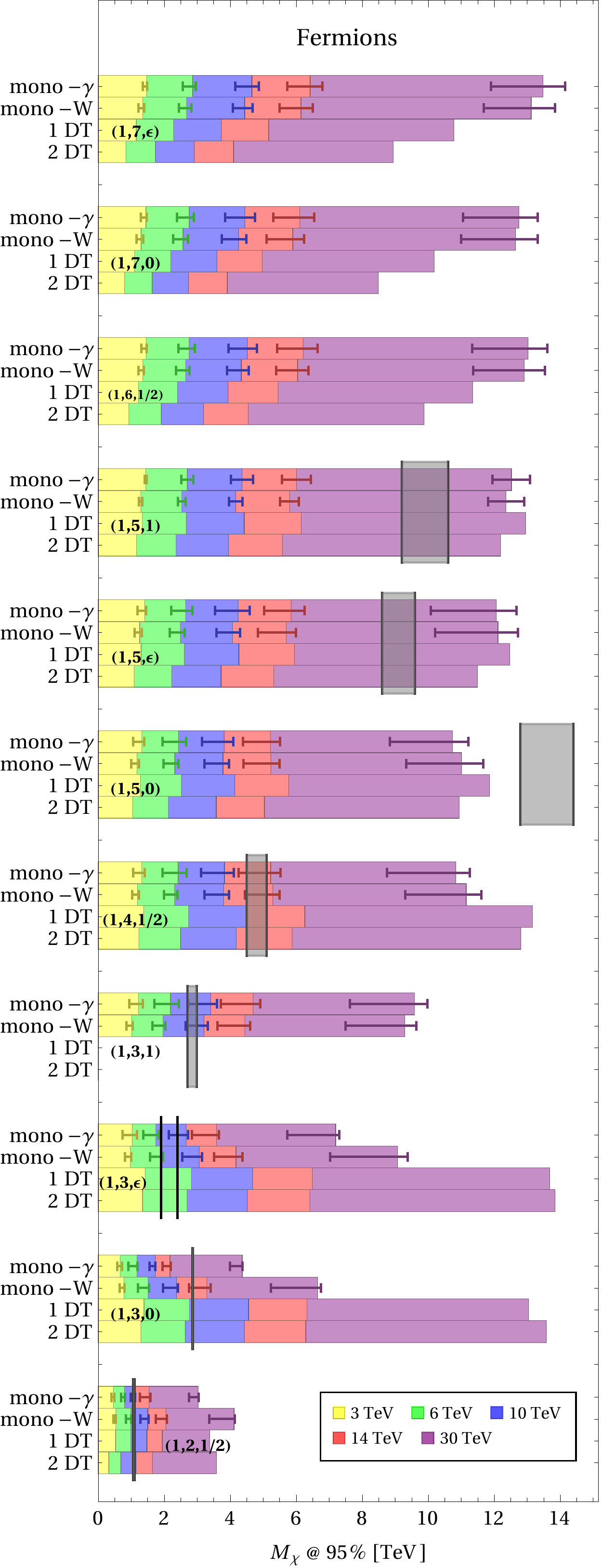}\hfill
        \includegraphics[width=0.46\textwidth]{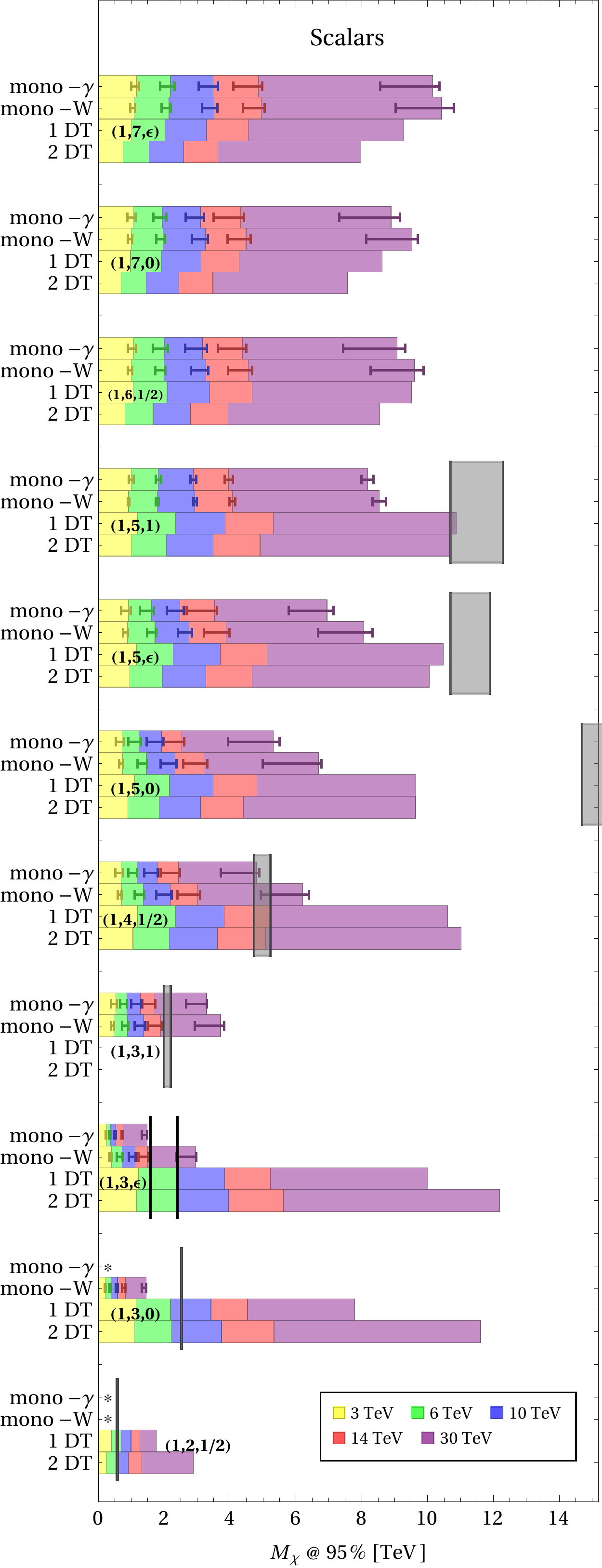}
    \caption{Mass reach in the mono-$\gamma$, mono-$W$ and DT channels for fixed luminosity as per \eqref{eq:bench} at $\sqrt{s}$ 3~TeV (yellow), 6~TeV (green), 10~TeV (light blue), 14~TeV (red), and 30~TeV (purple). In the mono-$W$ and mono-$\gamma$ searches we show an error bar, which covers the range of possible exclusion as the systematic uncertainties are varies from 0 to 1\%. The colored bars are for an intermediate choice of systematics at 0.1\%. 
    %All results consider only the masses range above $0.1\sqrt{s}$, where the Drell-Yan production mode is dominant. 
    Missing bars denoted by an asterisk * correspond to cases where no exclusion can be set in the mass range $M_\chi>0.1\sqrt{s}$. For such cases it is worth considering VBF production modes at the fixed luminosity \eqref{eq:bench} or higher luminosity at potentially smaller $\sqrt{s}$ as illustrated in Fig.~\ref{fig:energy+lumi+plane}}
    \label{fig:summary_all_colliders}
\end{figure*}
\bibliography{wimp}

\end{document}